\documentclass[%
 reprint,
superscriptaddress,
 amsmath,amssymb,
 aps,
]{revtex4-2}

\usepackage{graphicx}
\usepackage{dcolumn}
\usepackage{bm}
\usepackage[mathlines]{lineno}
\usepackage{braket}
\usepackage[dvipsnames]{xcolor}
\usepackage{amsmath,amssymb,amsfonts} 
\usepackage{appendix}
\usepackage{makecell}
\usepackage{subfigure}
\usepackage{float}
\usepackage{bbm}
\usepackage{nicefrac}
\usepackage{wasysym}
\usepackage[normalem]{ulem}

\usepackage{hyperref}
\hypersetup{
	colorlinks=true,
	linkcolor=blue,
	filecolor=blue,
	urlcolor=blue,
	citecolor=blue,
}

\begin{document}

\preprint{APS/123-QED}

\title{Unconventional Altermagnetism in Quasicrystals: A Hyperspatial Projective Construction}

\author{Yiming Li}
\thanks{These authors contributed equally to this work.}
\author{Mingxiang Pan}
\thanks{These authors contributed equally to this work.}
\author{Jun Leng}
\thanks{These authors contributed equally to this work.}
\author{Yuxiao Chen}
\thanks{These authors contributed equally to this work.}
\affiliation{School of Physics, Peking University, Beijing 100871, China}

\author{Huaqing Huang}
\email[Contact author: ]{huaqing.huang@pku.edu.cn}
\affiliation{School of Physics, Peking University, Beijing 100871, China}
\affiliation{Collaborative Innovation Center of Quantum Matter, Beijing 100871, China}
\affiliation{Center for High Energy Physics, Peking University, Beijing 100871, China}

\date{\today}

\begin{abstract}
Altermagnetism, a novel magnetic phase characterized by symmetry-protected, momentum-dependent spin splitting and collinear compensated magnetic moments, has thus far been explored primarily in periodic crystals.
In this Letter, we extend the concept of altermagnetism to quasicrystals---aperiodic systems with long-range order and noncrystallographic rotational symmetries. Using a hyperspatial projection framework, we construct decorated Ammann-Beenker and Penrose quasicrystalline lattices with inequivalent sublattices and investigate a Hubbard model with anisotropic hopping. We demonstrate that interaction-induced N\'eel order on such lattices gives rise to alternating spin-polarized spectral functions that reflect the underlying quasicrystalline symmetry, revealing the emergence of unconventional $g$-wave (octagonal) and $h$-wave (decagonal) altermagnetism. Our symmetry analysis and low-energy effective theory further reveal unconventional altermagnetic spin splitting, which is compatible with quasicrystalline rotational symmetry. Our work shows that quasicrystals provide a fertile ground for realizing unconventional altermagnetic phases beyond crystallographic constraints, offering a platform for novel magnetisms and transport phenomena unique to quasiperiodic systems.
\end{abstract}

\maketitle


\textit{Introduction}---Altermagnetism, a recently identified magnetic phase, exhibits collinear antiferromagnetic order that breaks both time-reversal and spin-rotation symmetries while preserving their combination, leading to distinctive momentum-dependent spin splitting in electronic band structures without net magnetization \cite{PhysRevX.12.031042,PhysRevX.12.040501,sciadv.aaz8809,JunweiLiu2021Natcomm, Song2025NatureRevMater,adfm.202409327}. This unconventional symmetry breaking has enabled striking transport phenomena, such as giant anisotropic magnetoresistance and spontaneous spin polarization, sparking intensive theoretical and experimental interest~\cite{PhysRevLett.126.127701,Naka2019spincurrent,PhysRevX.12.011028,Tanaka_2025, PhysRevX.14.011019, PhysRevLett.133.106701,Kremps2024Nature,Takagi2025Spontaneous, Reichlova2024Observation, PhysRevLett.132.036702, PhysRevLett.134.106802,PhysRevX.15.021083,PhysRevLett.131.256703,pnas.2108924118}. To date, however, studies of altermagnetism have been largely confined to periodic crystalline systems, where the symmetry-protected spin-splitting patterns are directly linked to conventional crystal rotation symmetries. Whether such altermagnetic behavior can be realized and generalized to quasiperiodic systems remains an open question. In this Letter, we address this challenge by extending the concept of altermagnetism to quasicrystals and propose a generic higher-dimensional projection framework for constructing quasicrystalline altermagnetic models.

The possibility of long-range magnetic order in quasicrystals has posed a longstanding puzzle and continues to attract wide theoretical and experimental attention~\cite{PhysRevLett.78.4637, Yokoyama1992newFerrimag, FISHER2000223, Hippert01052008, Deguchi2012critical, stadnik2013magnetic, goldman2013NatMat_binary, Goldman_2014, Labib_2020}. Magnetic orders such as ferromagnetic, ferrimagnetic, and antiferromagnetic phases have been theoretically explored in quasicrystalline lattices using Ising, XY, Heisenberg, and Hubbard-type models \cite{PhysRevB.55.8045,PhysRevLett.90.177205,PhysRevLett.92.047202,PhysRevB.71.115101,PhysRevB.75.212407,PhysRevB.77.104427,jagannathan2009antiferromagnetism,PhysRevB.79.172406, jagannathan2012quasiperiodic,PhysRevLett.115.036403,
PhysRevLett.90.137203,PhysRevLett.93.076407,VEDMEDENKO01012004,Vedmedenko_S0217984905009390,Vedmedenko21022006,
PhysRevLett.86.1331,PhysRevB.107.144202, PhysRevB.96.214402,PhysRevB.102.115125,koga2021antiferromagnetically, PhysRevB.96.214402,coates2021high,sakai2021effect}. In particular, the magnetic Ruderman-Kittel-Kasuya-Yosida interactions in quasiperiodic systems have been shown to support stable magnetic textures~\cite{Thiem_2015,PhysRevB.60.322,PhysRevB.64.024203,MATHO1993905, PhysRevB.68.134402,PREJEAN2004E107,PhysRevB.92.224409,PhysRevMaterials.4.024417,PhysRevB.93.024416}. Experimentally, long-range antiferromagnetic ordering has been observed in rare-earth-based icosahedral quasicrystals, such as $\text{Au}_{56}\text{In}_{28.5}\text{Eu}_{15.5}$ \cite{s41567-025-02858-0}, Au$_{65}$Ga$_{20}$(Tb/Gd)$_{15}$ \cite{jacs.1c09954}, revealing that magnetism in quasiperiodic systems is not only theoretically viable but also experimentally realizable. Motivated by these advances, we raise two key questions: can altermagnetism be extended to quasicrystals while respecting their unique rotational symmetries? Furthermore, are there any unconventional magnetism compatible with noncrystallographic symmetries beyond the known categories of crystalline magnetic orders?

In this Letter, we establish the theoretical foundation for unconventional altermagnetism in quasicrystals by introducing a hyperspace projection framework to construct quasicrystalline altermagnetic models from higher-dimensional parent crystals. Using this approach, we construct an anisotropic Hubbard model based on decorated Ammann-Beenker and Penrose tilings with inequivalent sublattices and uncover alternating spin-polarization in spectral functions, reflecting the underlying quasiperiodic symmetry. These results demonstrate the emergence of novel octagonal $g$-wave and decagonal $h$-wave altermagnetism in quasicrystals. Further supported by symmetry analysis and low-energy effective theory, we identify clear $k$-direction-dependent spin-splitting patterns near the pseudo-Brillouin zone center, that are compatible with quasicrystalline rotational symmetry, offering a clear signature of unconventional altermagnetism in quasicrystals.

\begin{figure*}
\centering
\includegraphics[width=0.8\linewidth]{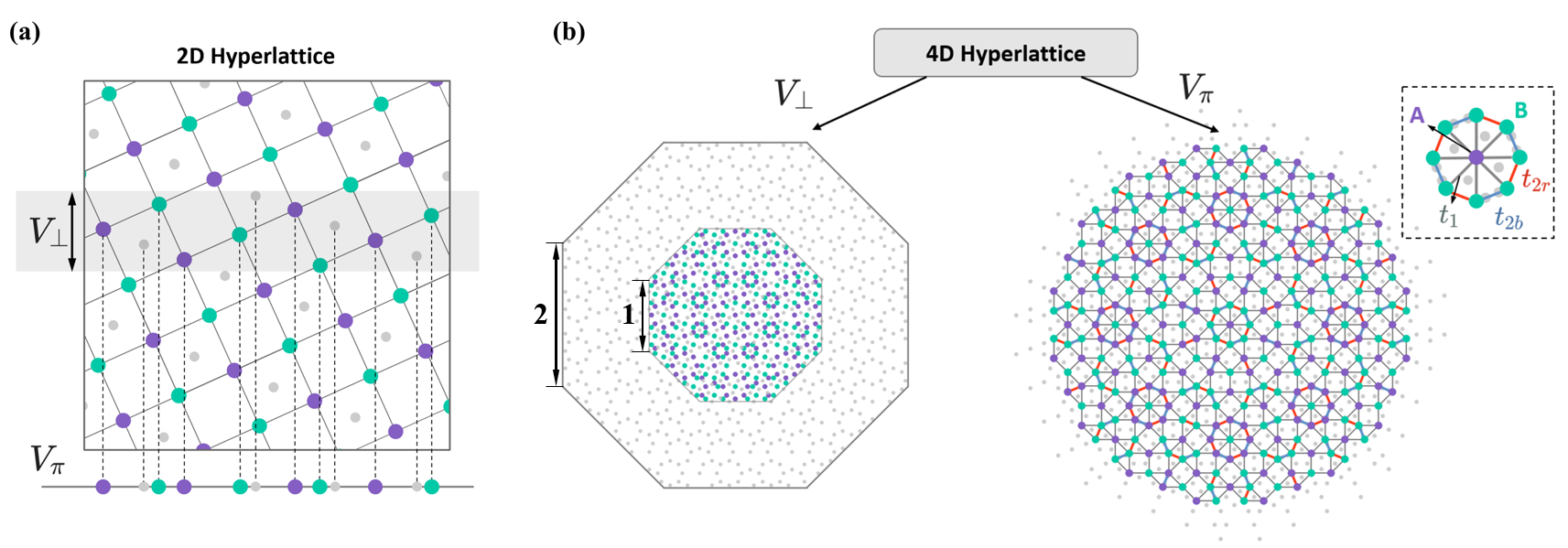}
\caption{\label{fig:lat} Schematic illustration of the hyperspatial projective construction of the quasicrystal lattice for altermagnetism. (a) Construction of 1D bipartite Fibonacci quasicrystal from a 2D square lattice with AB sublattices. (b) Construction of a 2D Ammann-Beenker quasicrystal from the hypercubic lattice with two sublattices in 4D. The projective selection window in perpendicular space ($V_\perp$) is an origin-centered regular octagon with a side length of 1 and 2 for hypercubic sites and decorations, respectively. The 2D Ammann-Beenker quasicrystal exhibits two sublattices in physical space ($V_\parallel$). The inset shows the AB sublattices and different hoppings.}
\end{figure*}

\textit{Hyperspatial projective construction of quasicrystals.}---Since altermagnetism typically requires alternating aligned spins on two inequivalent sublattices, we first construct a quasicrystal lattice with two sublattices from the higher-dimensional periodic lattice by the cut-and-project method, as illustrated in Fig.~\ref{fig:lat}(a). As a preliminary example, we demonstrate how the one-dimensional (1D) bipartite Fibonacci quasicrystal emerges from a projection of a 2D checkboard lattice onto the physical axis $V_\parallel$, selecting only those vertices that fall within a strip of the slope $1/\tau=(\sqrt{5}-1)/2$. The range of this strip projected onto the perpendicular axis $V_\perp$ defines the selection window \cite{RevModPhys.93.045001}.
Extending this approach to 4D, we consider a hypercubic lattice formed by integer combinations of basis vectors $\{\mathbf{e}_1, \mathbf{e}_2, \mathbf{e}_3, \mathbf{e}_4\}$: $X=\sum_{i=1}^4 n_i\mathbf{e}_i$ $(n_i\in \mathbb{Z})$. A checkboard-type AB sublattice distinction is imposed by the parity of $(\sum_in_i\mod 2)$ [see Supplemental Material (SM) for details~\footnote{\label{fn}See Supplemental Material at http://link.aps.org/supplemental/xxx, for more details about the construction of the decorated quasicrystals, the mean-field treatment for the Hubbard model, the derivation of the effective Hamiltonian, the realization of higher-order topology utilizing quasicrystalline altermagnets, which includes Refs.~\cite{wang2022effective,JIANG2014428,PhysRevLett.133.106601,PhysRevD.13.3398,PhysRevB.109.224502,PhysRevLett.124.227001,PhysRevLett.121.096803,PhysRevLett.123.167001}.}]. 
We construct the 2D Ammann-Beenker quasicrystal by projecting this hypercubic lattice into the physical and perpendicular spaces ($V_\pi$ and $V_\perp$), using the projection matrices~\cite{PhysRevB.107.144202,jagannathan2024propertiesammannbeenkertilingsquare}
\begin{equation}
\begin{split}
	\mathcal{S}=
	\begin{pmatrix}
		1 & 0 & \frac{1}{\sqrt{2}} & \frac{1}{\sqrt{2}}\\
		0 & 1 & \frac{-1}{\sqrt{2}} & \frac{1}{\sqrt{2}}
	\end{pmatrix},\quad
	\mathcal{S}_{\perp}=
	\begin{pmatrix}
		-1 & 0 & \frac{1}{\sqrt{2}} & \frac{1}{\sqrt{2}}\\
		0 & 1 & \frac{1}{\sqrt{2}} & \frac{-1}{\sqrt{2}}
	\end{pmatrix}.
\end{split} \label{projection}
\end{equation}
By choosing a regular octagonal selection window of unit side length centered at the origin in $V_\perp$, we generate an Ammann-Beenker quasicrystal with two sublattices [see Fig.~\ref{fig:lat}(b)].
We then determine the hopping vectors in physical space by projecting first and second-order neighbor pairs from the 4D lattice. This yields nearest inter- and intra-sublattice hoppings with amplitudes $t_1$ and $t_2$, which correspond to next-nearest-neighbor (NNN) and nearest-neighbor (NN) connections in the resulting quasicrystal. We denote the associated hopping vectors as $\mathcal{R}_\mathrm{NNN}$ and $\mathcal{R}_\mathrm{NN}$, respectively.

An antiparallel arrangement of spin up and down on two sublattices only realizes the N\'{e}el antiferromagnetic order in the quasicrystal \cite{LIFSHITZ2005219,PhysRevB.102.115125}. A crucial requirement for altermagnetism is that the sublattices must be globally inequivalent. In periodic crystals, this is achieved via differences in local coordination environments~\cite{PhysRevLett.132.263402, PhysRevLett.134.096703}, staggered orbital orders~\cite{PhysRevLett.132.236701}, or external fields~\cite{PhysRevB.109.144421, zhou2024AFM2AM, mazin2023inducedmonolayeraltermagnetismmnpsse3, sun2025multifieldinduced}. In our quasicrystalline system, we induce inequivalence by decorating the lattice with additional nonmagnetic sites. These decorations are introduced via projection of points shifted by $\bm{v}=\frac{1}{2}(1,1,1,1)$ from sublattice B into the physical space (see Sec. I in SM for details~\footnotemark[\value{footnote}]). As a result, we obtain anisotropic and sublattice-dependent intra-sublattice hopping amplitudes $t_{2r}$ and $t_{2b}$, modulated by whether the hopping paths are obstructed by decorations [highlighted in red and blue in Fig.~\ref{fig:lat}(b)]. Consequently, the NN hopping vectors $\mathbf{r}\in \mathcal{R}_\mathrm{NN}$ are partitioned into $\mathcal{R}_\mathrm{NN}^r$ and $\mathcal{R}_\mathrm{NN}^b$. For sublattice A, vectors in $\mathcal{R}_\mathrm{NN}^r$ satisfy $\theta=\pi/8 + n\pi/2$, while those in $\mathcal{R}_\mathrm{NN}^b$ satisfy $\theta=-\pi/8 + n\pi/2$ with $n\in \mathbb{Z}$. The assignment is reversed for sublattice B, reflecting their distinct crystallographic environments in the decorated structure.

\textit{Altermagnetic Hubbard model on 2D quasicrystal lattices.}---
We now consider two-component fermions with spin index $s=\uparrow,\downarrow$ on the quasicrystalline lattice, described by the following altermagnetic Hubbard model:
\begin{equation} \begin{split}
	H = -\sum_{\langle i\alpha, j\beta\rangle}t_{\beta\alpha}(\mathbf{r}_{ji})c_{j\beta }^{\dagger} c_{i\alpha }+U\sum_{i\alpha} n_{i\alpha\uparrow}n_{i\alpha\downarrow},\label{H_hubbard}\\
\end{split} \end{equation}
where $i\alpha$ labels site $i$ on sublattice $\alpha=A,B$, located at position $\mathbf{R}_{i}$. $c_{i\alpha}=(c_{i\alpha \uparrow},c_{i\alpha \downarrow})$ with $c_{i\alpha s}$ ($c^\dagger_{i\alpha s}$) annihilating (creating) an electron with spin $s$ at site $i\alpha$. 
$U$ denotes the on-site Hubbard interaction. The nearest inter-sublattice hopping $t_{AB}=t_1$ is uniform, while nearest intra-sublattice hoppings $t_{\alpha\alpha}=t_{2r}$ or $t_{2b}$ depend on whether the hopping vector $\mathbf{r}_{ji}=\mathbf{R}_{j}-\mathbf{R}_{i}$ belongs to $\mathcal{R}_\mathrm{NN}^r$ or $\mathcal{R}_\mathrm{NN}^b$, respectively.

\begin{figure}
    \centering
    \includegraphics[width=1\linewidth]{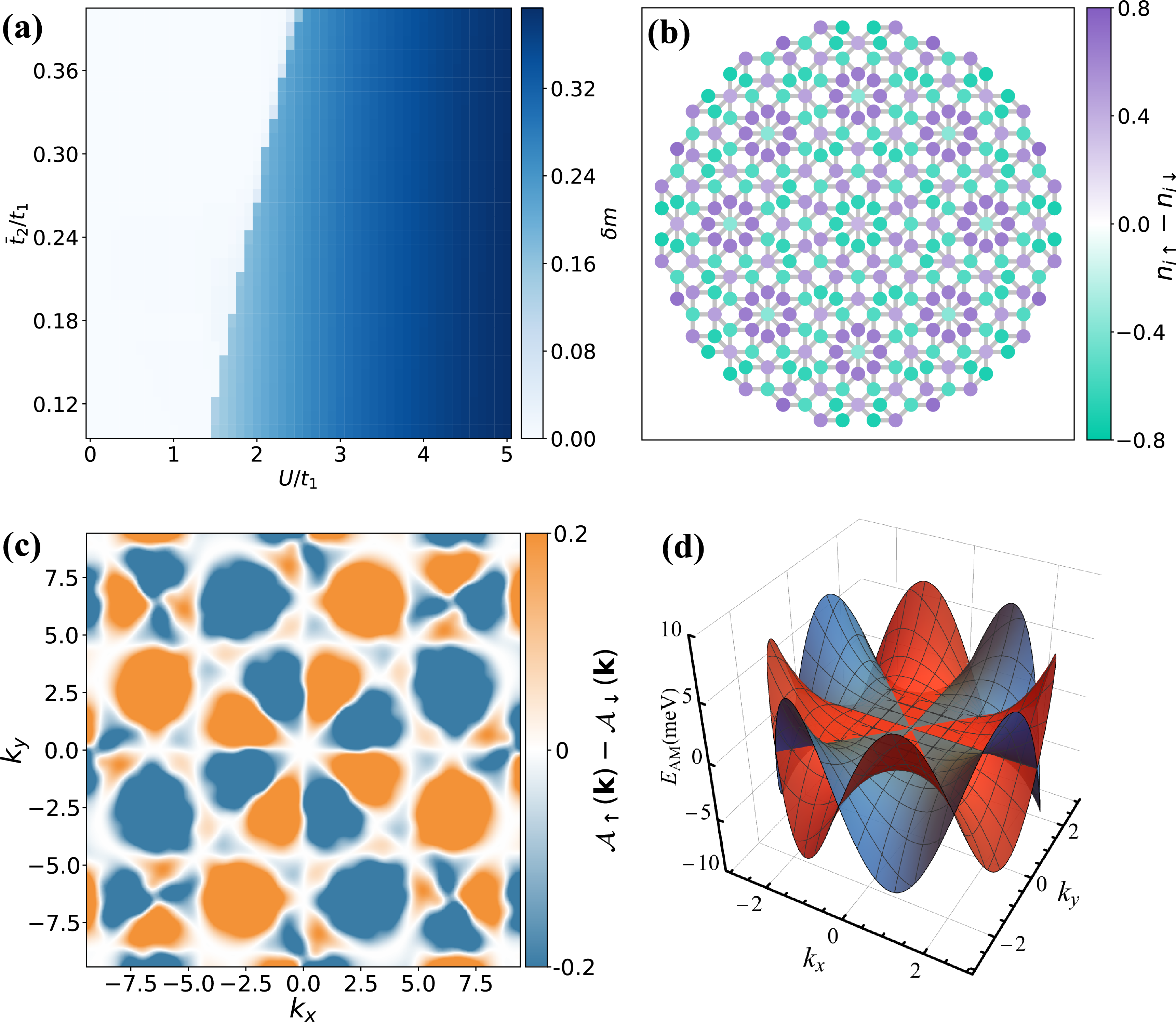}
    \caption{(a) Order parameter $\delta m$ in the plane of [$U,\bar{t}_2=(t_{2r}+t_{2b})/2$] for Eq.~\eqref{H_hubbard} from the self-consistent Hartree-Fock calculations on the Ammann-Beenker quasicrystal lattice with $N=329$ sites {under the open boundary condition}. The parameters used are $t_1=1$ and $\delta_2=0.2$. (b) Real-space magnetization distribution at $U=3$ and $\bar{t}_2=0.3$. (c) Spin-difference spectral function $\mathcal{A}_\uparrow-\mathcal{A}_\downarrow$ calculated from $H_\mathrm{MF}$ in Eq.~\eqref{H_MF}, 
    where parameters are  $|\bm{N}|=3, t_1=1, \bar{t}_2=0, \delta_2=2$, {$\iota=0.01$, and $\omega$ is set at the lowest eigenenergy of the spectrum for better visualization.} (d) Spin-polarized eigenvalues around $\Gamma$ of the $g$-wave altermagnetic term $H_\mathrm{AM}^{(g-wave)}$ in Eq.~\eqref{H_g-wave}.
    }
    \label{fig:phase_diagram}
\end{figure}

We perform a self-consistent Hartree-Fock calculation for the model~\eqref{H_hubbard} at half filling (see Sec. II in SM~\footnotemark[\value{footnote}]). We define the order parameter $\delta m =(1/2N) \sum_{i\alpha}(-1)^\eta\langle n_{i\alpha\uparrow}-n_{i\alpha\downarrow}\rangle$ where $\eta=0,1$ for $\alpha=$ A or B. Assuming the occupation number ansatz $\langle n_{i\alpha s}\rangle=n+\delta m(-1)^{\eta+s}$ at filling $n$, associating $s$ with 0 or 1 for $\uparrow,\downarrow$, we determine the phase diagram shown in Fig.~\ref{fig:phase_diagram}(a), plotting $\delta m$ as a function of $U/t_1$ and $\bar{t}_2/t_1$ for fixed $\delta_2/\bar t_2=0.2$, where $\bar{t}_2, \delta_2=(t_{2r}\pm t_{2b})/2$. Our calculations reveal that the ground state exhibits a nonzero $\delta m$ over a wide parameter range, indicating a robust sublattice N\'{e}el ordering with staggered magnetization $\boldsymbol{M}_\alpha=(-1)^{\eta-1} U\delta m$ in the decorated quasicrystal [see Fig.~\ref{fig:phase_diagram}(b)]. {Notably, the quasicrystalline N\'{e}el ordered phase persists even without the occupation number ansatz at a weak hopping anisotropy ($\delta_2/\bar{t}_2\ll 1$), and remains stable under finite temperature, slight doping, and various disorders, as discussed in the Sec. III of SM~\footnotemark[\value{footnote}].}

{\textit{Spin-resolved spectral function.}}---The mean-field Hamiltonian for our decorated quasicrystalline Hubbard model can be written as
\begin{equation}
\begin{split}
H_\mathrm{MF}=\sum_{i\alpha}(-1)^{\eta-1} \boldsymbol{N}\cdot\boldsymbol{\sigma}c_{i\alpha}^\dagger c_{i\alpha}-\sum_{\langle i\alpha, j\beta\rangle}t_{\beta\alpha}(\mathbf{r}_{ji})c_{j\beta }^{\dagger} c_{i\alpha },\label{H_MF}
\end{split}
\end{equation}
where itinerant electrons experience an effective Kondo-like coupling to the collinear local magnetic moments, which have staggered magnetization $\boldsymbol{N}=\boldsymbol{M}_A-\boldsymbol{M}_B$. To gain intuitive insight into the electronic structure, we calculate the momentum-dependent spin-resolved spectral function, which can be experimentally probed by spin-polarized angle-resolved photoemission spectroscopy (ARPES)~\cite{pnas.1720865115,Rogalev2015FermiState}. The spin-resolved spectral function can be expressed as
\begin{equation}
    \mathcal{A}_s(\omega,\mathbf{k})=\frac{\iota}{\pi}\sum_{n}\frac{|\langle n| \mathbf{k}s\rangle|^2}{(\omega-E_n)^2+\iota^2},
\end{equation}
where $\iota$ is a small spectral broadening parameter, $|n\rangle$ is the eigenstate with energy $E_n$, and $| \mathbf{k}s\rangle \sim\frac{1}{\sqrt{N}} \sum_j e^{i\mathbf{k} \cdot\mathbf{R}_j} |\mathbf{R}_j,s\rangle$ is a set of plane-waves state of spin $s$. Figure~\ref{fig:phase_diagram}(c) shows the spin-difference spectral function $\mathcal{A}_\uparrow-\mathcal{A}_\downarrow$, which exhibits alternating sign changes as a function of momentum direction $\mathbf{k}$, indicating pronounced directional spin polarization.
Both $\mathcal{A}_\uparrow$ and $\mathcal{A}_\downarrow$ preserve the underlying $C_4$ rotational symmetry, but their angular distributions are offset by $\pi/4${, resulting in 8 sign changes of the spin-difference spectral function upon a full $2\pi$ rotation in momentum space}. This reveals an emergent $C_{8z}\mathcal{T}$ symmetry in the spectral function and indicates the formation of a quasicrystalline $g$-wave altermagnetic state~\cite{PhysRevX.12.031042, PhysRevX.12.040501, PhysRevB.110.054406}. We emphasize that this altermagnetic spin polarization vanishes in the absence of decoration ($\delta_2=0$), underscoring the crucial role of symmetry-breaking anisotropic hoppings in realizing quasicrystalline altermagnetism.

\textit{{Symmetry analysis}}.---For quasiperiodic crystals, the spin density field can be expressed as a Fourier sum with a countable infinity of wavevectors
\begin{equation}
    \boldsymbol{S}(\mathbf{r})=\sum_{\mathbf{k}\in \mathcal{L}}\boldsymbol{S}(\mathbf{k})e^{i\mathbf{k}\cdot\mathbf{r}},
\end{equation}
where $\mathcal{L}$ denotes the set of all integral linear combinations of the basis wavevectors. Different from conventional symmetry analysis in crystals, symmetry operations in the magnetic point group $G_M$ of quasiperiodic lattice do not leave the crystal invariant but rather take it into one that contains the same spatial distribution of bounded structures on any scale, which is termed \textit{indistinguishability}~\cite{
lifshitz1996symmetry, RevModPhys.69.1181, PhysRevLett.80.2717, Lifshitz:pz5002}. The indistinguishability requires that every symmetry operation $g$ and antisymmetry operation $g^\prime$ in $G_M$ is associated with a phase function $\Phi_g(\mathbf{k})$ or $\Phi_{g^\prime}(\mathbf{k})$, satisfying \cite{LIFSHITZ2005219},
\begin{equation}
    \boldsymbol{S}(g\mathbf{k})=\begin{cases}
e^{2\pi i\Phi_g(\mathbf{k})} \boldsymbol{S}(\mathbf{k}) &  g\in G_M, \\
e^{2\pi i\Phi_{g^\prime}(\mathbf{k})} [-\boldsymbol{S}(\mathbf{k})] &  g^\prime\in G_M, \\
\end{cases}
\end{equation}
For the undecorated Ammann-Beenker quasicrystal with staggered spin order on AB sublattices, the corresponding gray magnetic point group $8mm1'$~\cite{LIFSHITZ2005219}, enforcing the constraint $\boldsymbol{S}(C_{8z}\mathbf{k})=\boldsymbol{S}(\mathbf{k})$, thus precluding altermagnetism. Decoration breaks this rotational symmetry, reducing the magnetic point group to $8'm'm$, thereby allowing altermagnetism in collinear spin configurations.

To substantiate the origin of this symmetry reduction and resulting altermagnetism, we invoke a superspace crystallographic perspective~\cite{jansseninternational, JANNER197947, janssen2018aperiodic, deloudi2009crystallography}, tracing the symmetry from the 4D parent hyperlattice. In the undecorated hypercubic lattice with opposite spins on AB sublattices, time reversal $\mathcal{T}$ ($1'$), exchanging the two types of spins, can be combined with a unit lattice translation $\tau=\mathbf{e}_i$ to recover the original hyperlattice, prohibiting spin splitting. However, the inclusion of decoration vertices breaks the combined $\tau\mathcal{T}=\{1|\tau;-1\}$ symmetry but preserves the composite operation $\mathcal{T}\tau C_{8z}=\{C_{8z}|\tau;-1\}$, which combines an 8-fold rotation, a translation $\tau$ and time reversal $\mathcal{T}$. Explicitly, the 4D $C_{8z}$ rotation acts on the hypercubic basis as: $\mathrm{e}_1\to-\mathrm{e}_3$, $\mathrm{e}_3\to\mathrm{e}_2$, $\mathrm{e}_2\to-\mathrm{e}_4$, $\mathrm{e}_4\to-\mathrm{e}_1.$ Under this transformation, a vertex $X_B=(n_1, n_2, n_3, n_4)$ in sublattice B and its nearby decoration $X_D=X_B+\bm{v}$, transforms to $\widetilde{X}_B=(-n_4, n_3, -n_1, -n_2)$ and $\widetilde{X}_D=\widetilde{X}_A+\bm{v}$ with $\widetilde{X}_A=(-n_4-1, n_3, -n_1-1, -n_2-1)$ in sublattice A. While $\widetilde{X}_B$ remains within the B sublattice, $\widetilde{X}_D$ does not map to a decorated site, thus breaking the original hyperlattice symmetry. Nonetheless, the combined operation $\{C_{8z}|\tau;-1\}$ maps the decorated hyperlattice onto itself up to a translation (see SM~\footnotemark[\value{footnote}]). This residual composite symmetry is responsible for the emergent $C_{8}\mathcal{T}$ invariance of the spin-resolved spectral function.

\begin{figure}
    \centering
    \includegraphics[width=8cm]{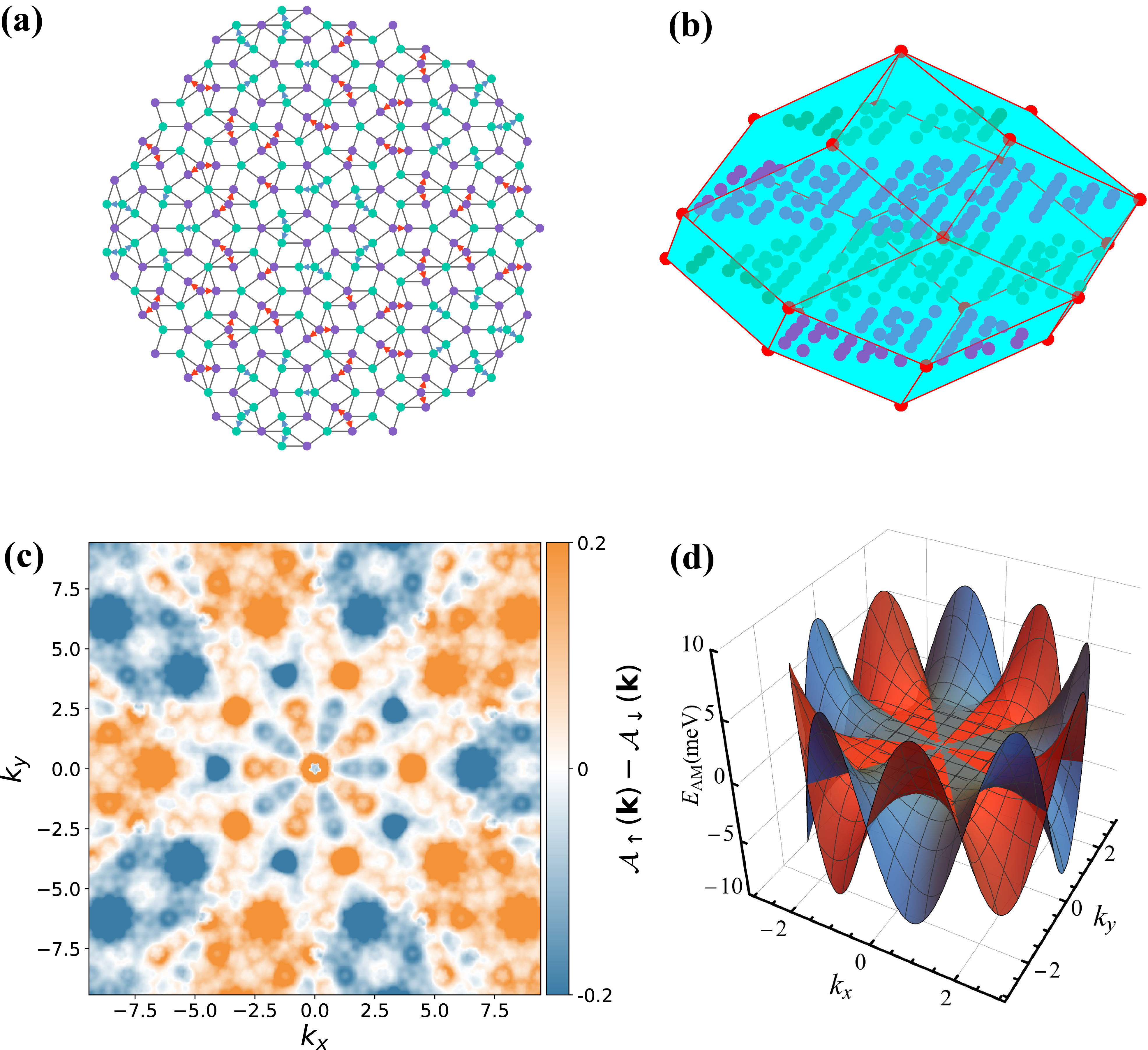}
    \caption{\label{fig:penrose} (a) Penrose tiling lattice with two sublattices (A and B), where the red and blue arrows represent nearest intra-sublattice hopping vectors with positive imaginary part within each sublattice, respectively. (b) Rhombic icosahedron as the selection window in 3D $V_\perp$. (c) Spin-difference spectral function $\mathcal{A}_\uparrow-\mathcal{A}_\downarrow$, 
    where parameters are $|\bm{N}|=3, t_1=1, t_2=0, \delta_2=10$, {$\iota=0.01$, and $\omega$ is set at the lowest eigenenergy of the spectrum for better visualization.} (d) Spin-polarized eigenvalues around $\Gamma$ of the unconventional $h$-wave term $H_\mathrm{AM}^{(h-wave)}$ in Eq.~\eqref{H_h-wave}.}
\end{figure}

\textit{{Low-energy effective theory for altermagnetic spin-splitting.}}---To further elucidate the quasicrystalline altermagnetism, we construct a low-energy effective theory to describe the momentum-dependent spin splitting near the center $\Gamma$ of the pseudo-Brillouin zone. Starting from the mean-field Hamiltonian~\eqref{H_MF}, we apply a Fourier transformation and approximate the hopping term in the long-wavelength limit using the projection method developed in Refs.~\cite{JIANG2014428, wang2022effective}:
\begin{equation}
        H_\mathrm{hop}\approx-\sum_{\langle \alpha \beta\rangle,\mathbf{r}\in\mathcal{R}}\mathcal{P}_{\beta\alpha}(\mathbf{r})t_{\beta\alpha}(\mathbf{r}) e^{i\mathbf{k}\cdot \mathbf{r}}c_{\mathbf{k},\beta}^\dagger c_{\mathbf{k},\alpha}.
\end{equation}
Here, $\mathbf{k}$ is restricted to lie near the $\Gamma$ point, where the plane-wave basis remains approximately orthogonal (see SM~\footnotemark[\value{footnote}]).
The summation runs over all types of intersite hopping vectors $\mathbf{r}\in\mathcal{R} = \mathcal{R}_\mathrm{NNN} \cup\mathcal{R}_\mathrm{NN}$, and $\mathcal{P}_{\beta\alpha}(\mathbf{r})$ denotes the average number of intersite vectors for a given type of hopping, which is also known as the Patterson function~\cite{PhysRev.46.372}. We evaluate $\mathcal{P}_{\beta\alpha}(\mathbf{r})$ numerically by integrating the allowed area of their perpendicular projection in the selection window (see SM~\footnotemark[\value{footnote}]).

Expanding the Hamiltonian around $\Gamma$, we derive the effective Hamiltonian of the decorated Ammann-Beenker quasicrystal:
\begin{equation} \begin{split}
{H}_\Gamma(\mathbf{k}) =&  -\bar{t}_2f_2(\mathbf{k})\lambda_0-t_1f_1(\mathbf{k})\lambda_x -[\boldsymbol{N}\cdot \boldsymbol{\sigma}+\delta_2\phi_2(\mathbf{k})] \lambda_z,
\end{split}
\end{equation}
where $\lambda_i$ are Pauli matrices acting on the sublattice space, and $f_{1}(\mathbf{k})=4-k^2+{k^4}/{16}$,
    $f_{2}(\mathbf{k})=2\sqrt{2}\gamma-\frac{2\sqrt{2}\gamma^3}{1+\gamma^2}k^2+\frac{\gamma^5}{\sqrt{2}(1+\gamma^2)^2}k^4$, 	
    $\phi_{2}(\mathbf{k})=\frac{2\sqrt{2}\gamma^5}{3(1+\gamma^2)^2}k_xk_y(k_x^2-k_y^2)$, where $\gamma=\sqrt{2}-1$.
Importantly, its bands exhibit alternating spin splitting consistent with $C_{8z}\mathcal{T}$ symmetry, as observed in the spin-resolved spectral function [see Fig.~\ref{fig:phase_diagram}(c)] and further confirm the quasicrystalline altermagnetism.
By directly extracting the spin splitting effect, we obtain the effective altermagnetic term:
\begin{equation} \begin{split}
    H^{(g-wave)}_\mathrm{AM}=(\boldsymbol{J}\cdot \boldsymbol{\sigma}) k_xk_y(k_x^2-k_y^2),\label{H_g-wave}
\end{split} \end{equation}
where $\boldsymbol{J}\sim (t_{2r}-t_{2b})\boldsymbol{N}$ quantifies the altermagnetic strength. As shown in Fig.~\ref{fig:phase_diagram}(d), this term captures the octagonal symmetric spin splitting in the Ammann-Beenker quasicrystal and encapsulates the essential physics of what we identify as quasicrystalline $g$-wave altermagnetism---a novel form of altermagnetic ordering not dictated by conventional crystalline symmetry classifications~\cite{PhysRevB.110.054406}.

\textit{{Unconventional $h$-wave magnetism in Penrose tiling.}}---Beyond octagonal quasicrystals, our construction scheme naturally extends to the decagonal Penrose quasicrystal. As shown in Fig.~\ref{fig:penrose}(a,b), the Penrose quasicrystal is generated by projecting a 5D hypercubic lattice with AB sublattices into 2D physical space, with a selection window defined by a rhombic icosahedron with 22 vertices in the 3D perpendicular space~\cite{DEBRUIJN198153}. To model altermagnetism in this system, we introduce a Hubbard model with uniform nearest inter-sublattice hopping $t_1$ but with complex nearest intra-sublattice hopping, adopting $t_{2+/2-}=t_2\pm i\delta_2$ for hopping vector $\mathbf{r}$ with its polar angle $\theta=2n\pi/5$ or $\pi+2n\pi/5$ ($n\in\mathbb{Z}$) for sublattice A. The hopping assignment is reversed for sublattice B, distinguishing the two sublattices and ensuring the Hermiticity of the system, while remaining consistent with the pentagonal symmetry locally. These complex hoppings, reminiscent of those in the famous Haldane model \cite{PhysRevLett.61.2015}, could be feasible via magnetic flux, spin-orbit coupling, or synthetic gauge fields.

The magnetic point group of the decorated Penrose quasicrystal is $5m1'$, indicating that the spin density satisfies $\bm{S}(C_{5z}\mathbf{k})=\bm{S}(\mathbf{k})$ and $\bm{S}(-\mathbf{k})=-\bm{S}(\mathbf{k})$. As shown in Fig.~\ref{fig:penrose}(c), the spectral function exhibits an alternating spin polarization pattern with an approximate $C_{5z}\mathcal{T}$ symmetry. 
Because fivefold rotations are forbidden in 2D by the crystallographic restriction theorem, the resulting altermagnetism has no crystalline counterpart. We thus identify it as an unconventional $h$-wave magnetism, which is a new form of odd-parity magnetism extending beyond the $p$-wave and $f$-wave ones \cite{PhysRevX.12.031042, PhysRevX.12.040501, PhysRevLett.133.236703, Song2025p-wave, jungwirth2024altermagnetsbeyondnodalmagneticallyordered, hellenes2024pwavemagnets,
yu2025oddparitymagnetismdrivenantiferromagnetic, jungwirth2025altermagnetismunconventionalspinorderedphase, yu2025oddparitymagnetismdrivenantiferromagnetic}.

Employing a similar derivation, we derive a low-energy effective Hamiltonian for the decorated Penrose quasicrystal (see SM~\footnotemark[\value{footnote}]), and extract the altermagnetic contribution as
\begin{equation} \begin{split}
    H^{(h-wave)}_\mathrm{AM}=(\boldsymbol{J}\cdot \boldsymbol{\sigma}) (5k_xk_y^4 - 10k_x^3k_y^2 + k_x^5),\label{H_h-wave}
\end{split} \end{equation}
which describes the alternating spin splitting around $\Gamma$, satisfies the rotational symmetry of the magnetic Penrose quasicrystal, as shown in Fig.~\ref{fig:penrose}(d). This further verifies the realization of unconventional $h$-wave magnetism in the Penrose quasicrystal.

\textit{{Discussion and summary}}.---Quasicrystalline altermagnetism manifests as spin-polarized Fermi surfaces rotational anisotropies forbidden in periodic lattices. This opens the door to a wide range of novel spin-resolved phenomena, such as highly anisotropic spin conductivity with 8- or 10-fold anisotropy~\cite{PhysRevLett.133.236703}, nonlinear or nonreciprocal spin-current generations \cite{PhysRevB.111.125420}, and higher-order quasicrystalline topological states \cite{wang2022effective, PhysRevLett.123.196401, PhysRevLett.124.036803}. For example, as we show in SM~\footnotemark[\value{footnote}], proximity coupling between quasicrystalline altermagnets and topological insulators or superconductors leads to higher-order topological corner states or Majorana zero modes whose spatial arrangement matches the underlying quasicrystalline symmetry, offering a clear signature of quasicrystalline altermagnetism and its topological consequences. {Experimentally, the proposed mechanism for altermagnetism in quasicrystals may be realized in twisted heterostructures formed from an antiferromagnetic quasicrystal and a nonmagnetic quasicrystal (or an incommensurate bilayer crystal)~\cite{PhysRevLett.59.1010, PhysRevB.103.045408, PhysRevB.106.075121, Jin:20, PhysRevB.111.195161, PhysRevB.107.235425, PhysRevB.107.144202, peng2018effect}, where the twist-induced quasiperiodicity leads to the required modulation of electronic hopping. Related physics can also be explored in artificial platforms such as optical quasicrystals in ultracold-atom systems~\cite{xu2025neutral, Jagannathan_2013, Corcovilos:19,cryst6100124,mazurenko2017cold, doi:10.1126/science.1239873} and quasicrystalline metal-organic frameworks~\cite{urgel2016quasicrystallinity, SHI20242464}. Further details are provided in the Supplemental Material [68].}

In summary, we have proposed a generic strategy to realize unconventional altermagnetism in quasicrystals using a hyperspatial projection construction. We have demonstrated the emergence of quasicrystalline altermagnetism, which exhibits unconventional $g$-wave and $h$-wave altermagnetism in decorated Ammann-Beenker and Penrose tilings, respectively. Due to the inherent self-similarity and scale equivalence of quasicrystals, such altermagnetic features can emerge across scales, provided the effective ``sub-cluster"-dependent spin and hopping distributions persist. 
Our work greatly extends the scope of altermagnetism to symmetry regimes inaccessible in crystals and suggests broader possibilities for engineering more exotic magnetism in quasicrystalline materials by including noncollinear and higher multipolar spin textures.

\begin{acknowledgments}
This work is supported by the National Key R\&D Program of China (Grant No. 2021YFA1401600), the National Natural Science Foundation of China (Grant No. 12474056), and the 2022 basic discipline top-notch students training program 2.0 research project (Grant No. 20222005). The work was carried out at the National Supercomputer Center in Tianjin, and the calculations were performed on Tianhe new generation supercomputer. The high-performance computing platform of Peking University supported the computational resources.

\textit{Note added:} During the final stage of preparing our manuscript, we became aware of a recent work on arXiv \cite{chen2025quasicrystallinealtermagnetism}, which deals with similar topics but uses a distinct model \cite{dornellas2025altermagnetismcrystalsymmetry}.

\end{acknowledgments}


\begin{widetext}

\appendix

\section{Hyperspatial projective construction of altermagnetic quasicrystals}

\subsection{Lattice construction and Hopping Geometry for Quasicrystals \label{sec:construct}}
In this section, we will show that both the vertices and edges on the Ammann-Beenker and Penrose tilings, which correspond to the orbital sites and hoppings in the 2D quasicrystal lattice, can be generated from the hypercubic lattice by the cut-and-project method.

\subsubsection{Ammann-Beenker tiling}
Ammann-Beenker tiling (ABT) lattice can be constructed by projecting a 4D hypercubic lattice. Let $\{\mathbf{e}_1,\mathbf{e}_2,\mathbf{e}_3,\mathbf{e}_4\}$ be a basis set of the 4D hyperspace, and define the hypercubic lattice as the set of their integer combinations, i.e., $\mathbf{R}=\sum_{i=1}^4 n_i\mathbf{e}_i$ ($n_i\in \mathbb{Z}$). We artificially classify the vertices of the hypercubic lattice into two groups: A (B) sublattices contain only those combinations in which the sum of the four integers ($\sum_{i=1}^4 n_i$) is odd (even). By doing so, we actually arrange the two sublattices alternatively in the hypercubic lattice in analogy with the checkboard lattice in 2D and the cubic rocksalt structure in 3D.

We construct the conventional ABT quasicrystal by projecting a 4D hypercubic lattice with sublattices onto two orthogonal subspaces: the physical and perpendicular space ($\pi$ and $\perp$), as shown in Fig.~\ref{fig:ABT_proj}. The basis vectors of the two subspaces are expressed as
\begin{equation}
\begin{split}
	\mathcal{S}\begin{pmatrix}
		\mathbf{e}_1,\mathbf{e}_2,\mathbf{e}_3,\mathbf{e}_4
	\end{pmatrix} &=
	\begin{pmatrix}
		1 & 0 & \frac{1}{\sqrt{2}} & \frac{1}{\sqrt{2}}\\
		0 & 1 & \frac{-1}{\sqrt{2}} & \frac{1}{\sqrt{2}}
	\end{pmatrix},\\
	\mathcal{S}_{\perp}\begin{pmatrix}
		\mathbf{e}_1,\mathbf{e}_2,\mathbf{e}_3,\mathbf{e}_4
	\end{pmatrix} &=
	\begin{pmatrix}
		-1 & 0 & \frac{1}{\sqrt{2}} & \frac{1}{\sqrt{2}}\\
		0 & 1 & \frac{1}{\sqrt{2}} & \frac{-1}{\sqrt{2}}
	\end{pmatrix}.
\end{split} \label{projection}
\end{equation}
where $\mathcal{S}$ and $\mathcal{S}_\perp$ are projection operators into the physical and perpendicular spaces, respectively.

The specific pattern of the quasicrystalline tiling is determined by the selection window (or the acceptance window) in the perpendicular space. We choose the selection window as an origin-centered regular octagon with a side length of 1 in the perpendicular space. For vertices whose perpendicular projection falls in the selection window, we accept their projection on the physical space. Otherwise, we reject the physical projection. Thus,  we arrive at the conventional ABT quasicrystal with two sublattices.

Next, we discuss the hopping vectors that are projected from the hyperspace. We divide the neighboring pairs in the hyperspace into several orders. For the first-order neighbors, they are connected by eight vectors $\pm\{ \mathbf{h}_1^{(1)}, \mathbf{h}_2^{(1)}, \mathbf{h}_3^{(1)},\mathbf{h}_4^{(1)}\}= \pm\{(1,0,0,0),(0,1,0,0),(0,0,1,0),(0,0,0,1)\}$, which coincide with the basis vectors $\pm\{\mathbf{e}_i\}$ of the hypercubic lattice. They can be projected onto the physical and perpendicular space, as displayed in Eq.~\eqref{projection}. In ABT, they correspond to the next-nearest-neighbor (NNN) hopping. Consequently, we define $\mathcal{R}_\mathrm{NNN}=\mathcal{S}\{\pm\mathbf{e}_i\}$.
It is worthwhile noting that the first-order neighboring pairs connecting AB sulattice sites in the hypercubic lattice, which correspond to the nearest inter-sublattice hopping (A-B or B-A) in the real-space quasicrystal lattice (i.e., the edges of squares or rhombuses), as shown in Fig.~\ref{fig:ABT_proj}(b).

For the second-order neighboring pairs, there are 24 different categories, which can be notably divided into three classes according to their projection in the physical space:
\begin{itemize}
	\item Short diagonals of rhombi:\\
	 $\pm\{(0,1,1,0),(1,0,-1,0),(-1,0,0,1), (0,1,0,-1)\}$;
	\item Long diagonals of rhombi:\\
	 $\pm\{(1,0,0,1),(0,1,0,1),(0,1,-1,0),(-1,0,-1,0)\}$;
	\item Diagonals of squares:\\
	 $\pm\{(1,1,0,0),(0,0,-1,1),(-1,1,0,0),(0,0,-1,-1)\}$;
\end{itemize}
which are in hypercubic coordinates. Here we only focus on the first class, which corresponds to the nearest neighbor (NN) hopping in ABT, and project them into physical and perpendicular subspaces, which can be expressed as:
\begin{equation} \begin{split}
	\mathcal{S}\begin{pmatrix}
		\mathbf{h}^{(2)}_1,\mathbf{h}^{(2)}_2,\mathbf{h}^{(2)}_3,\mathbf{h}^{(2)}_4
	\end{pmatrix} &=
	\begin{pmatrix}
		\frac{1}{\sqrt{2}} & \frac{\gamma}{\sqrt{2}} & -\frac{\gamma}{\sqrt{2}} & -\frac{1}{\sqrt{2}} \\
		\frac{\gamma}{\sqrt{2}} & \frac{1}{\sqrt{2}} & \frac{1}{\sqrt{2}} & \frac{\gamma}{\sqrt{2}}
	\end{pmatrix},\\
	\mathcal{S}_{\perp}\begin{pmatrix}
		\mathbf{h}^{(2)}_1,\mathbf{h}^{(2)}_2,\mathbf{h}^{(2)}_3,\mathbf{h}^{(2)}_4
	\end{pmatrix} &=
	\begin{pmatrix}
		\frac{1}{\sqrt{2}} & -\frac{1}{\sqrt{2}\gamma} & \frac{1}{\sqrt{2}\gamma} & -\frac{1}{\sqrt{2}}\\
		\frac{1}{\sqrt{2}\gamma} & -\frac{1}{\sqrt{2}} & -\frac{1}{\sqrt{2}} & \frac{1}{\sqrt{2}\gamma}
	\end{pmatrix},
\end{split} \end{equation}
where $\gamma=\sqrt{2}-1$. We define $\mathcal{R}_\mathrm{NN}=\mathcal{S}\{\pm \mathbf{h}_i^{(2)}\}$. In ABT, they correspond to the nearest intra-sublattice hopping (A-A or B-B) in the real-space quasicrystal lattice (i.e., the short diagonals of the rhombuses). For convenience, we further divide NN neighboring vectors to two subgroups: $\mathcal{R}_\mathrm{NN}=\mathcal{R}_\mathrm{NN}^1 \cup \mathcal{R}_\mathrm{NN}^2$. By defining $\theta$ as the polar angle of a vector to the $x$ axis, we classify the vector with $\theta=\pi/8\pm n\pi/2$ and $-\pi/8\pm n\pi/2$ for B-B with a integer $n$ into $\mathcal{R}_\mathrm{NN}^1$ and $\mathcal{R}_\mathrm{NN}^1$, respectively [see Fig.~\ref{fig:ABT_proj}(c)].

\begin{figure}
    \centering
        \includegraphics[width=1\linewidth]{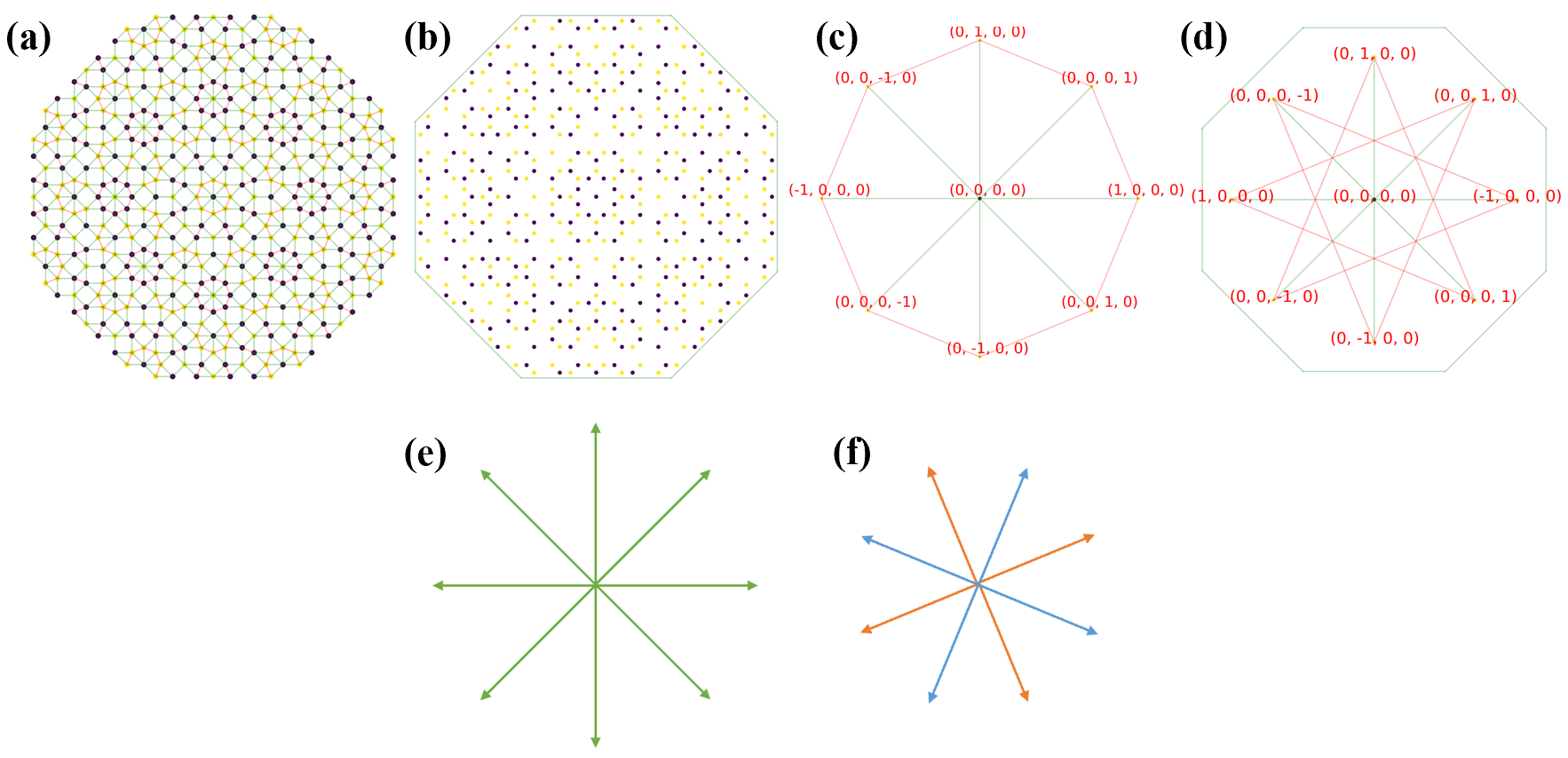}
    \caption{\label{fig:ABT_proj} (a) (b) Initial ABT in (a) the physical and (b) perpendicular space. (c) (d)
	The 1st order hopping (green lines) and the first class of the 2nd order hopping (red lines) at the center of ABT in (c) the physical and (d) perpendicular space. The octagon represents the selection window.
    (e) All possible 1st hopping vectors $\mathcal{R}_\mathrm{NNN}$ for vertices in the physical space.
    (f) All possible 2nd hopping vectors $\mathcal{R}_\mathrm{NN}=\mathcal{R}_\mathrm{NN}^1\cup\mathcal{R}_\mathrm{NN}^2$
    for vertices in physical space, where orange vectors $\mathcal{R}_\mathrm{NN}^1$
    with polar angles $\theta=\pi/8\pm n\pi /2$, while blue vectors $\mathcal{R}_\mathrm{NN}^2$
    with $\theta=-\pi/8\pm n\pi /2$ ($n\in\mathbb{Z}$).}
\end{figure}

So far, we have derived the mapping for 1st and 2nd neighboring pairs between physical and perpendicular spaces. Given that for any vertex in the ABT tiling, its corresponding perpendicular projection falls in the selection window, we can easily estimate whether a neighboring pair can be projected to the physical space, by checking whether the perpendicular projection of both the starting and ending sites falls in the selection window simultaneously. Namely, for any vertex in ABT, whether a hopping vector in physical space started from this vertex exists or not can be diagnosed by examining its coordinates and the hopping vector in perpendicular space.

\subsubsection{Penrose tiling}
Similarly, the Penrose tiling can be constructed by projecting a 5D hypercubic lattice. Let $\{\mathbf{e}_1,\mathbf{e}_2,\mathbf{e}_3,\mathbf{e}_4,\mathbf{e}_5\}$ be the basis vectors of the hypercubic lattice
expressed as:
\begin{equation} \begin{split}
	\mathcal{S}\begin{pmatrix}
		\mathbf{e}_1,\mathbf{e}_2,\mathbf{e}_3,\mathbf{e}_4,\mathbf{e}_5
	\end{pmatrix} &= \frac{1}{\sqrt{10}}
	\begin{pmatrix}
		2 &-\sigma &-\tau &-\tau & -\sigma\\
		0 &\beta\tau & \beta &-\beta &-\beta\tau
	\end{pmatrix},\\
	\mathcal{S}_{\perp}\begin{pmatrix}
		\mathbf{e}_1,\mathbf{e}_2,\mathbf{e}_3,\mathbf{e}_4,\mathbf{e}_5
	\end{pmatrix} &= \frac{1}{\sqrt{10}}
	\begin{pmatrix}
		2 &-\tau &-\sigma &-\sigma &-\tau \\
		0 & \beta & -\beta\tau &\beta\tau & -\beta\\
		\sqrt{2} &\sqrt{2}&\sqrt{2}&\sqrt{2}&\sqrt{2}
	\end{pmatrix},
\end{split} \end{equation}
where $\tau=\frac{1+\sqrt{5}}{2}$, $\sigma=\frac{1-\sqrt{5}}{2}$, and $\beta=\sqrt{3-\tau}$. Here, these basis vectors (and their opposite) also represent 1st hopping vectors in the 5D hypercubic lattice, which correspond to the NNN hoppings in Penrose tiling when projected to the physical space.

For Penrose tiling, the acceptance window we choose is a 3D convex hull of 22 points, expressed by cylindrical coordinates as:
$(0,0,\frac{3}{\sqrt{5}})$, $(r_2,\theta_1,\frac{2}{\sqrt{5}})$, $(r_1,\theta_2,\frac{1}{\sqrt{5}})$, $(r_1,\theta_1,\frac{0}{\sqrt{5}})$,
$(r_2,\theta_2,\frac{-1}{\sqrt{5}})$, $(0,0,\frac{-2}{\sqrt{5}})$,
in which $r_1=\frac{1+\sqrt{5}}{\sqrt{10}},r_2=\sqrt{2/5}$, and $\theta_1=\{\pm\frac{\pi}{5},\pm\frac{3\pi}{5},\pi\},\theta_2=\{0,\pm\frac{2\pi}{5},\pm\frac{4\pi}{5}\}$ are sets of degrees.
This makes vertices in perpendicular space arranged on 4 planes $z=-\frac{1}{\sqrt{5}},0,\frac{1}{\sqrt{5}},\frac{2}{\sqrt{5}}$, shown as Fig.~\ref{fig:penrose_proj}.

\begin{figure}
    \centering
        \includegraphics[width=1\linewidth]{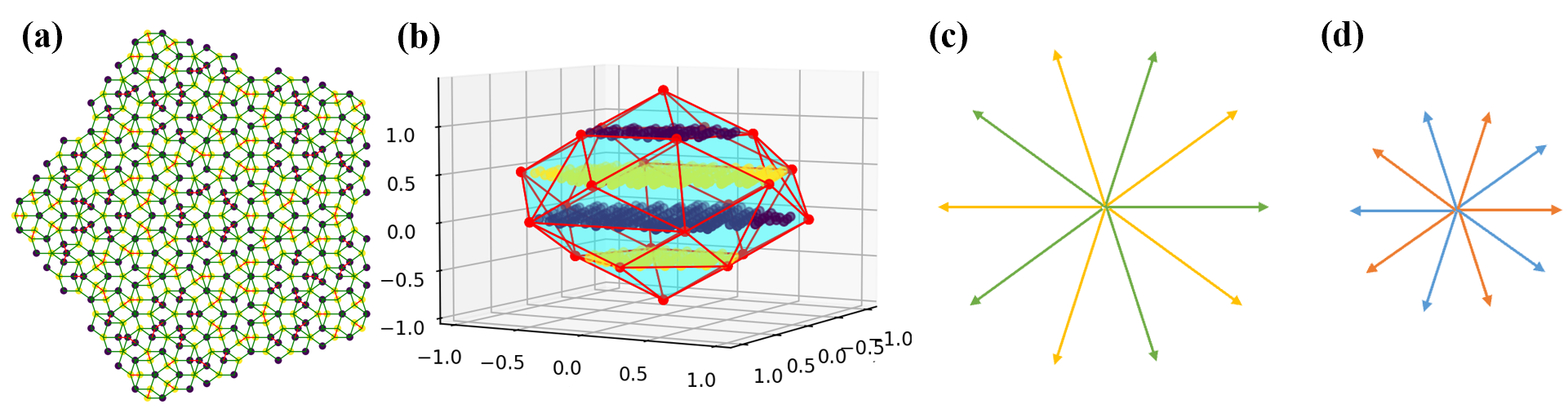}
    \caption{\label{fig:penrose_proj} (a) (b) Initial Penrose tiling in (a) the physical space and (b) the perpendicular space. 
	(c) All possible 1st-order hopping vectors for vertices in the physical space. (d) All possible 2nd-order hopping vectors for the vertices in the physical space.
	}
\end{figure}

We focus on short diagonals in thin rhombi, which correspond to NN hoppings in the Penrose tiling. They are expressed as:
$\pm\{(1,0,0,1,0),(0,1,0,1,0),(0,1,0,0,1),(0,0,1,0,1),$
$(1,0,1,0,0)\}$ in hypercubic coordinates. By projecting them into two subspaces, we arrive at:
\small
\begin{equation} \begin{split}
\displaystyle
	\mathcal{S}\begin{pmatrix}\mathbf{h}^{(2)}_1,\mathbf{h}^{(2)}_2,\mathbf{h}^{(2)}_3,\mathbf{h}^{(2)}_4,\mathbf{h}^{(2)}_5
	\end{pmatrix} &= \frac{1}{\sqrt{10}}
	\begin{pmatrix}
		2-\tau &-1 &-2\sigma &-1 & 2-\tau\\
		-\beta &-\beta\sigma & 0 &\beta\sigma &-\beta
		\end{pmatrix},\\
	\mathcal{S}_{\perp}\begin{pmatrix}
		\mathbf{h}^{(2)}_1,\mathbf{h}^{(2)}_2,\mathbf{h}^{(2)}_3,\mathbf{h}^{(2)}_4,\mathbf{h}^{(2)}_5
	\end{pmatrix} &=
	\frac{1}{\sqrt{10}}\begin{pmatrix}
		2+\tau &\sigma-\tau &0 &\tau-\sigma &-2-\tau \\
		-\beta & -\beta(\tau+1) &-2\beta\tau &\beta(\tau+1) & -\beta\\
		2\sqrt{2} &2\sqrt{2}&2\sqrt{2}&2\sqrt{2}&2\sqrt{2}
	\end{pmatrix}.
\end{split} \end{equation}
\normalsize

For convenience, here we also define $\mathcal{R}_\mathrm{NN} = \mathcal{R}^1_\mathrm{NN}\cup\mathcal{R}^2_\mathrm{NN}$, where $\mathcal{R}^1_{NN}$ contains NN hopping vectors with polar angle $\theta \in \{0,+2\pi/5,+4\pi/5,-2\pi/5, -4\pi/5\}$, while $\mathcal{R}^2_\mathrm{NN}$ contains NN hopping vectors with $\theta\in \{\pi,+\pi/5,+3\pi/5,-\pi/5, -3\pi/5\}$. In Penrose tiling, we perform the same classification for NNN hopping $\mathcal{R}_\mathrm{NNN} = \mathcal{R}^1_\mathrm{NNN}\cup\mathcal{R}^2_\mathrm{NNN}$, for the difference between their distributions in perpendicular space.

\subsection{Decoration induced anisotropic intra-sublattice hoppings}
\subsubsection{Decoration from hyperspace}
To realize the altermagnetism in the quasicrystal lattice with two sublattices, we added several decoration sites to make the A and B sublattices globally inequivalent.
Taking the ABT quasicrystal as an example, we translate all B vertices by $(1/2,1/2,1/2,1/2)$ in hypercubic coordinates and add decorating vertices on these intersite positions. Then, we project these decorating vertices into physical ($V_\pi$) and perpendicular ($V_\perp$) spaces and set the selection window in $V_\perp$ as an origin-centered octagon with edge length of 2.
This makes half of all unit cells in the hyperlattice containing a body-center decorating vertex. By doing so, we arrive at a decorated quasicrystal lattice with decorating sites. We assume that the hopping amplitudes are $t_{2r}$ and $t_{2b}$ for intra-sublattice hoppings, depending on whether the neighboring pairs are blocked by the decorated nonmagnetic sites. These intra-sublattice hoppings are represented by red ($t_{2r}$) and blue ($t_{2b}$) lines in Fig.~\ref{fig:decoration}, which belong to $\mathcal{R}_\mathrm{NN}^r$ or $\mathcal{R}_\mathrm{NN}^b$, respectively.

Due to the distinct local environments of A and B sublattices, the intra-sublattice hopping $t_2$ depends on both the sublattice type and the direction of neighboring pairs. Specifically, we define the polar angle $\theta$ as the angle formed by the intra-sublattice hopping vector to the horizontal direction and find that for sublattice A, intra-sublattice hopping vectors $\mathbf{r}\in \mathcal{R}_\mathrm{NN}^1$ (i.e., the polar angle $\theta \in \{\pi/8, -3\pi/8,-7\pi/8,5\pi/8\}$) indicate $t_{2r}$, and $\mathbf{r}\in \mathcal{R}_\mathrm{NN}^2$ (i.e., $\theta\in \{-\pi/8,3\pi/8,7\pi/8,-5\pi/8\}$) indicates $t_{2b}$. The correspondence for the sublattice B is exactly the opposite.

To sum up, we classify vertices into 2 categories, A and B, through their parity of the hyperlattice coordinate summation. As a result, A and B sublattices are distinguishable when projected into the physical space. Notably, in the quasicrystalline lattice in the physical space, the NNN hoppings $\mathcal{R}_\mathrm{NNN}$ (which correspond to 1st hopping vectors in the 4D hypercubic lattice) connect nearest inter-sublattice pairs (A-B or B-A). The NN hoppings $\mathcal{R}_\mathrm{NN}$ (the first class of the 2nd hopping vectors in 4D) connect nearest intra-sublattice pairs (A-A or B-B), which are divided into two subgroups: $\mathcal{R}^r_\mathrm{NN}$ and $\mathcal{R}^b_\mathrm{NN}$ depending on the sublattice type and polar angle of the hopping vector.

\begin{figure}[h]
    \centering
    \includegraphics[width=10cm]{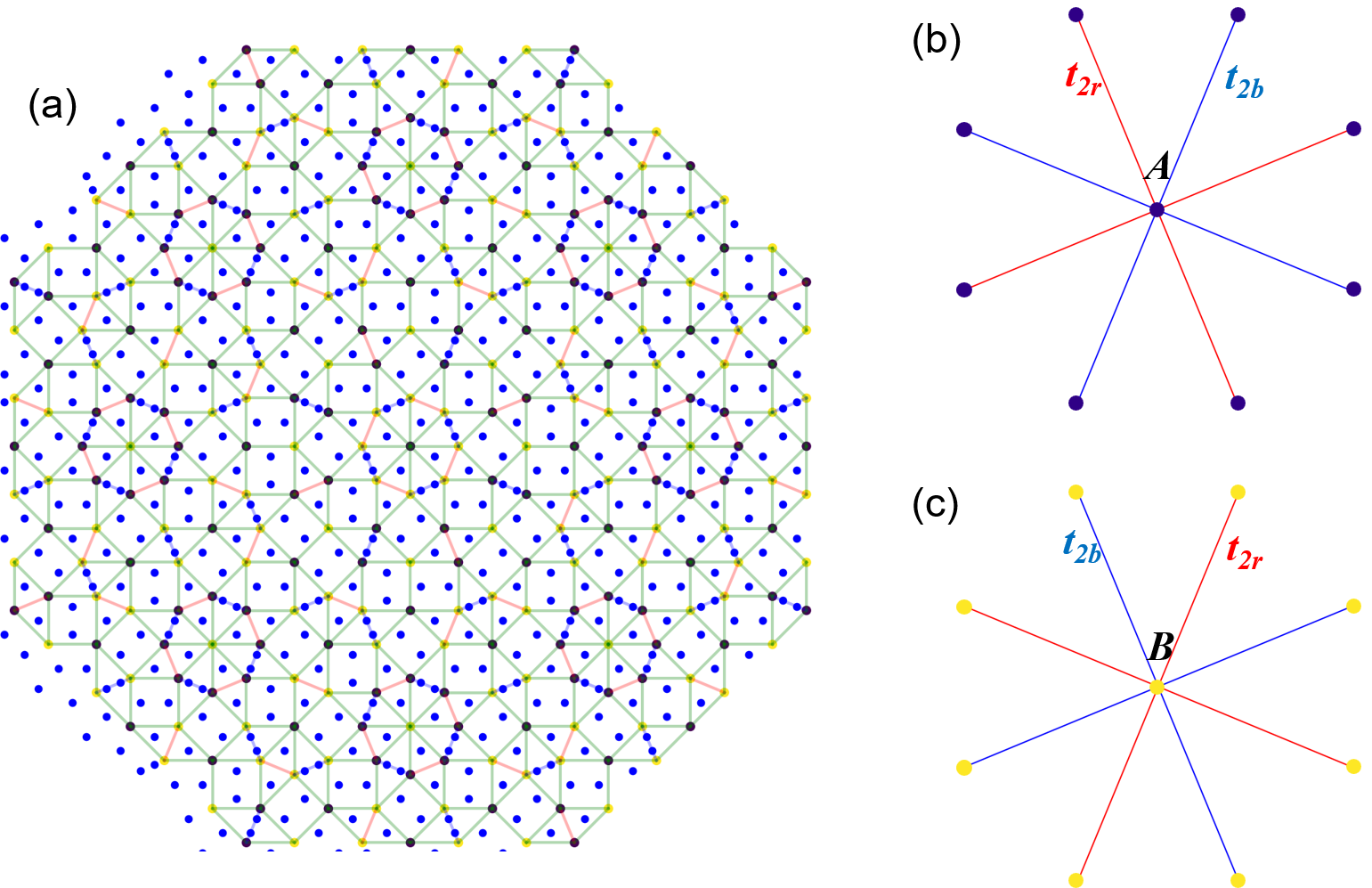}
    \caption{\label{fig:decoration} Decorated ABT quasicrystalline lattice constructed via the cut-and-project method. Purple and yellow points denote vertices belonging to sublattices A and B, respectively, while blue points represent decorated vertices. Green lines indicate NNN hoppings with uniform amplitude $t_1$. Red and blue lines denote NN hoppings with amplitudes $t_{2r}$ and $t_{2b}$, respectively, determined by the sublattice type and the polar angle of the hopping vector. (b) (c) All possible NN hopping for (b) A or (c) B vertex in the decorated ABT quasicrystalline lattice.}
\end{figure}

\subsubsection{Hopping terms in the Hamiltonian}
We can express the hopping terms in the full tight-binding Hamiltonian as
\begin{equation} \begin{split}
	H_\mathrm{hop} = -\sum_{\mathbf{r} \in \mathcal{R}}
    \sum_{\langle\mathbf{R}_\alpha \mathbf{R}'_\beta\rangle_\mathbf{r}}t_{\beta\alpha}(\mathbf{r})c_{\mathbf{R}'_\beta}^{\dagger} c_{\mathbf{R}_\alpha },\label{h_hop}
\end{split} \end{equation}
where $c_{\mathbf{R}_\alpha}=(c_{\mathbf{R}_\alpha\uparrow},c_{\mathbf{R}_\alpha\downarrow})$ with $c_{\mathbf{R}_\alpha,s}$ ($c^{\dagger}_{\mathbf{R}_\alpha,s}$) representing the annihilation (creation) operator of an electron at the vertex whose position is $\mathbf{R}_\alpha$, while $\alpha=A, B$ represents category of the vertex, and spin $s = \uparrow,\downarrow$.
$\langle\mathbf{R}_\alpha\mathbf{R}'_\beta\rangle$ denotes the start and end vertices of the hopping $t_{\beta\alpha}(\mathbf{r})$, whose categories ($\alpha,\beta=A,B$) are also regulated by the hopping.

$\mathcal{R}=\mathcal{R_\mathrm{NNN}}\cup\mathcal{R}^r_\mathrm{NN}\cup\mathcal{R}^b_\mathrm{NN}$ is a set that contains vectors describing the lengths and directions of hoppings in physical space.
Here, we consider nearest-neighbor (NN) and next-nearest-neighbor (NNN) hoppings, which makes $\mathcal{R}$ a finite set.
In both ABT and Penrose tiling quasicrystals, NNN and NN hopping vectors correspond to the edges of fundamental quadrilateral tiles and the short diagonals of thin rhombi, respectively. Therefore, for a hopping $t_{\beta\alpha}(\mathbf{r})$, we have $\alpha=\beta$ if $\mathbf{r}$ is NN hopping, and $\alpha \neq \beta$ if $\mathbf{r}$ is NNN hopping.

In the decorated ABT quasicrystal lattice, the hoppings are given by:
\begin{equation} \begin{split}
	t_{\beta\alpha}(\mathbf{r}) = \begin{cases}
		t_1 &(\mathbf{r} \in \mathcal{R}_\mathrm{NNN})\\
		t_{2r} &(\mathbf{r} \in \mathcal{R}^r: \mathbf{r} \in \mathcal{R}^1_\mathrm{NN} \text{ and } \alpha = \beta = A;\\
		&\text{or }\mathbf{r} \in \mathcal{R}^2_\mathrm{NN} \text{ and } \alpha = \beta = B)\\
		t_{2b} &(\mathbf{r} \in \mathcal{R}^b: \mathbf{r} \in \mathcal{R}^1_\mathrm{NN} \text{ and } \alpha = \beta = B;\\
		&\text{or }\mathbf{r} \in \mathcal{R}^2_\mathrm{NN} \text{ and } \alpha = \beta = A)\\
	\end{cases},\label{t_ba_ABT}
\end{split} \end{equation}
where $\mathcal{R}_\mathrm{NNN},\mathcal{R}^r_\mathrm{NN},\mathcal{R}^b_\mathrm{NN}$ are parts of $\mathcal{R}$, which are shown in Fig.~\ref{fig:decoration}.

Due to the $C_5$ symmetry of the Penrose tiling, we introduce NN hoppings with an imaginary part in the decorated Penrose tiling quasicrystal:
\begin{equation} \begin{split}
	t_{\beta\alpha}(\mathbf{r}) = \begin{cases}
		t_1 &(\mathbf{r} \in \mathcal{R}_\mathrm{NNN})\\
		t_{2}+i\delta_2 &(\mathbf{r} \in \mathcal{R}^r: \mathbf{r} \in \mathcal{R}^1_\mathrm{NN} \text{ and } \alpha = \beta = A;\\
		&\text{or }\mathbf{r} \in \mathcal{R}^2_\mathrm{NN} \text{ and } \alpha = \beta = B)\\
		t_{2}-i\delta_2 &(\mathbf{r} \in \mathcal{R}^b: \mathbf{r} \in \mathcal{R}^1_\mathrm{NN} \text{ and } \alpha = \beta = B;\\
		&\text{or }\mathbf{r} \in \mathcal{R}^2_\mathrm{NN} \text{ and } \alpha = \beta = A)\\
	\end{cases}.
\end{split} \end{equation}
Since opposite hoppings are represented by the same connection, $t_{\beta\alpha}(\mathbf{r})=t^*_{\alpha\beta}(-\mathbf{r})$ should be satisfied to ensure the Hermiticity of the system.

\section{Mean-field approximation for the Hubbard model in the decorated quasicrystal}

\subsection{The altermagnetic Hubbard model}
We consider two species of fermionic atoms labeled by spin s in an ABT quasicrystalline lattice described by the following altermagnetic Hubbard model
\begin{equation}
\begin{split}
    \hat{H}=&-\sum_{\langle i,j\rangle,s}(t_{ij}\hat c_{is}^\dagger \hat c_{js}+h.c.)+U\sum_i \hat n_{i\uparrow}\hat n_{i\downarrow}\\
    =&-\sum_{\langle iA, jB\rangle,s}t_1(\hat c_{iAs}^\dagger \hat c_{jBs}+\hat c_{jBs}^\dagger \hat c_{iAs})-\sum_{\langle i\alpha,j\alpha\rangle,s}t_{\alpha\alpha}(\mathbf{r})\hat c_{i\alpha s}^\dagger \hat c_{j\alpha s}+U\sum_{i\alpha} \hat n_{i \alpha\uparrow}\hat n_{i\alpha\downarrow},
\end{split}
\end{equation}
where $\hat c^{(\dagger)}_{is}=\hat c^{(\dagger)}_{i\alpha s}$ is the annihilation (creation) operator of electrons with spin $s=\uparrow\downarrow$ at the $i$-th site which belongs to sublattice $\alpha=A, B$. $\hat n_{is}=\hat c_{is}^\dagger\hat c_{is}$ is the number operator, $U$ is the on-site Hubbard interaction, and $t_{ij}$ is the hopping matrix element, which is uniform and of strength $t$ for nearest inter-sublattice hoppings, sublattice-dependent anisotropic for nearest intra-sublattice hoppings, and zero otherwise. The nearest intra-sublattice hopping $t_{\alpha\alpha}(\mathbf{r})=t_{2r}$ or $t_{2b}$ for $\mathbf{r}\in \mathcal{R}_\mathrm{NN}^r$ or $\mathbf{r}\in \mathcal{R}_\mathrm{NN}^b$ as presented in Eq.~\eqref{t_ba_ABT}.

\subsection{Order parameter and ansatz}
We apply a self-consistent Hartree-Fock (HF) method to characterize the magnetic order of the system. For this purpose, we introduce the N\'{e}el order parameter,
\begin{equation}
    \delta m =\frac{1}{2N}\left(\sum_{i\in A}\langle\hat n_{i\uparrow}-\hat n_{i\downarrow}\rangle-\sum_{i\in B}\langle\hat n_{i\uparrow}-\hat n_{i\downarrow}\rangle \right),
\end{equation}
where capital letters $A (B)$ in the summation denote the set of sublattices A (B) and $N$ denotes the number of sites. The altermagnetic order parameter is proportional to the staggered magnetization.
A non-vanishing order parameter $\delta m$ reveals the emergence of sublattice N\'eel ordering, which, together with the rotational symmetries of the quasicrystal, yields the altermagnetic phase.

Furthermore, at a filling of $\frac{1}{2N}\sum_i\langle \hat n_{i\uparrow}+\hat n_{i\downarrow}\rangle=n$, we assume the occupation takes the following staggered form:
\begin{equation}
    \langle \hat n_{is}\rangle =n+
    \left\{
    \begin{array}{cc}
         \delta m (-1)^s& i\in A \\
          -\delta m (-1)^s& i\in B
    \end{array}
    \right.\label{ansatz}
\end{equation}
where we associate $s$ in the term $(-1)^s$ with $0(1)$ for $\uparrow(\downarrow)$. Namely, for a fixed spin index, we assume that the occupation numbers at different sites depend solely on the sublattice type to which they belong. This assumption is conventional in crystalline systems, yet appears to constitute a crude approximation for quasicrystals owing to their lack of translational symmetry. However, since an electron is only influenced by a very short-range environment, its specific position is irrelevant, whereas the sublattice type represents the most significant feature. This assertion is corroborated by numerical results, as shown in Fig. \ref{phase_diagram}(b), which demonstrates that the order parameter under the ansatz Eq. (\ref{ansatz}) deviates only slightly from that obtained without the ansatz.

\subsection{The mean-field treatment}
With the ansatz \eqref{ansatz} of the occupation number and the mean-field approximation $\hat n_{i\uparrow}\hat n_{i\downarrow}\approx \langle\hat n_{i\uparrow}\rangle\hat n_{i\downarrow}+\hat n_{i\uparrow}\langle\hat n_{i\downarrow}\rangle-\langle\hat n_{i\uparrow}\rangle\langle\hat n_{i\downarrow}\rangle$, the interacting term can be reduced to a decoupled form in terms of the order parameter:
\begin{equation}
    \begin{split}
        U\sum_{i}\hat n_{i\uparrow}\hat n_{i\downarrow}
        &\approx U\sum_{i}\left[\langle\hat n_{i\uparrow}\rangle\hat n_{i\downarrow}+\hat n_{i\uparrow}\langle\hat n_{i\downarrow}\rangle\right]+\text{const.}\\
        &=U\left\{
        \sum_{i\in A}\left[(n+\delta m)\hat n_{i\downarrow}+\hat n_{i\uparrow}(n-\delta m)\right]+\sum_{i\in B}\left[(n-\delta m)\hat n_{i\downarrow}+\hat n_{i\uparrow}(n+\delta m)\right]
        \right\}+\text{const.}\\
        &=-U\delta m\left[\sum_{i\in A}(\hat n_{i\uparrow}-\hat n_{i\downarrow})-\sum_{i\in B}(\hat n_{i\uparrow}-\hat n_{i\downarrow})\right]+Un\sum_i(\hat n_{i\uparrow}+\hat n_{i\downarrow})+\text{const.}\label{decouple}
    \end{split}
\end{equation}
where the second term $Un\sum_i(\hat n_{i\uparrow}+\hat n_{i\downarrow})=Un\sum_i\hat n_i$ merely changes the chemical potential, and the third term is an irrelevant constant. Neither term affects the Hamiltonian's spectrum and can thus be neglected.

With the mean-field approximation $\ref{decouple}$, the Hamiltonian now becomes a spin block-diagonal form, which can be readily shown by introducing a concise notation $\hat \Psi=(\hat c_{1\uparrow},\hat c_{2\uparrow},\ldots,\hat c_{1\downarrow},\hat c_{2\downarrow},\ldots)^\mathrm{T}$ for the real-space basis, under which the mean-field Hamiltonian is expressed as:
\begin{equation}
    \hat H_{\mathrm{MF}}=\hat \Psi^\dagger
    \begin{bmatrix}
     H_{\uparrow}& 0\\
    0&  H_{\downarrow}
    \end{bmatrix}
    \hat\Psi\label{meanfield}
\end{equation}
where the $N\times N$ matrices $H_{\uparrow}$ and $H_{\downarrow}$ have a compact expression: $H_s=H_0+ (-1)^sH^\mathrm{HF}_{int}$. The non-interacting matrix $H_0$ reads $(H_0)_{ij}\equiv t_{ij}$ where $t_{ij}$ is defined by the first term of the altermagnetic Hubbard model [Eq. ($\ref{H_hubbard}$) in the main text], and $H^\mathrm{HF}_{int}$ is a sublattice-dependent diagonal matrix with
\begin{equation}
     (H^\mathrm{HF}_{int})_{ii}=
    \left\{
    \begin{array}{cc}
         -U\delta m & i\in A \\
          U\delta m & i\in B
    \end{array}
    \right.\label{diagonal}
\end{equation}
In accordance with the form of $H_s$, a non-zero order parameter $\delta m$ results in a spin-polarized Hamiltonian, which gives rise to a spin-split band structure.

\subsection{Self-consistent process}
We solve the real-space mean-field equations by self-consistently determining the order parameter $\delta m$ as well as the chemical potential $\mu$, which is set by solving the equation $n=\frac{1}{2N}\sum_i \langle \hat n_{i\uparrow}+\hat n_{i\downarrow}\rangle $ under a fixed particle number.
The self-consistent process can be expressed in the following steps:

\begin{enumerate}
    \item Make an initial guess on $\delta m$ and obtain the initial mean-field Hamiltonian $H_s=H_0+(-1)^s H^\mathrm{HF}_{int}$ where $H^\mathrm{HF}_{int}$ is defined by Eq. (\ref{diagonal})
    \item Diagonalize the mean-field Hamiltonian to obtain $2N$ wavefunctions, $\psi_a$, where $a=1,2,\ldots,2N$ labels $2N$ eigenstates for the mean-field Hamiltonian. Each wavefunction $\psi_a$ is a $2N$-dimensional vector, which reads $\psi_a=(c_{a1\uparrow},c_{a2\uparrow},\ldots, c_{a1\downarrow}, c_{a2\downarrow},\ldots)^\mathrm{T}$, where $c_{ais}$ are c-numbers.
    \item Update chemical potential $\mu$ and order parameter $\delta m$. Explicit expressions of them can be derived as follows: with given wavefunctions, the occupation numbers can be expressed by:
\begin{equation}
    \langle \hat n_{is}\rangle
    =\langle \hat c^\dagger_{is}\hat c_{is}\rangle
    =\sum_{a=1}^{2N} f(\epsilon_a-\mu)\langle \psi_a|\hat c^\dagger_{is}\hat c_{is}|\psi_a\rangle
    =\sum_{\alpha=1}^{2N}c^*_{ai s}c_{ais}f(\epsilon_a-\mu)\label{occupation}
\end{equation}
where $f(\epsilon-\mu)=\left[e^{-\beta(\epsilon-\mu)+1}\right]^{-1}$ with $\beta=1/k_BT$ is the Fermi-Dirac distribution. With Eq. (\ref{occupation}), the expression for filling number $n$ reads:
\begin{equation}
    n=\frac{1}{2N}\sum_{is}\langle \hat n_{is}\rangle
    =\frac{1}{2N}\sum_{a=1}^{2N}\sum_{is}c^*_{ais}c_{ai s}f(\epsilon_a-\mu)
    =\frac{1}{2N}\sum_{a=1}^{2N} f(\epsilon_a-\mu)
\end{equation}

By solving the equation for the inverse, the chemical potential $\mu$ is obtained. Besides, the order parameter can be calculated directly from the following expression:
\begin{equation}
\begin{aligned}
    \delta m
    &=\frac{1}{2N}\left(\sum_{i\in A}\langle\hat n_{i\uparrow}-\hat n_{i\downarrow}\rangle
    -\sum_{i\in B}\langle\hat n_{i\uparrow}-\hat n_{i\downarrow}\rangle
    \right)\\
    &=\frac{1}{2N}\sum_{a=1}^{2N}
    \left[\sum_{i\in A} (c_{ai\uparrow}^*c_{ai \uparrow}-c_{ai \downarrow}^*c_{ai \downarrow})
    \right.\left.
    -\sum_{i\in B}(c_{ai \uparrow}^*c_{ai \uparrow}-c_{ai \downarrow}^*c_{ai\downarrow})
    \right]f(\epsilon_a-\mu)\\
\end{aligned}
\end{equation}
    \item Calculate the deviation between the new $\delta m$ and the old one, if the derivation is small enough, end the iteration and output $\delta m$, otherwise, go back to step-1 and use the updated $\delta m$ as the new input.
\end{enumerate}

\subsection{Numerical results}
\paragraph{Mean-field phase diagram of the Hubbard model in the ABT quasicrystal.}
Our calculations are performed at half-filling, $n = \frac{1}{2N} \sum_i \langle \hat{n}_{i\uparrow} + \hat{n}_{i\downarrow} \rangle = 1/2$, although the resulting phase diagram remains stable under slight doping.
We evaluate the order parameter $\delta m$ as a function of $U/t_1$ and $\bar{t}_2/t_1$ for $\delta_2/\bar t_2 = 0.2$ in an ABT quasicrystal with $N = 329$ sites {under the open boundary condition}. The results are presented in Fig.~\ref{phase_diagram}, with $t_1 = 1$, $\bar{t}_2 = (t_{2r} + t_{2b})/2$, and $\delta_2 = (t_{2r} - t_{2b})/2$. Our computations indicate that for a given $\bar{t}_2$, the N\'{e}el ordering emerges as the on-site interaction $U$ increases. 

Figure~\ref{phase_diagram}(a) displays the order parameter $\delta m$ as a function of $U$ and $\bar{t}_2$ for $\delta_2/\bar t_2 = 0.2$ at zero temperature ($T = 0$). The calculation is initially performed using the ansatz in Eq.~(\ref{ansatz}), which assumes uniform occupation numbers within each sublattice for a given spin orientation. To validate this assumption, we compare the $\delta m$-$U$ curves at $\bar{t}_2/t_1 = 0.3$ with and without the ansatz, as shown in Fig.~\ref{phase_diagram}(d). Additionally, we present the spatial distributions of spin magnetization at $U/t_1 = 3.0$, $\delta_2/\bar t_2 = 0.2$, and $\bar{t}_2/t_1 = 0.3$ in Figs.~\ref{phase_diagram}(e) and \ref{phase_diagram}(f) with and without the ansatz, respectively. The two distributions are nearly indistinguishable, confirming the validity of the approximation.

\paragraph{Temperature effect.} We also investigate the influence of finite temperature and electron doping on the mean-field results for the decorated ABT quasicrystal. Comparing the phase diagrams at $T = 0$ and $T = 0.2$ [Figs.~\ref{phase_diagram}(a) and \ref{phase_diagram}(b)], we find that the N\'eel order remains robust against thermal fluctuations. Moreover, Fig.~\ref{phase_diagram}(c) presents the order parameter $\delta m$ as a function of electron filling and interaction strength $U$ at $\delta_2/\bar t_2=0.2$ and $\bar t_2/t_1=0.3$, revealing that the N\'eel ordering phase persists as long as the electron filling $n$ does not deviate significantly from the half-filling. These results demonstrate the robustness of the N\'eel order against both finite temperature and moderate electron doping.

{\paragraph{Size effect.}
We conducted a multi-scale analysis to evaluate the possible influence of boundary effects. Specifically, we calculate the N\'eel order parameter $\delta m$ for various system sizes $N$, while keeping the parameters fixed at $\delta_2/t_1=0.2,\bar t_2/t_1=0.3$ and $U/t_1=4.0$. The $\delta m$-$N$ curve is shown in Fig.~\ref{multi_scale}(a), the resulting size dependence demonstrates that $\delta m$ saturates for $N>1000$, indicating that the system effectively reaches the thermodynamic limit. Even for a moderately sized system of $N\approx300$, the deviation from the large-size limit remains below $2\%$, i.e.,
\begin{equation}
\begin{aligned}
\frac{\left|\delta m_{N=329}-\delta m_{N=3801}\right|}{\delta m_{N=3801}}<2\%,
\end{aligned}
\end{equation}
confirming that boundary effects are minimal in this size range. Furthermore, we plot the $\delta m$-$U$ curves for various system sizes $N$, with the parameters fixed at $\delta_2/t_1=0.2$ and $\bar t_2/t_1=0.3$, as presented in Fig.~\ref{multi_scale}(b). The curves show nearly identical transition behavior and critical interaction strength, indicating that bulk properties dominate and boundary effects are negligible. In summary, our multi-scale analysis confirms that the system size mainly used ($N\approx300$) is sufficiently large to capture the intrinsic altermagnetic phase and that the obtained results are robust against boundary effects.}

\paragraph{Decorated Penrose lattice.} Furthermore, we perform the self-consistent mean-field calculation for the Penrose lattice. Figure~\ref{phase_diagram_penrose} shows the mean-field results for the decorated Penrose quasicrystal, which also support the N\'eel order within a large range of the parameter space.

\begin{figure} [htb]
    \centering
     \includegraphics[width=1\linewidth]{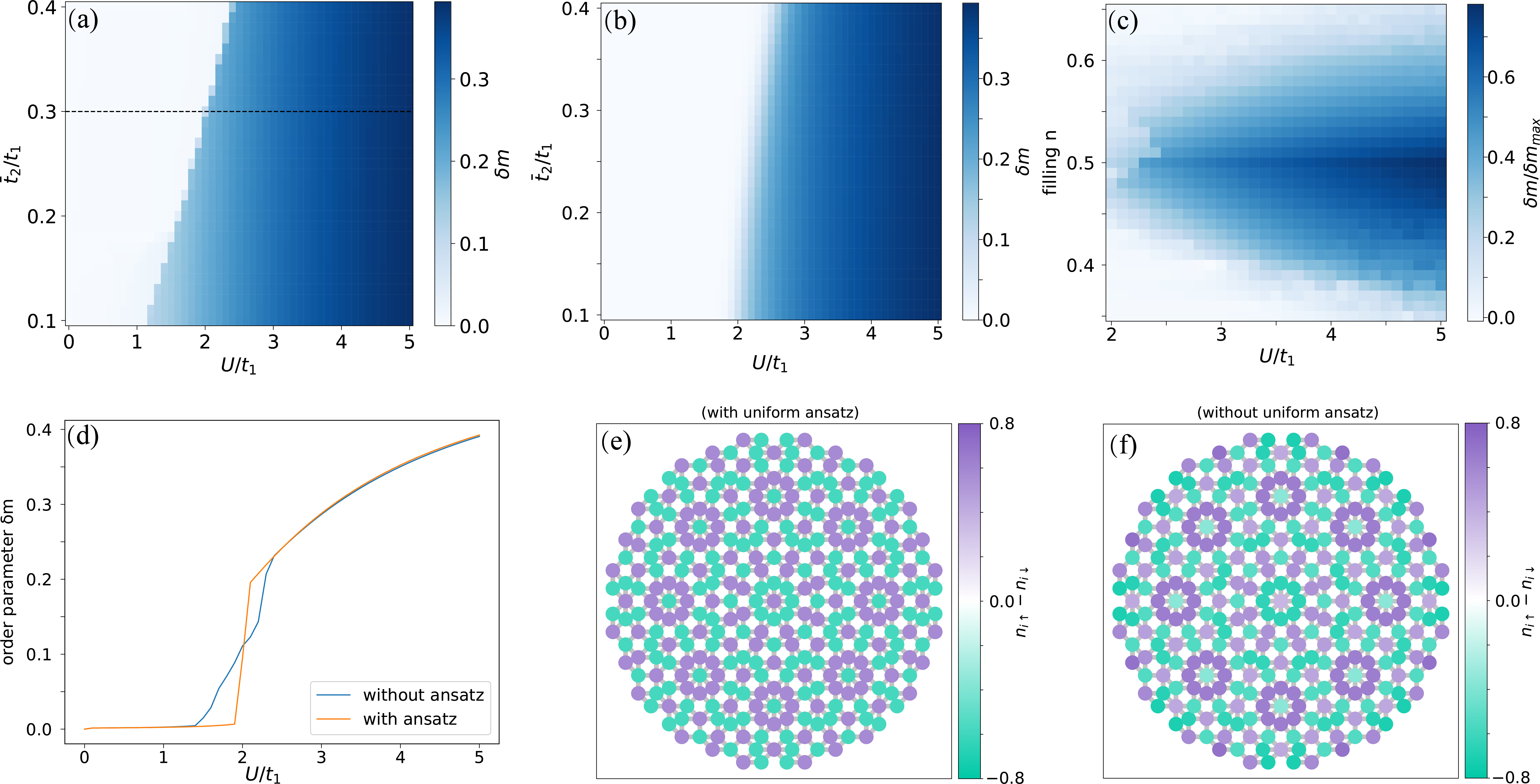}
    \caption{Mean-field results for the decorated ABT quasicrystal. The parameters are $N=329, t_1=1$, and $\delta_2/\bar t_2=0.2$. {The open boundary condition is adopted.} (a,b) Phase diagram in the $U-\bar{t}_2$ parameter space at half-filling $n=1/2$ for (a) zero temperature $T=0$ and (b) finite temperature $T=0.2$.
    (c) The distribution of order parameter $\delta m$ in the parameter space of filling $n$ and Hubbard $U$ from the calculation without uniform occupation ansatz at $\bar t_2/t_1=0.3$.
    (d) $U-\delta m$ curves with and without uniform occupation ansatz at $\bar{t}_2=0.3$. (e,f) Spin magnetic distribution (e) with and (f) without uniform occupation ansatz at $U=3, \bar{t}_2=0.3$, and $\delta_2/\bar t_2=0.2$. }
    \label{phase_diagram}
\end{figure}

\begin{figure} [htb]
    \centering
     \includegraphics[width=0.7\linewidth]{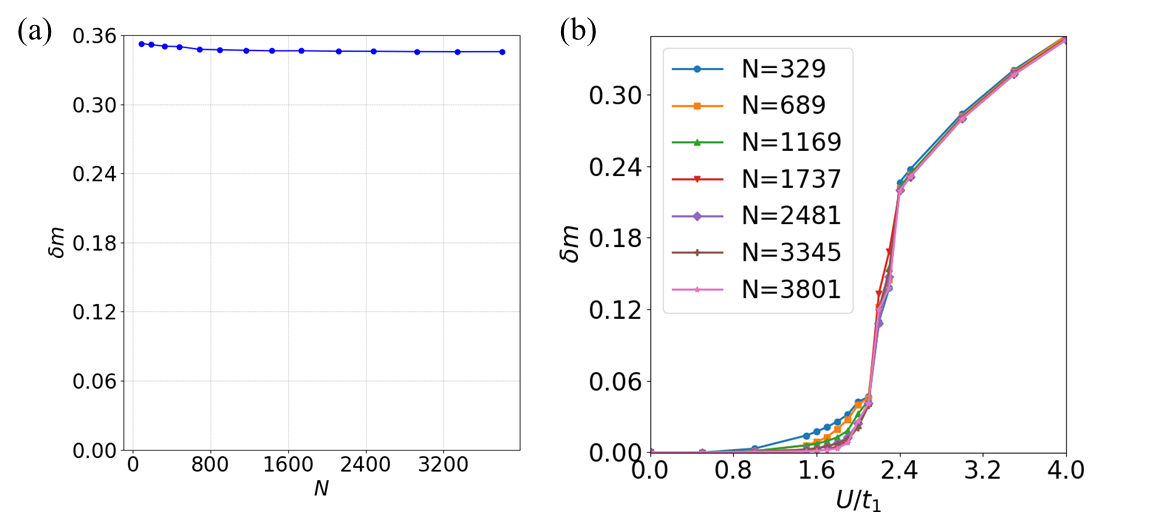}
    \caption{{Mean-field results for the decorated ABT quasicrystals with different system size $N$ at half-filling $n=1/2$ and zero temperature $T=0$. (a) $\delta m$-$N$ curve at $t_1=1,\bar t_2=0.3,\delta_2=0.2$, and $U=4$.
    (b) $\delta m$-$U$ curves for different system sizes $N$. Parameters are $t_1=1,\bar t_2=0.3$, and $\delta_2=0.2$.  }}
    \label{multi_scale}
\end{figure}

\begin{figure}
    \centering
        \includegraphics[width=0.7\linewidth]{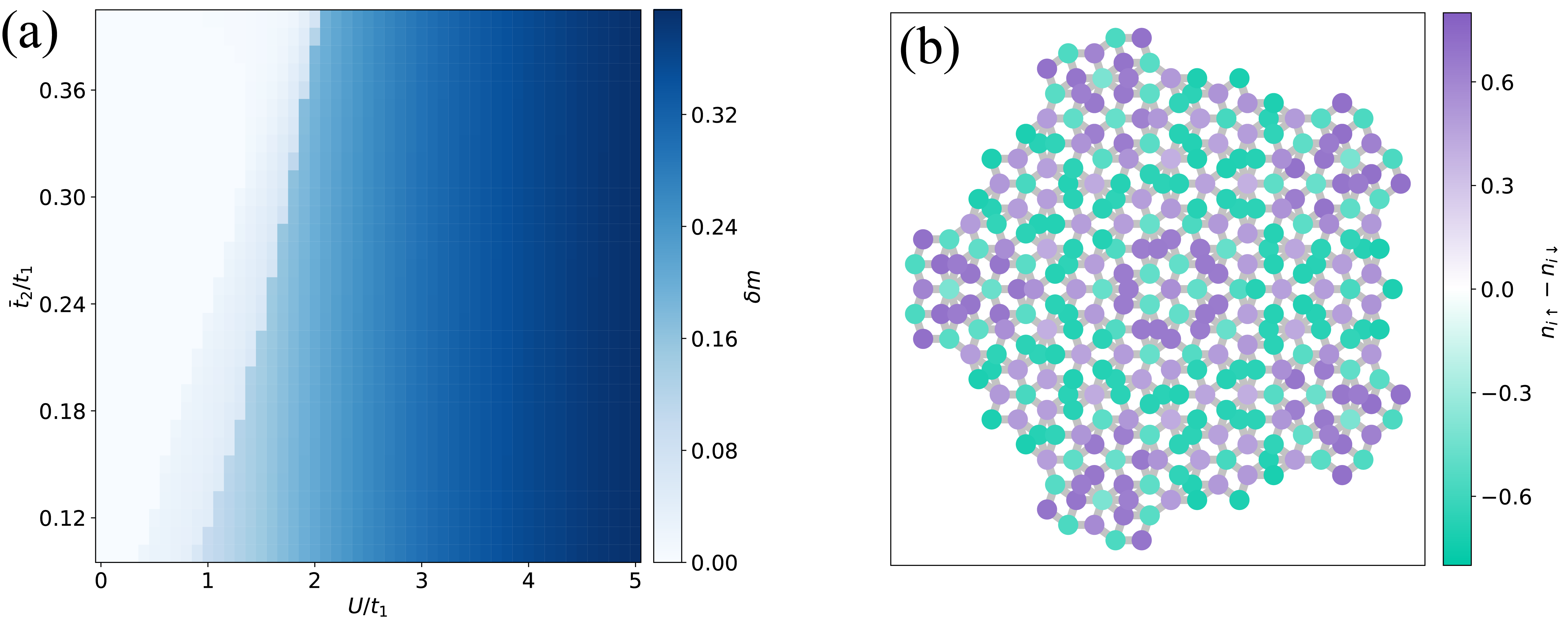}
    \caption{Mean-field results for the decorated Penrose tiling. (a) Phase diagram in the $U-t_2$ parameter space at half-filling and $T=0$. {The open boundary condition is adopted.}
    (b) Spin magnetic distribution without uniform occupation ansatz.
    Parameters are $N=326, U=3,t_1=1, t_2=0.3$, and $\delta_2/ t_2=0.2$. In the model the 2nd hoppings are $t_{2r/2b}=t_2\pm i\delta_2$.}
    \label{phase_diagram_penrose}
\end{figure}

\subsection{{Discussion on the Hartree-Fock analysis for the mean-field calculations}}
{The above mean-field implementation is based on the Hartree approximation. We also study the effect of the Fock term using a full Hartree-Fock method in the self-consistent calculations. Specifically, the interaction term is expressed as:
\begin{equation}
    \begin{aligned}
        H_{int}^\mathrm{HF}=U\sum_{i}\left(\left\langle n_{i\uparrow}\right\rangle n_{i\downarrow}+n_{i\uparrow}\left\langle n_{i\downarrow}\right\rangle-\Delta_{i-}c_{i\uparrow}^\dagger c_{i\downarrow}-\Delta_{i+}c_{i\downarrow}^\dagger c_{i\uparrow}\right)+\mathrm{const},
    \end{aligned}
\end{equation}
where $\langle n_{is}\rangle=\langle c_{is}^\dag c_{is}\rangle$ represents the Hartree field, and $\Delta_{i+}=\Delta_{i-}^\ast=\langle c_{i\uparrow}^\dag c_{i\downarrow}\rangle$ captures the Fock term.
In the self-consistent iteration, we diagonalize the Hartree-Fock Hamiltonian $H_0+H_{int}^\mathrm{HF}$ to obtain the eigenenergies $\epsilon_l$ and the corresponding eigenfunctions $\left|\psi^l\right\rangle=\sum_{is}\psi_{is}^lc_{is}^\dag\left|0\right\rangle$, where $l=1,2,\ldots,2N$ labels the eigenenergies and $N$ is the number of lattice sites. Subsequently, the updated coefficients are computed as:
\begin{equation}
    \begin{aligned}
        \left\langle n_{is}\right\rangle=\langle c_{is}^\dagger c_{is}\rangle=\sum_{l}\psi_{is}^{l\ast}\psi_{is}^lf\left(\epsilon_l-\mu\right),\\
        \Delta_{i+}=\Delta_{i-}^\ast=\langle c_{i\uparrow}^\dagger c_{i\downarrow}\rangle=\sum_{l}\psi_{i\uparrow}^{l\ast}\psi_{i\downarrow}^lf\left(\epsilon_l-\mu\right),
    \end{aligned}
\end{equation}
where $f\left(x\right)=1/[\exp{(x})+1]$ is the Fermi-Dirac distribution. The N\'eel order parameter---staggered magnetization between sublattices---is then defined as $\delta m=\frac{1}{2N}\left(\sum_{i\in A}\left\langle n_{i\uparrow}-n_{i\downarrow}\right\rangle-\sum_{i\in B}\left\langle n_{i\uparrow}-n_{i\downarrow}\right\rangle\right)$.
}

To specifically evaluate the influence of the Fock term, we performed two sets of control calculations. Both sets adopted identical initial values for the Hartree term $\left\langle n_{is}\right\rangle$ and a uniform initial value of $\delta m=0.05$ for the order parameter. Specifically, one set was initialized with a zero initial value for the Fock term ($\Delta_{i\pm}=0$), while the other set employed initial values of the Fock term following a random Gaussian distribution with a standard deviation of 0.1, comparable to the initial magnetization amplitude.

The resulting phase diagrams obtained without and with Fock term are shown in Fig.~\ref{Fock}(a) and ~\ref{Fock}(b), respectively. Both diagrams exhibit nearly identical structures. The line-cut comparison at $\bar{t_2}/t_1=0.3$, shown in Fig.~\ref{Fock}(c), reveals only minimal quantitative deviations between the two cases.
In conclusion, our full self-consistent Hartree-Fock calculations confirm that the inclusion of the Fock term does not qualitatively affect the stability or nature of the collinear magnetic order. Thus, the collinear N\'eel phase obtained in the Hartree approximation remains valid and robust against the inclusion of exchange-induced fluctuations.

\begin{figure} [htb]
    \centering
     \includegraphics[width=1\linewidth]{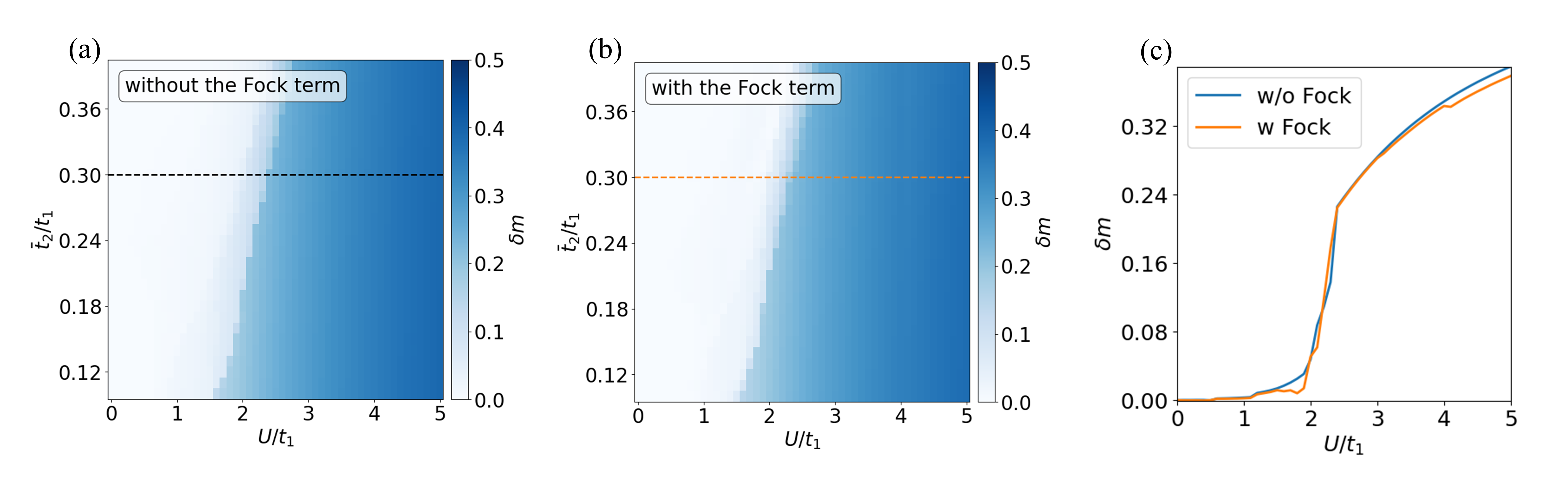}
    \caption{{(a) (b) Order parameter $\delta m$ in the plane of $\left[U,\bar{t_2}=\left(t_{2r}+t_{2b}\right)/2\right]$ from the self-consistent mean-field calculations without and with the Fock term, respectively. The parameter used is $\delta_2/t_1=0.2$. (c) Comparison of the $\delta m$-$U$ curves between the cases with and without Fock term, under the conditions of $\bar{t}_2/t_1=0.3$, and $\delta_2/t_1=0.2$. All calculations are performed on an ABT quasicrystal with a system size of $N=329$, at zero temperature and half-filling.}}
    \label{Fock}
\end{figure}



\newpage
\section{More results on the altermagnetism}

\subsection{{Robustness of the altermagnetism in quasicrystals}}
{\paragraph{Large hopping anisotropy.}
We further examine the robustness of the altermagnetic phase under large hopping anisotropy. Specifically, we consider the regime where $\delta_2/t_1$ is less than but close to unity. Figures~\ref{moderate_delta}(a) and \ref{moderate_delta}(b) present the phase diagram and the corresponding spin magnetic distribution for $\delta_2/t_1 = 0.8$, respectively. Although the N\'{e}el ordering persists in this strongly anisotropic regime, the ordered region in the phase diagram shifts toward larger Hubbard interaction $U$, and an intermediate region with a relatively weak order parameter $\delta m$ emerges. The spin-difference spectral function $\mathcal{A}_\uparrow-\mathcal{A}_\downarrow$, shown in Fig.~\ref{moderate_delta}(c), further confirms that the characteristic $C_{8z}\mathcal{T}$-symmetric altermagnetic feature remains robust even in the presence of pronounced hopping anisotropy.}

\begin{figure}
    \centering
        \includegraphics[width=1\linewidth]{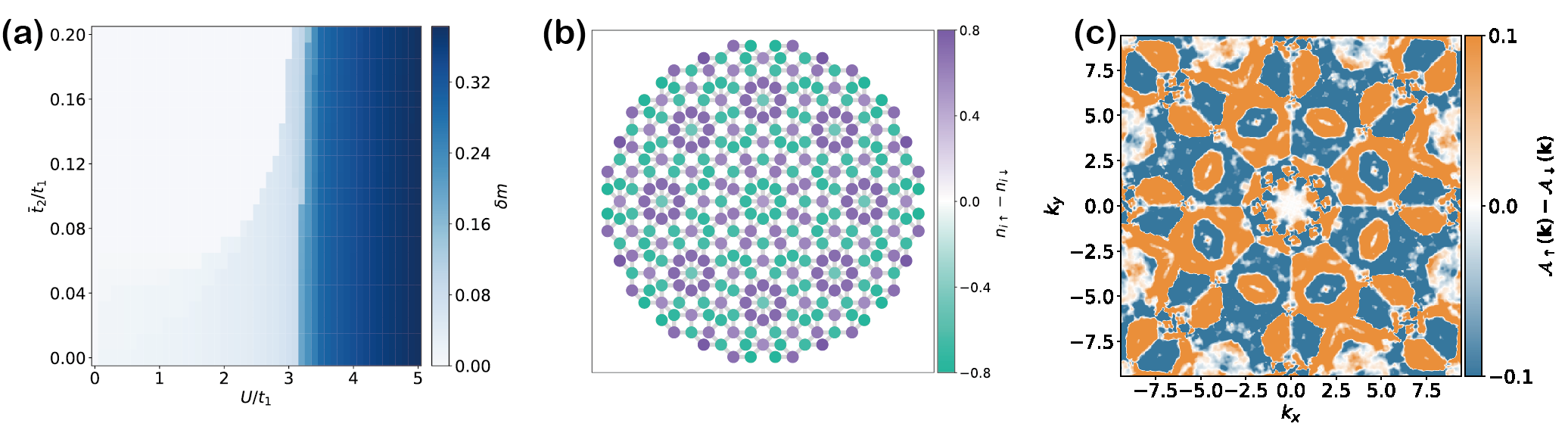}
    \caption{{Mean-field results for the decorated ABT quasicrystal with $\delta_2\lesssim1$. (a) Phase diagram in the $U-\bar t_2$ parameter space under uniform occupation ansatz obtained at half-filling $n=1/2, T=0, N=329,t_1=1$ and $\delta_2=0.8$. (b) Spin magnetic distribution at $U=4$ and $\bar t_2=0$. (c) Spin-difference spectral function $\mathcal{A}_\uparrow-\mathcal{A}_\downarrow$ in a larger lattice ($N=3801$) at $\bar t_2=0,\delta_2=0.8$, staggered magnetization $\bm{N}=1.5$ and $\omega=-2.59$.}}
    \label{moderate_delta}
\end{figure}

{\paragraph{Phason-flip-induced lattice displacement.}
We consider disorder originating from local phason flips, which cause atomic displacements that are known to occur between neighboring structural domains in quasicrystals. A representative example is illustrated in Fig.~\ref{displacement}(a), where an F-type site (coordination number 3) shifts from one side of a small hexagon to the other. In the undistorted configuration, gray lines represent the next nearest-neighbor (NNN) hopping $t_1$, while blue and red lines denote the two types of nearest-neighbor (NN) hoppings, $t_{2b}$ and $t_{2r}$, respectively. The difference between $t_{2b}$ and $t_{2r}$, determined by the presence of the decoration sites (gray dots), is essential for breaking $C_{8z}$ and $P\mathcal{T}$ while preserving the composite $C_{8z}\mathcal{T}$ symmetry---key to realizing the altermagnetic phase.}

{After displacement as shown in Fig.~\ref{displacement}(a), the original NNN bonds become first-neighbor bonds, and two initial first-neighbor bonds are converted into new second-neighbor bonds (orange lines). This displacement introduces a mismatch between the decoration pattern and the lattice geometry, locally disturbing the $C_{8z} \mathcal{T}$ symmetry. To evaluate the resulting impact, we assign the averaged value $\bar t_2$ to the new NN bonds, reflecting their approximately equal probability of being influenced by the decoration sites. We randomly select one half of possible F-type site in the system and shift the phason-flip displacement. Then, we perform self-consistent mean-field calculations on the modified Hamiltonian $H^\prime=H^\prime_0+U\sum_i n_{i\uparrow} n_{i\downarrow}$where $H_0^\prime$ represents the new hopping configuration.}

{Figures~\ref{displacement}(d) show that the staggered magnetization $\delta m$ remains nearly unchanged across the parameter space, indicating that N\'eel ordering persists despite the local displacements. The comparison of $\delta m$-$U$ curves before and after the displacement confirms the quantitative robustness of the order parameter [Fig.~\ref{displacement}(e)]. Our results indicate that the N\'eel ordering persists in a broad range of the phase diagram.}

{To further examine the electronic structure, we computed the spin-resolved spectral function $\mathcal{A}_s(\omega,\mathbf{k})$. As shown in Fig.~\ref{displacement}(f), the characteristic spin-splitting structure consistent with $C_{8z}\mathcal{T}$ symmetry is preserved, demonstrating that the altermagnetic feature survives the phason-flip–induced disorder.}

\begin{figure}
    \centering
        \includegraphics[width=1\linewidth]{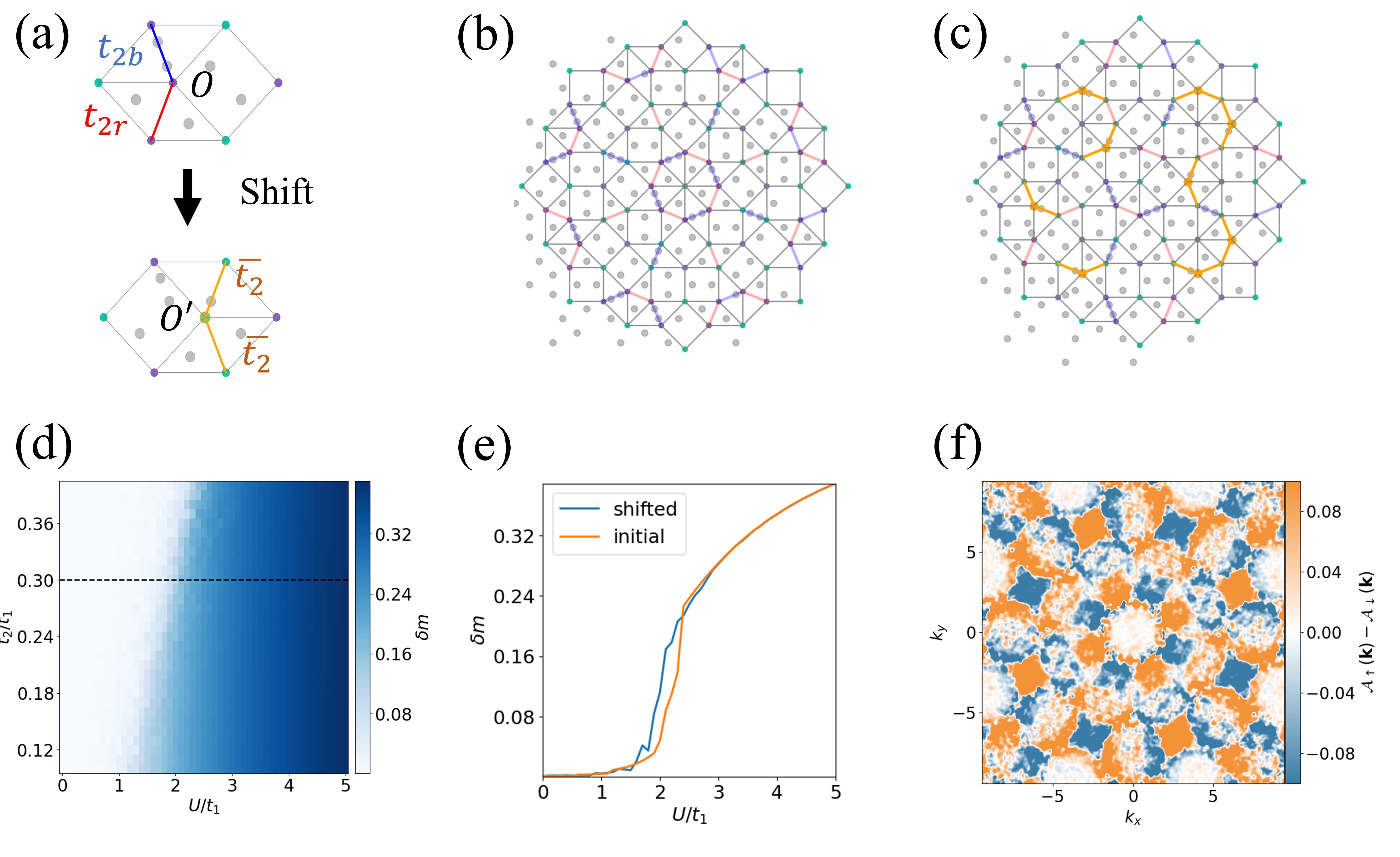}
    \caption{{(a) Illustration for the shift of an F-type point. After the shift, the initial second order hoppings $t_{2b}$ and $t_{2r}$ become first order hoppings ($t_1$), and the new second order hoppings (orange bonds) are assigned with $\bar{t_2}=\left(t_{2b}+t_{2r}\right)/2$. (b)(c) show the quasicrystal before and after half of the F-type points are shifted, respectively. (d) The staggered magnetization $\delta m$ in the $\left[U,\bar{t_2}=\left(t_{2r}+t_{2b}\right)/2\right]$ plane obtained from the self-consistent mean-field calculations after half of the F-type points are shifted. It is performed at zero temperature and half-filling for a system size of $N=329$.\ Parameter used is  $\delta_2/t_1=0.2$. (e) Comparison of the $\delta m$-$U$ curves between the cases before and after displacement, under the conditions of $\bar{t_2}/t_1=0.3$, and $\delta_2/t_1=0.2$. Both are performed at zero temperature and half-filling for a system size of $N=329$. (f) The spin-difference spectral function of the mean-field Hamiltonian. Parameters used are $N=3801,\left|\boldsymbol{N}\right|=3,t_1=1,\bar{t_2}=0,\delta_2=0.8$, and $\omega=-3.20$.}}
    \label{displacement}
\end{figure}


{\paragraph{Random on-site disorder.}
We next introduce an effective random on-site potential $H_{random}=\sum_{is}\epsilon_i n_{is}$, with $\epsilon_i$ uniformly distributed within $[-W/2,W/2]$. Setting $W=1.5t_1$, comparable to the on-site interaction strength $U$, we performed self-consistent mean-field calculations to determine $\delta m$. The phase diagram in the $[U, \bar t_2 ]$ plane [Fig.~\ref{onsite_disorder}(a)] and the $[U,W]$ plane [Fig.~\ref{onsite_disorder}(b)] shows that N\'eel order persists for moderate $U$ even under relatively strong randomness. Furthermore, the spin-difference spectral function in Fig.~\ref{onsite_disorder}(c) exhibits a clear spin splitting consistent with $C_{8z}\mathcal{T}$ symmetry, confirming that the altermagnetic phase remains stable against random disorder.}

\begin{figure}
    \centering
        \includegraphics[width=1\linewidth]{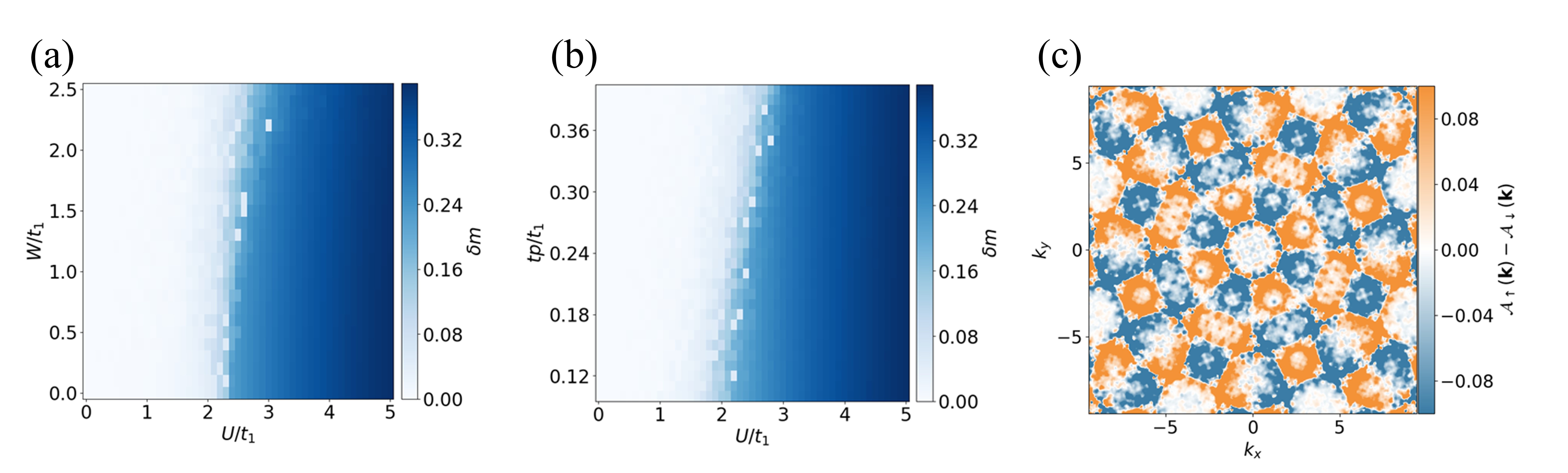}
    \caption{{(a) The staggered magnetization $\delta m$ in the $\left[U,\bar{t_2}=\left(t_{2r}+t_{2b}\right)/2\right]$ plane obtained from the self-consistent mean-field calculations for the Hamiltonian with a random onsite term of strength$ W/t_1=1.5$. Parameters used is $\delta_2/t_1=0.2$. (b) The staggered magnetization $\delta m$ in the $\left[U,W\right]$ plane. Parameters used are $\bar{t_2}/t_1=0.3$ and $\delta_2/t_1=0.2$. All calculations on $\delta m$ are performed at zero temperature and half-filling for a system size of $N=329$. (c) The spin-difference spectral function of the mean-field Hamiltonian. Parameters used are $N=3801,|\boldsymbol{N}|=3,t_1=1,\bar{t_2}=0,\delta_2=0.8,W=5$ and $\omega=\left(E_{1190\uparrow}+E_{1190\downarrow}\right)/2$, where $E_{ns}$ denotes the 1190th energy of the spin-s state. }}
    \label{onsite_disorder}
\end{figure}

{\paragraph{Absence of uniform N\'eel order.}
Finally, we examine the altermagnetic feature of a more general magnetic configuration without enforcing a uniform staggered magnetization. We determined the magnetic configuration using a self-consistent mean-field approach that does not impose a uniform occupation ansatz. The calculation was performed for a system near the phase boundary with a moderate Hubbard $U=3$ and a large $\delta_2=0.8$. In this case, the spin distribution [Fig.~\ref{general_phase}(a)] exhibits sub-cluster-dependent magnetization patterns localized around octagonal patches, deviating from perfect N\'eel order. Nevertheless, due to the inherent self-similarity and scale equivalence of quasicrystals, an effective ``cluster-level'' staggered order persists. The resulting spin-difference spectral function [Fig.~\ref{general_phase}(b)] continues to display the symmetry-protected spin-splitting structure characteristic of $C_{8z}\mathcal{T}$ altermagnetism, indicating that our proposed altermagnetism remains robust in quasicrystals across different scales.}

\begin{figure}
    \centering
        \includegraphics[width=0.85\linewidth]{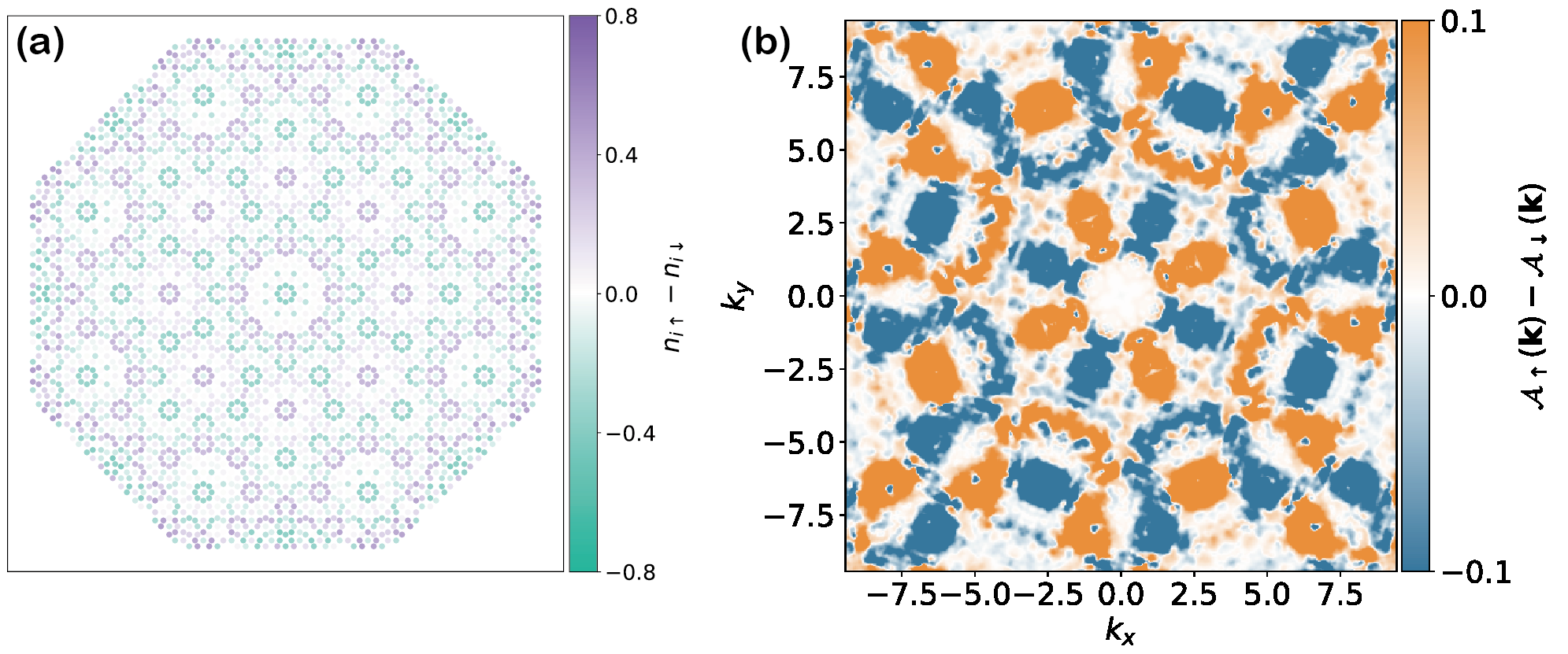}
    \caption{{Mean-field results for the decorated ABT quasicrystal with $\delta_2\lesssim1$ at half-filling $n=1/2$ and zero temperature $T=0$ without uniform occupation ansatz. Parameters are $N=3801, U=3, t_1=1, \bar t_2=0$, and $\delta_2=0.8$. (a) Spin magnetic distribution shows sub-cluster-dependent staggered magnetization. (b) Spin-difference spectral function $\mathcal{A}_\uparrow-\mathcal{A}_\downarrow$ at $\omega=0.56$, where the altermagnetic feature persist. }}
    \label{general_phase}
\end{figure}

\subsection{Spectral function for N\'eel orders without decorations}
Figure~\ref{t2r=t2b} and~\ref{t2r=t2b_penrose}  show the calculated spectral function in the absence of hopping anisotropy (i.e., $\delta_2=0$) for ABT and Penrose quasicrystals, respectively.

\begin{figure}[h] 
    \centering
    \subfigure[]{
    \includegraphics[width=8cm]{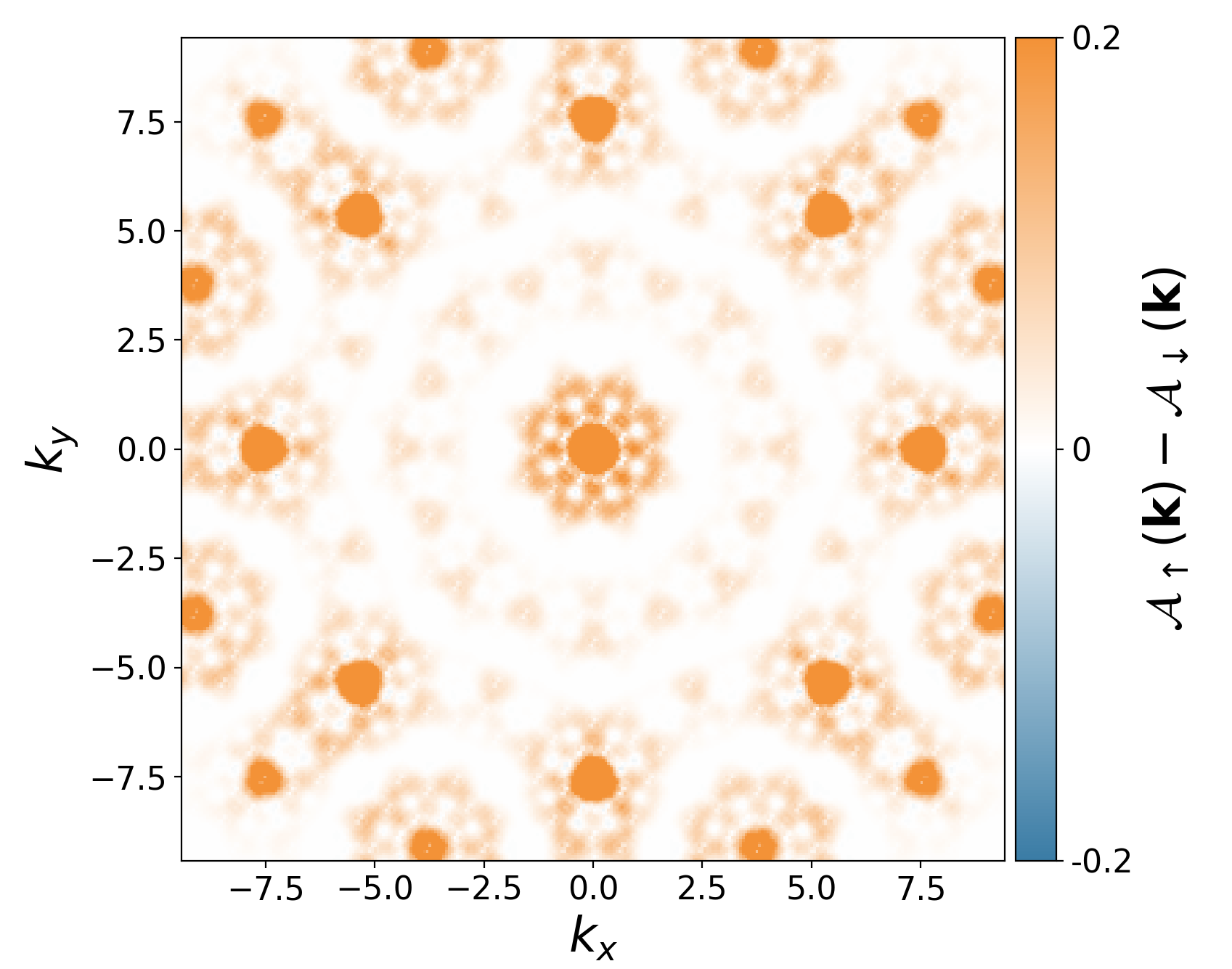}
    }
    \subfigure[]{
    \includegraphics[width=8cm]{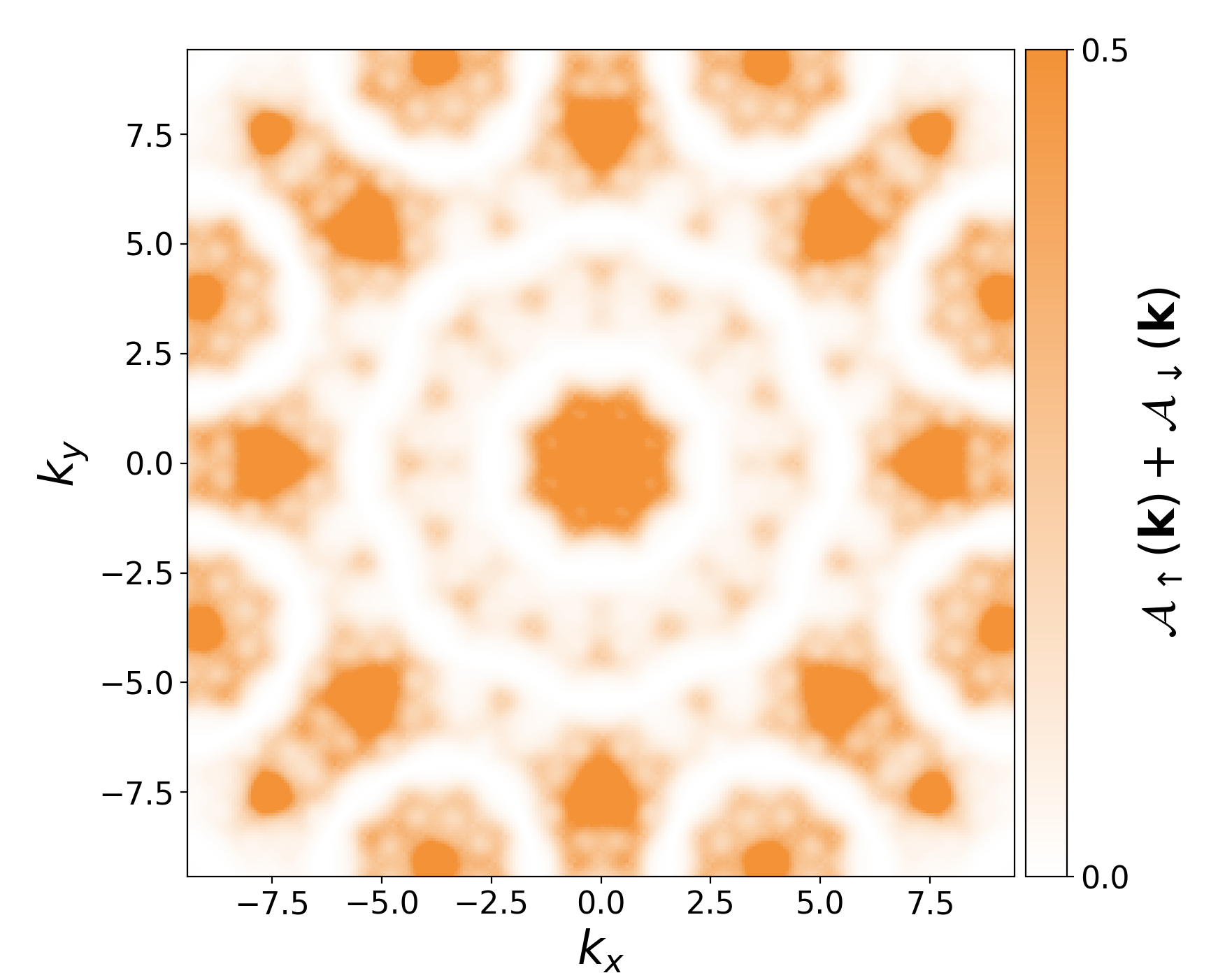}
    }
    \caption{Spectral functions of the ABT quasicrystal with nonzero staggered magnetization but without decorations: $\bm{N}\neq0, \delta_2=0$ $(i.e., t_{2r}=t_{2b})$. (a) Spin-difference spectral function $\mathcal{A}_\uparrow-\mathcal{A}_\downarrow$ and (b) Total spectral function $\mathcal{A}_\uparrow+\mathcal{A}_\downarrow$.
    The spectral function exhibits a uniform $C_{8z}$ rotational symmetry without alternating spin polarization.}
    \label{t2r=t2b}
\end{figure}

\begin{figure}[h] 
    \centering
    \subfigure[]{
    \includegraphics[width=8cm]{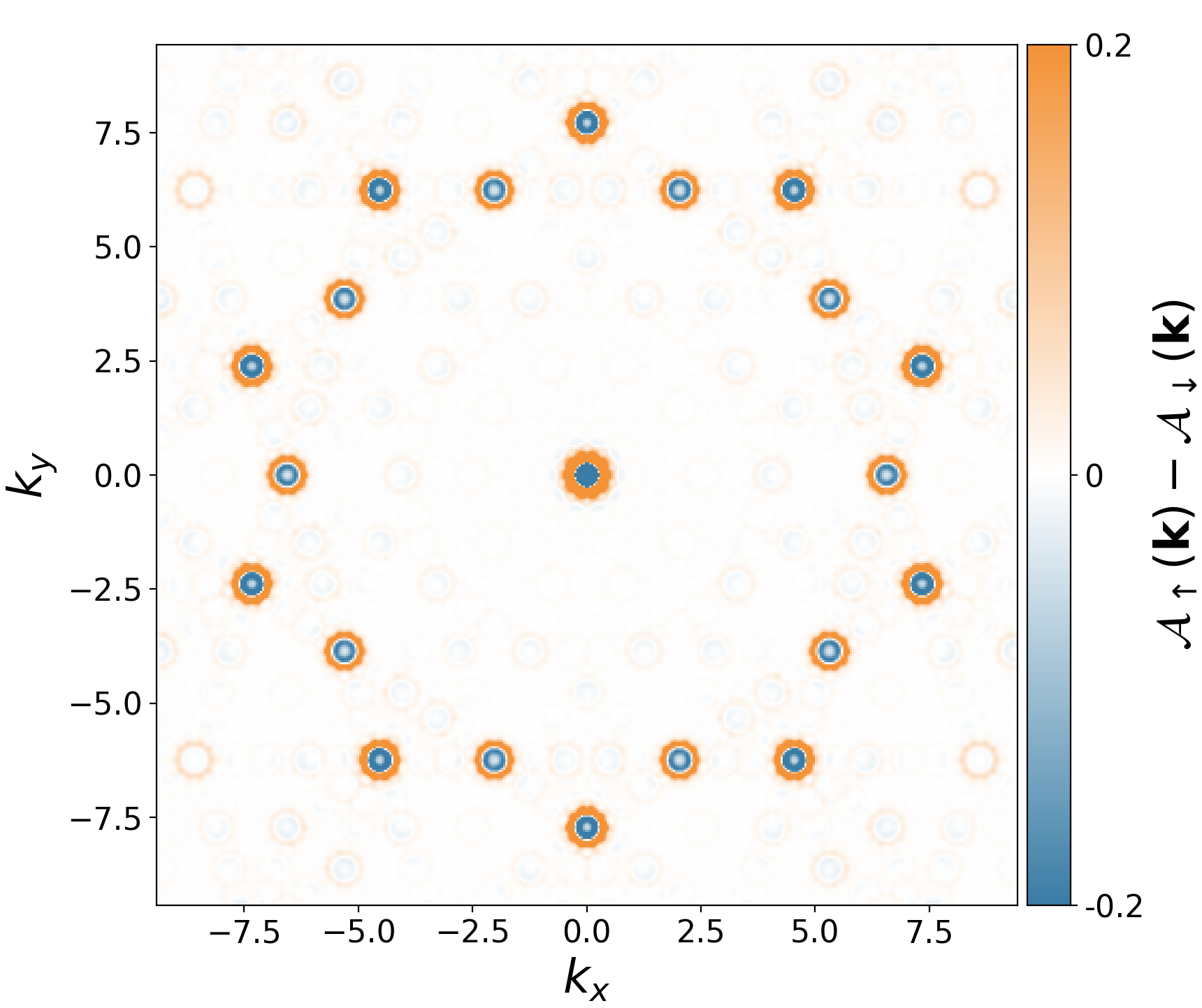}
    }
    \subfigure[]{
    \includegraphics[width=8cm]{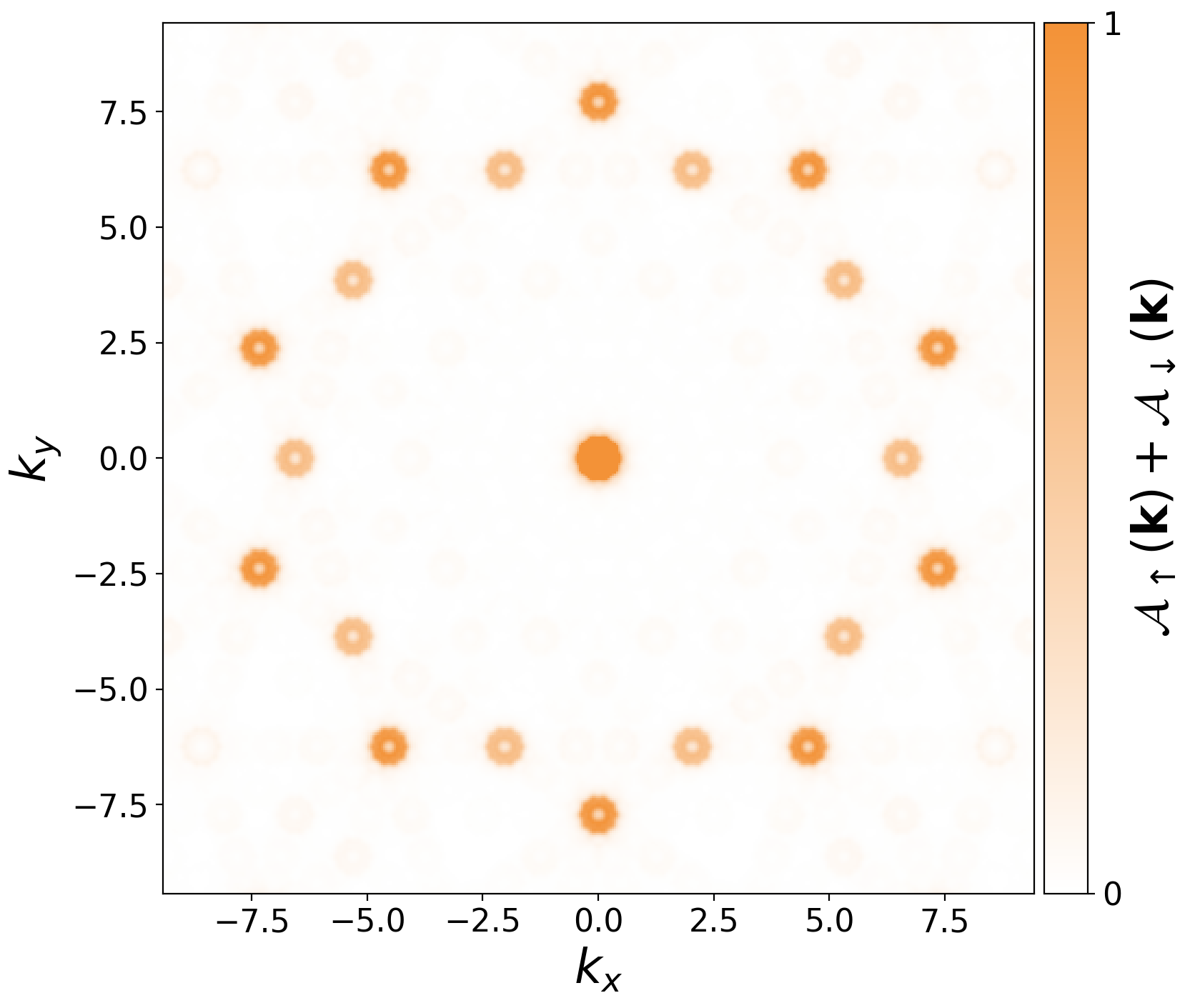}
    }
    \caption{Spectral function of the Penrose quasicrystal with nonzero staggered magnetization but without decorations: $\bm{N}\neq0, \delta_2=0$ $(i.e., t_{2r}=t_{2b})$.  (a) Spin-difference spectral function $\mathcal{A}_\uparrow-\mathcal{A}_\downarrow$ and (b) Total spectral function $\mathcal{A}_\uparrow+\mathcal{A}_\downarrow$.
    The spectral function exhibits a uniform $C_{10z}$ rotational symmetry without alternating spin polarization.}
    \label{t2r=t2b_penrose}
\end{figure}

\newpage
\section{Symmetry analysis from superspace crystallography}
\subsection{Decorated Ammann-Beenker tiling}
By directly considering the spin arrangement on the ABT quasicrystal, the $C_{8z}$ rotational symmetry only correlates vertices within sublattice A or B, meaning the corresponding operation is \{$C_{8z};1$\}, and no operation includes time reversal. Therefore, it does not support altermagnetism based on the previous analysis of magnetic point groups in crystals. Moreover, when decorations are added, the anisotropy of hopping occurs (i.e., $t_{2r}\neq t_{2b}$). Consequently, the $C_{8z}$ symmetry is broken, and the lattice point group reduces to $4mm$.

Instead, we analyze the system's symmetry from a 4D perspective, which may recover the symmetry operations lost during projection. Since the ABT lattice can be constructed by projecting a 4D hypercubic lattice, we define \{e$_1$, e$_2$, e$_3$, e$_4$\} as the basis vector set. The coordinates of the A and B sublattices are denoted as:
\begin{equation}
\begin{split}
    X=&(n_1,n_2,n_3,n_4)\equiv \sum_{i=1}^4n_i\mathrm{e}_i,\\
    X\in&
    \begin{cases}
        A & \text{if } \sum_in_i\in 2\mathbb{Z},\\
        B & \text{if } \sum_in_i\in 2\mathbb{Z}+1.\\
    \end{cases}
\end{split}
\end{equation}
Namely, sublattices A and B are distinguished by the parity of the sum of their four integer coordinates.

The decoration pattern is constructed by translating all vertices of sublattice B by the vector ${v}=\frac{1}{2}(1, 1, 1, 1)$. As a result, half of the body centers of the hypercubic cells are occupied by nonmagnetic decoration vertices that do not carry spins. By adding these vertices, the $\{1|\tau;-1\}$ (i.e., $\tau \mathcal{T}$) symmetry of the original hypercubic lattice with sublattice is broken, where $\tau=\mathrm{e}_i$ is the lattice translation operation. This is crucial because altermagnetism requires the breaking of $\tau \mathcal{T}$ symmetry. In this 4D hyperspace, the $C_{8z}$ operator acts as follows:
\begin{equation}
\begin{split}
    &\mathrm{e}_1\to-\mathrm{e}_3,\quad \mathrm{e}_3\to\mathrm{e}_2,\quad\mathrm{e}_2\to-\mathrm{e}_4,\quad\mathrm{e}_4\to-\mathrm{e}_1,\\
    &X=(n_1,n_2,n_3,n_4)\to \widetilde{X}=(-n_4,n_3,-n_1,-n_2),\\
    &X_D=X+v=(n_1,n_2,n_3,n_4)+\frac1 2(1,1,1,1)\\
    \to &\widetilde{X}_D=X'+v=(-n_4-1,n_3,-n_1-1,-n_2-1)+\frac1 2(1,1,1,1)\\
\end{split}
\end{equation}
Note that $n_1+n_2+n_3+n_4= -n_4+n_3-n_1-n_2\neq -n_4+n_3-n_1-n_2-3$ (mod 2), so $X$ and $\widetilde{X}$ share the same parity, but differ from $X'$. Thus, $C_{8z}$ does not exchange vertices between sublattices A and B, but transforms the decoration vertices $X_D$ (originally shifted from B by $v$) into sites shifted from A by $v$. However, a subsequent translation $\tau$ combined with time reversal $\mathcal{T}$ leaves the crystal invariant to within a translation.
Therefore, this system has the symmetry \{$C_{8z}|\tau;-1$\}, which acts like a glide rotation $\widetilde{C}_{8z}$ combined with time reversal $\mathcal{T}$. It is precisely this 4D hyperspace symmetry that endows the spectral function with $C_{8z}\mathcal{T}$ symmetry.

\subsection{Decorated Penrose tiling}
Let \{e$_1$, e$_2$ ,e$_3$, e$_4$, e$_5$\} be the basis vectors of a 5D hypercubic lattice. The coordinates of the A and B sublattices are denoted as :
\begin{equation}
\begin{split}
    X=&(n_1,n_2,n_3,n_4,n_5)\equiv \sum_{i=1}^5n_i\mathrm{e}_i,\\
    X\in&
    \begin{cases}
        A & \text{if } \sum_in_i\in 2\mathbb{Z},\\
        B & \text{if } \sum_in_i\in 2\mathbb{Z}+1,\\
    \end{cases}
\end{split}
\end{equation}
In this 5D space, the $C_{10z}$ operator act as follows:
\begin{equation}
\begin{split}
    &\mathrm{e}_1\to-\mathrm{e}_4, \mathrm{e}_2\to-\mathrm{e}_5,\mathrm{e}_3\to-\mathrm{e}_1,\mathrm{e}_4\to-\mathrm{e}_2,\mathrm{e}_5\to-\mathrm{e}_3,\\
    &(n_1,n_2,n_3,n_4,n_5)\to(-n_3,-n_4,-n_5,-n_1,-n_2).\\
\end{split}
\end{equation}
Note that $\sum_{i=1}^5n_i=-\sum_{i=1}^5n_i$ (mod 2), so $C_{10z}$ does not exchange vertices between sublattices A and B. The NN hoppings considered are separated into two groups. One group $\mathcal{R}_\mathrm{NN}^1$ consists of the set \{(1,0,0,1,0), (0,1,0,0,1), (1,0,1,0,0), (0,1,0,1,0), (0,0,1,0,1)\}. The other group is defined as $\mathcal{R}_\mathrm{NN}^2=-\mathcal{R}_\mathrm{NN}^1$. These two parts are connected via $C_{10z}$. From Eq.~(A7), we can get $t_{AA}(\mathcal{R}_\mathrm{NN}^1)=t_2+i\delta_2, t_{AA}(\mathcal{R}_\mathrm{NN}^2)=t_2-i\delta_2$. $C_{10z}:\mathcal{R}_\mathrm{NN}^1\to\mathcal{R}_\mathrm{NN}^2,A\to A$. When $\delta_2\neq0$, $t_{AA}(\mathcal{R}_\mathrm{NN}^1)\neq t_{AA}(\mathcal{R}_\mathrm{NN}^2)$. Therefore, the $C_{10z}$ symmetry of the original hypercubic lattice is broken.

Similarly, $C_{10z}\mathcal{T}:\mathcal{R}_\mathrm{NN}^1\to\mathcal{R}_\mathrm{NN}^2,A\to B$. However, it should be noted that $\mathcal{T}$ will change $t_2+i\delta_2\to t_2-i\delta_2$. When $\delta_2\neq 0$, $t^*_{AA}(\mathcal{R}_\mathrm{NN}^1)\neq t_{BB}(\mathcal{R}_\mathrm{NN}^2)$. Therefore, the $C_{10z}\mathcal{T}$ symmetry is also broken.

Since $C_{5z}$ and $\tau \mathcal{T} : \mathcal{R}_\mathrm{NN}^1\to\mathcal{R}_\mathrm{NN}^1,\mathcal{R}_\mathrm{NN}^2\to\mathcal{R}_\mathrm{NN}^2,A\to A,B\to B$, then the system still possesses $C_{5z}$ and $\tau \mathcal{T}$ symmetry. As a result, the spectral function respects a $C_{5z}\mathcal{T}$ symmetry.

\section{Low-energy effective treatment of the mean-field Hamiltonian}

\subsection{Mean-field Hamiltonian}
Now, we express the mean-field Hamiltonian as
\begin{equation}
\begin{split}
H_\mathrm{MF}=&\sum_{i\alpha}(-1)^{\eta-1} \boldsymbol{N}\cdot\boldsymbol{\sigma}c_{i\alpha}^\dagger c_{i\alpha}-\sum_{\langle i\alpha, j\beta\rangle}(t_{\beta\alpha}(\mathbf{r}_{ji})c_{j\beta }^{\dagger} c_{i\alpha }+h.c.)\\
=&\boldsymbol{N}\cdot\boldsymbol{\sigma}(\sum_{\mathbf{R}_A}c_{\mathbf{R}_A}^\dagger c_{\mathbf{R}_A}-\sum_{\mathbf{R}_B}c_{\mathbf{R}_B}^\dagger c_{\mathbf{R}_B})
-\sum_{\langle \mathbf{R}_A,\mathbf{R}_B\rangle}t_1\left( c_{\mathbf{R}_A}^\dagger c_{\mathbf{R}_B}+c_{\mathbf{R}_B}^\dagger c_{\mathbf{R}_A} \right)\\
&-\sum_{(\mathbf{R}'_\alpha-\mathbf{R}_\alpha)\in\mathcal{R}^r_\mathrm{NN}} t_{2r} c_{\mathbf{R}_\alpha}^\dagger c_{\mathbf{R}'_\alpha}-\sum_{(\mathbf{R}'_\alpha-\mathbf{R}_\alpha)\in\mathcal{R}^b_\mathrm{NN}} t_{2b} c_{\mathbf{R}_\alpha}^\dagger c_{\mathbf{R}'_\alpha}
,\\
\end{split}
\end{equation}
where $i\alpha$ denote the $i$-th site with position $\mathbf{R}_i$ belonging to sublattice $\alpha=A$ or $B$ according to the hyperspatial projection, $c_{i\alpha}=(c_{i\alpha,\uparrow},c_{i\alpha,\downarrow})$ with $c_{i\alpha,s}$ ($c^\dagger_{i\alpha,s}$) annihilates (creates) an electron with spin $s=\uparrow\downarrow$ at site $i\alpha$. The itinerant electrons have a Kondo coupling $J$ to the collinear local moments, which have staggered magnetization $\boldsymbol{N}=\boldsymbol{M}_A-\boldsymbol{M}_B$. Here, $\langle iA,jB\rangle$ and $\langle i\alpha,j\alpha \rangle$ represent the nearest inter-sublattice (intra-sublattice) hopping in the quasicrystal lattice, which correspond to NNN and NN vectors. $t_1$ is a uniform hopping parameter, and $t_{2r/2b}$ denotes anisotropic {intra-sublattice} hopping parameters due to different crystallographic environments of the $A$ and $B$ sublattices in the decorated quasicrystal, as presented in Eq.~\eqref{t_ba_ABT}.

\subsection{Low-energy effective theory}
Having obtained the mean-field Hamiltonian, we explore the momentum-dependent spin splitting in its electronic structure, which is a key characteristic of altermagnetism. However, since the QL possesses long-range orientational order but lacks translational symmetry, we cannot use the Bloch theorem as for the crystal calculations. Instead, we establish a low-energy effective theory under the long-wavelength approximation to capture the electronic structure at the center of the pseudo-Brillouin zone, where the important spin splitting occurs.

First, we analyze the mean-field model for the decorated ABT quasicrystal based on the projection method introduced in Ref.~\cite{wang2022effective}. Generally, a function $f(\mathbf{r})$ with quasicrystal periodicity in physical space can be expressed as a Fourier sum:
\begin{equation} \begin{split}
	f(\mathbf{r})=\sum_{\mathbf{G}\in \mathcal{L}} \tilde{f}(\mathbf{G})e^{i\mathbf{G}\cdot \mathbf{r}}, \label{FFT_g}
\end{split} \end{equation}
where $\mathcal{L}$ is a countable set of reciprocal wave vectors which contains $\mathbf{G} = \sum_{i=1}^4 n_i \mathbf{g}_i$ $(n_i \in \mathbb{Z})$, filling the reciprocal space densely. $\mathbf{g}_i = 2\pi(\cos(\frac{(i-1)\pi}{4}),\sin(\frac{(i-1)\pi}{4}))$, $(i=1,2,3,4)$ are four independent principal reciprocal vectors of the initial whole quasilattice. Here we set the length of NNN hopping vectors as unit, without loss of generality.

On the other hand, we can implement a novel but equivalent Fourier expansion, proposed by Jiang and Zhang (JZ) \cite{JIANG2014428},
\begin{equation} \begin{split}
	f(\mathbf{r})=\sum_{\mathbf{H}} \tilde{f}(\mathbf{H})e^{i(\mathcal{S}\mathbf{H})\cdot \mathbf{r}},
\end{split} \end{equation}
where $\cal S$ is the projection operator (Eq.~\eqref{projection}) acting on reciprocal vectors $\mathbf{H}\in\mathbb{R}^4$, from the 4D hyperspace to the 2D physical space. Here, $\mathbf{H} = \sum_{i} m_i \mathbf{Q}_{i=1}^4$ $(m_i \in \mathbb{Z})$, where $\mathbf{Q}_i$ are primitive reciprocal vectors of the (sub-)hyperlattice.
For the whole hypercubic lattice ignoring sublattice types, one set $\{\mathbf{Q}_1, \mathbf{Q}_2, \mathbf{Q}_3, \mathbf{Q}_4\}= 2\pi\{(1,0,0,0), (0,1,0,0), (0,0,1,0), (0,0,0,1)\}$ as the primitive reciprocal vectors of the whole hyperlattice. We find $\mathcal{S}\mathbf{Q}_i=\mathbf{g}_i$, which demonstrates that this expansion is equivalent to Eq.~\eqref{FFT_g}.
However, for the 4D hyperlattice with two sublattices A and B, we set the primitive reciprocal vectors as
$$\{\mathbf{Q}_1, \mathbf{Q}_2, \mathbf{Q}_3, \mathbf{Q}_4\} = \pi\{(1,1,0,0), (0,0,-1,1), (-1,1,0,0), (0,0,-1,-1)\}.$$

For the mean-field TB model, we express annihilation and creation operators as:
\begin{equation} \begin{split}
c_{\mathbf{R}_\alpha s}=\frac{1}{\sqrt{N_A}}\sum_{\mathbf{H}} c_{\mathbf{k}s,\alpha}  e^{-i (\mathcal{S} \mathbf{H})\cdot \mathbf{R}_\alpha},\\
c^\dagger_{\mathbf{R}_\alpha s}=\frac{1}{\sqrt{N_A}}\sum_{\mathbf{H}} c^\dagger_{\mathbf{k}s,\alpha}  e^{i (\mathcal{S}\mathbf{H})\cdot \mathbf{R}_\alpha},
\end{split} \end{equation}
Here we define the annihilation (creation) operator in the pseudo-k space
$c^{(\dagger)}_{\mathbf{H} s}=c^{(\dagger)}_{\mathbf{k} s}$, and $\mathbf{k}$ as the physical projection of $\mathbf{H}$, i.e. $\mathbf{k}=\mathcal{S} \mathbf{H}$. For its validity, $\mathbf{k}$ is restricted around the $\Gamma$ point.

Significantly, there is orthogonality in this set of basis function $e^{i(\cal S \mathbf{H})\cdot \mathbf{r}}$:
\begin{equation} \begin{split}
	\sum_{\mathbf{R}_i} e^{i(\mathcal{S} \mathbf{H})\cdot \mathbf{R}_i} \simeq N_i \delta_{\cal S \mathbf{H}},\label{orthogonality}
\end{split} \end{equation}
where $N_i$ is the number of vertices summed over. Remarkably, in the vicinity around $\Gamma$ point in the pseudo-k space,
this orthogonality is always valid for any quasilattice, regardless of the size and shape of the selection window, i.e., as long as the vertex set $\{\mathbf{R}_i\}$ forms a ``quasicrystal" constructed by the cut-and-project method with an arbitrary selection window in the perpendicular space. The detailed derivation and discussion of this approximate orthogonality are presented in Sec.\ref{sec:orth}.

Now we derive the effective Hamiltonian in the pseudo-k space. The mean-field Hamiltonian can be divided into the on-site and hopping terms $H_\mathrm{MF}=H_\mathrm{on-site}+H_\mathrm{hop}$.
For the on-site term in the A sublattice,
\begin{equation} \begin{split}
	H_\mathrm{on-site}^A=&\sum_{\mathbf{R}_A,ss'} c_{\mathbf{R}_A s'}^{\dag} (\bm{N}\cdot \boldsymbol{\sigma})_{s's} c_{\mathbf{R}_A s} \\
	&= \frac{1}{N_A} \sum_{\mathbf{k}\mathbf{k}',ss'} \sum_{\mathbf{R}_A} e^{i (\mathbf{k}'-\mathbf{k})\cdot \mathbf{R}_{A}}
	c_{\mathbf{k}'s',A}^{\dag} (\bm{N}\cdot \boldsymbol{\sigma})_{s's} c_{\mathbf{k}s,A} \\
	&= \sum_{\mathbf{k},ss'}c_{\mathbf{k}s',A}^{\dag} (\bm{N}\cdot \boldsymbol{\sigma})_{s's} c_{\mathbf{k}s,A}.\label{H_on-site}
\end{split} \end{equation}
Similarly, the on-site term in the B sublattice is $H_\mathrm{on-site}^B=\sum_{\mathbf{k}}c_{\mathbf{k},B}^{\dagger} (-\bm{N}\cdot \boldsymbol{\sigma})c_{\mathbf{k},B}$.

For the hopping terms, we first rewrite $H_\mathrm{hop}$ as
\begin{equation} \begin{split}
	H_\mathrm{hop}=-\sum_{\mathbf{r}\in \mathcal{R}}\sum_{\langle\mathbf{R}_\alpha\rangle_\mathbf{r}}t_{\beta\alpha} (\mathbf{r}) c_{\mathbf{R}_\alpha+\mathbf{r},s}^{\dagger} c_{\mathbf{R}_\alpha,s},
\end{split} \end{equation}
where $\mathcal{R}$ includes different hopping vectors, including nearest inter-sublattice ($\mathcal{R}_\mathrm{NNN}$) and nearest intra-sublattice ($\mathcal{R}_\mathrm{NN}$), $\mathbf{r}$ is the vector of the hopping from $\mathbf{R}_\alpha$ to $\mathbf{R}'_\beta=\mathbf{R}_\alpha+\mathbf{r}$ in the quasicrystalline lattice. Then, we transfer it into the pseudo-k space as (assuming $N_A=N_B$):
\begin{equation} \begin{split}
	H_\mathrm{hop}=&-\frac{1}{N_A}\sum_{\mathbf{r}\in\mathcal{R}}\sum_{\mathbf{k}\mathbf{k}'}\sum_{\langle\mathbf{R}_\alpha\rangle_\mathbf{r}}t_{\beta\alpha}(\mathbf{r}) e^{i(\mathbf{k}'\cdot (\mathbf{\mathbf{R}_\alpha}+\mathbf{r})-	\mathbf{k}\cdot \mathbf{\mathbf{R}_\alpha})} c_{\mathbf{k}'s,\beta}^{\dagger} c_{\mathbf{k}s,\alpha} \\
	=&-\frac{1}{N_A}\sum_{\mathbf{r}\in\mathcal{R}}\sum_{\mathbf{k}\mathbf{k}'}e^{i\mathbf{k}'\cdot \mathbf{r}}\sum_{\langle\mathbf{R}_\alpha\rangle_\mathbf{r}}t_{\beta\alpha}(\mathbf{r}) e^{i(\mathbf{k}'-\mathbf{k})\cdot
	\mathbf{\mathbf{R}_\alpha}} c_{\mathbf{k}'s,\beta}^{\dagger} c_{\mathbf{k}s,\alpha}
\end{split} \end{equation} 

\begin{figure}
    \centering
    \includegraphics[width=1\linewidth]{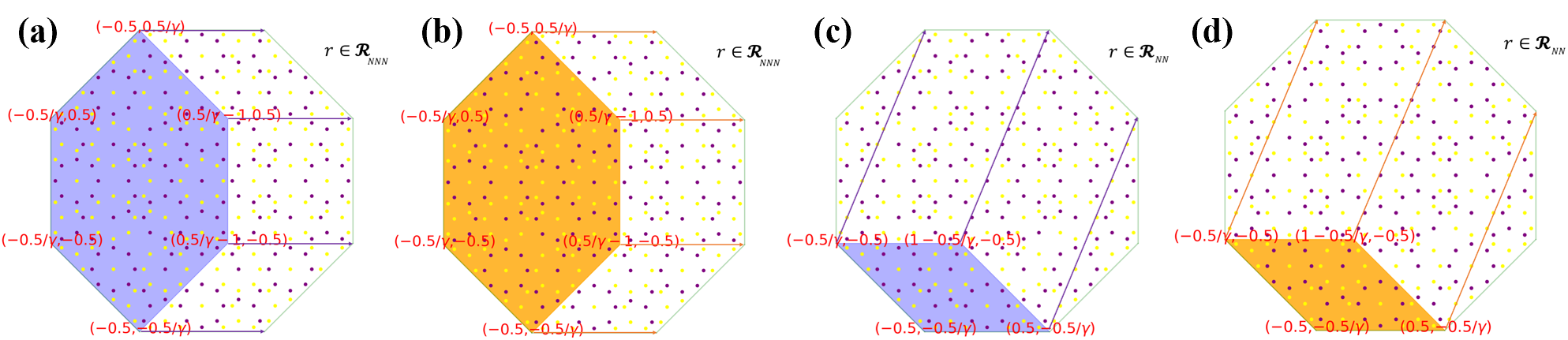}
    \caption{\label{fig:acceptance_hop} Acceptance windows for different hopping vectors $\mathbf{r}$ in the perpendicular space. For the blue area, only A (purple) vertices are selected. For the orange area, only B (yellow) vertices are selected.
	(a) (b) Hopping $t_{BA}(\mathbf{r})$ and $t_{AB}(\mathbf{r})$, where $\mathbf{r}=(1,0)$ is a NNN hopping vector.
	(c) (d) Hopping $t_{AA}(\mathbf{r})$ and $t_{BB}(\mathbf{r})$, where $\mathbf{r}=(\frac{1}{\sqrt{2}},1-\frac{1}{\sqrt{2}})\in\mathcal{R}^1_\mathrm{NN}$ is a NN hopping vector.
	}
\end{figure}

As discussed in Sec.~\ref{sec:construct}, the distribution of hopping can be easily expressed in the perpendicular space. A given hopping vector in the perpendicular space within the selection window represents the corresponding hopping in the physical space, as shown in Fig.~\ref{fig:acceptance_hop}. Consequently, the start point of this hopping is restricted in a patch within the acceptance window, which can be regarded as a new selection window with a new orthogonality \eqref{orthogonality}. Therefore, the hopping terms can be simplified:
\begin{equation} \begin{split}
	H_\mathrm{hop} \approx \sum_{\mathbf{r}\in\mathcal{R}} \sum_{\mathbf{k}} -\frac{N_{\beta\alpha}(\mathbf{r})}{N_A}t_{\beta\alpha}(\mathbf{r}) e^{i\mathbf{k}\cdot \mathbf{r}} c_{\mathbf{k}s,\beta}^{\dagger} c_{\mathbf{k}s,\alpha}\label{H_hop_approx}
\end{split} \end{equation}
where $N_{\beta\alpha}(\mathbf{r})$ denotes the number of intersite vectors in the quasicrystal lattice for the type of hopping between two sites belonging to sublattices $\alpha$ and $\beta$ and connected by vector $\mathbf{r}$.
Here $\mathcal{P}(\mathbf{r}) = \frac{N_{\beta\alpha}(\mathbf{r})}{N_A}$ is the statistical average distribution of intersite vectors in quasicrystals, which is also known as the Patterson function and can be extracted from diffraction data.

In our model, $\mathcal{P}(\mathbf{r})$ of each type of hopping can be estimated from its allowed area in perpendicular space. In the thermodynamic limit, the vertices of the quasicrystal are arranged densely and well-distributed within the selection window in the perpendicular space. Therefore, we can derive $\mathcal{P}(\mathbf{r})$ from the allowed area for the starting point of the hopping vector in the selection window, which yields:
\begin{equation} \begin{split}
	\frac{N_{\beta\alpha}(\mathbf{r})}{N_A} = \begin{cases}
		\frac{1}{2} &(\mathbf{r}\in\mathcal{R}_\mathrm{NNN}) \\
		\frac{\gamma}{2\sqrt{2}} &(\mathbf{r}\in\mathcal{R}_\mathrm{NN})
	\end{cases},
\end{split} \end{equation}
where $\gamma=\sqrt{2}-1$ is a constant.

Before proceeding, we have a few remarks on the long-wavelength approximation. Since Eq.~\eqref{orthogonality} is an approximate orthogonality, the rigorous expression of $H_\mathrm{hop}$ should include nondiagonal terms. Specifically, the hopping term is expressed as
\begin{equation} \begin{split}
	H_\mathrm{hop} = \sum_{\mathbf{r}\in\mathcal{R}}\sum_{\mathbf{k}\mathbf{k}'} -\frac{N_{\beta\alpha}(\mathbf{r})}{N_A}t_{\beta\alpha}(\mathbf{r}) e^{i\mathbf{k}\cdot \mathbf{r}} c_{\mathbf{k}'s,\beta}^{\dagger} c_{\mathbf{k}s,\alpha} \Delta(\mathbf{k},\mathbf{k}').
\end{split} \end{equation}
In Sec.~\ref{sec:orth}, we propose that $\Delta(\mathbf{k}-\mathbf{k}')$ is a sum of a set of delta functions. When the Hamiltonian is local in the pseudo-k space,
and especially when $\mathbf{k}\to 0$ (i.e., under the long-wavelength limit), we treat the delta function at the center $\delta_{\mathbf{k}-\mathbf{k}'}$ as the main contribution, while neglecting other delta functions as they are perturbations and they do not bring any new physical insights. In all, under the long-wavelength approximation, $\Delta(\mathbf{k}-\mathbf{k}')$ is approximated by a strict delta function $\delta_{\mathbf{k}-\mathbf{k}'}$ around the $\Gamma$ point.

Based on the approximate expressions in Eq.~\eqref{H_on-site} and \eqref{H_hop_approx}, we can derive the low-energy effective Hamiltonian around $\Gamma$:
\begin{equation} \begin{split}
	H = \sum_{\mathbf{k}ss'}
	\begin{pmatrix}
	c_{\mathbf{k}s',A}^{\dag} & c_{\mathbf{k}s',B}^{\dag}
	\end{pmatrix}
	\begin{pmatrix}
	\bm{N}\cdot \boldsymbol{\sigma}-\bar{t}_2f_2(\mathbf{k})+\delta_{2}\phi_{2}(\mathbf{k}) & -t_1f_1(\mathbf{k})\\
	-t_1f_1(\mathbf{k}) & -\bm{N}\cdot \boldsymbol{\sigma}-\bar{t}_2f_2(\mathbf{k})-\delta_{2}\phi_{2}(\mathbf{k})
	\end{pmatrix}_{s's}
	\begin{pmatrix}
	c_{\mathbf{k}s,A} \\
	c_{\mathbf{k}s,B}
	\end{pmatrix}, \\
\end{split} \end{equation}
where $\bar{t}_2=(t_{2a}+t_{2b})/2$ and $\delta_2 = (t_{2a}-t_{2b})/2$, and
\begin{equation} \begin{split}
	\begin{cases}
    f_{1}(\mathbf{k})=&4-k^2+\frac{k^4}{16}\\
    f_{2}(\mathbf{k})=&\frac{\gamma}{\sqrt{2}}(4-\frac{4\gamma^2}{1+\gamma^2}k^2+\frac{\gamma^4}{(1+\gamma^2)^2}k^4)\\
    \phi_{2}(\mathbf{k})=&\frac{2\sqrt{2}\gamma^5}{3(1+\gamma^2)^2}k_xk_y(k_x^2-k_y^2)
	\end{cases}.
\end{split} \end{equation}
Note that the effective Hamiltonian is valid only if $\mathbf{k}=\mathcal{S}\mathbf{H}$ is limited around the $\Gamma$ point.

Taking $\boldsymbol{\lambda}=(\lambda_x,\lambda_y,\lambda_z)$ as the Pauli matrices acting on the sublattice degree of freedom, the effective Hamiltonian can be expressed in a compact form:
\begin{equation}
{H}_\Gamma(\mathbf{k}) = -\bar{t}_2f_2(\mathbf{k})\lambda_0-t_1f_1(\mathbf{k})\lambda_x -[\boldsymbol{N}\cdot \boldsymbol{\sigma}+\delta_2\phi_2(\mathbf{k})] \lambda_z.
\end{equation}
After diagonalizing on the $\lambda$ subspace and expand the Hamiltonian
to the fourth order on $k$, we arrive at the effective Hamiltonian for the lower and higher bands are expressed as
\begin{equation} \begin{split}
	H_\mathrm{eff,\mp}=(m+tk^2-ak^4)\sigma_0 \mp (\bm{J}\cdot \boldsymbol{\sigma}) k_xk_y(k_x^2-k_y^2)+O(k^5),\\
	\begin{cases}
		m=-2\sqrt{2}\gamma t_2 \mp \sqrt{N^2+16t_1^2}\\
		t=\frac{2\sqrt{2}\gamma^3}{1+\gamma^2}t_2 \pm \frac{4t_1^2}{\sqrt{N^2+16t_1^2}}\\
		a=\frac{\gamma^5}{\sqrt{2}(1+\gamma^2)^2}t_2 \pm (\frac{3t_1^2}{4\sqrt{N^2+16t_1^2}}-\frac{8t_1^4}{(N^2+16t_1^2)^{3/2}})\\
		\bm{J} =\bm{N} \frac{\delta_2}{\sqrt{N^2+16t_1^2}}\cdot\frac{2\sqrt{2}\gamma^5}{3(1+\gamma^2)^2}
	\end{cases}.
\end{split} \end{equation}

Notably, the spin splitting effect can be directly extracted as the altermagnetic term:
\begin{equation} \begin{split}
	{H}_\mathrm{AM}^{(g-wave)}=(\bm{J}\cdot \boldsymbol{\sigma}) k_xk_y(k_x^2-k_y^2),
\end{split} \end{equation}
which describes a g-wave altermagnetism in the decorated ABT quasicrystal.

\begin{figure}
    \centering
        \includegraphics[width=0.8\linewidth]{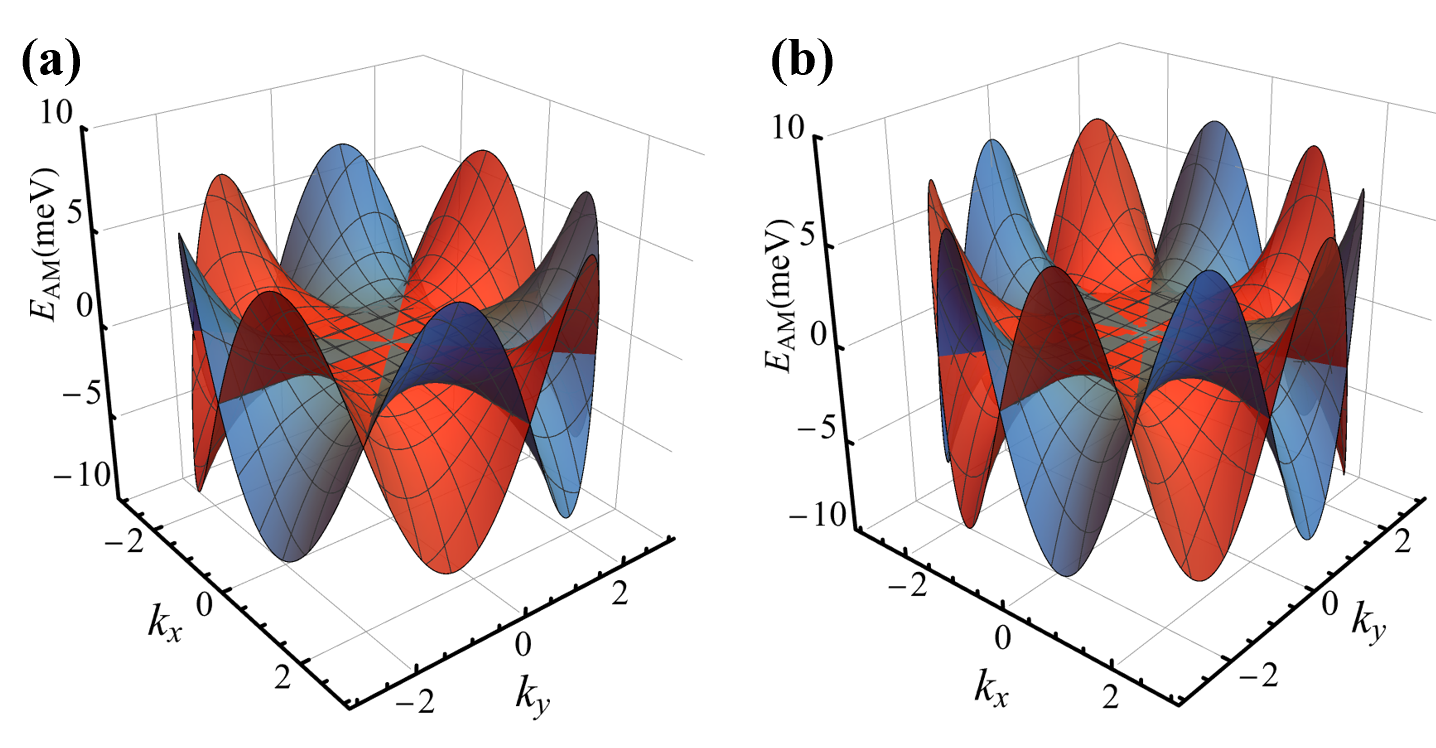}
    \caption{\label{fig:epsart} Spin-polarized eigenvalues around the $\Gamma$ point described by the AM terms (a) $\mathcal{H}_\mathrm{AM}^{(g-wave)}$ in the decorated ABT quasicrystal and (b) $\mathcal{H}_\mathrm{AM}^{(h-wave)}$ in the decorated Penrose tiling quasicrystal.}
\end{figure}

\subsubsection{Effective Hamiltonian for the decorated Penrose-tiling quasicrystal}
Next, we derive the low-energy effective Hamiltonian for the decorated Penrose-tiling quasicrystal based the a similar algebra. The Patterson function for the distribution of intersite hopping vectors in the Penrose tiling is estimated as:
\begin{equation} \begin{split}
	\frac{N_{\beta\alpha}(\mathbf{r})}{N_A} = \begin{cases}
		\frac{6}{5\tau^2} &(\mathbf{r} \in \mathcal{R}^1_\mathrm{NNN} \text{ and } \alpha = A, \beta = B;\\
		&\text{or }\mathbf{r} \in \mathcal{R}^2_\mathrm{NNN} \text{ and } \alpha = B, \beta = A)\\
		\frac{-2\sigma}{5} &(\mathbf{r} \in \mathcal{R}^1_\mathrm{NNN} \text{ and } \alpha = B, \beta = A;\\
		&\text{or }\mathbf{r} \in \mathcal{R}^2_\mathrm{NNN} \text{ and } \alpha = A, \beta = B)\\
		\frac{-2\sigma}{5\tau^2} &(\mathbf{r} \in \mathcal{R}_\mathrm{NN})
	\end{cases},
\end{split} \end{equation}
where $\tau=\frac{1+\sqrt{5}}{2}$ and $\sigma=\frac{1-\sqrt{5}}{2}$ are constants.

Under the long-wavelength approximation, the low-energy effective Hamiltonian is derived as
\begin{equation} \begin{split}
	H = \sum_{\mathbf{k}ss'}
	\begin{pmatrix}
	c_{\mathbf{k}s',A}^{\dag} & c_{\mathbf{k}s',B}^{\dag}
	\end{pmatrix}
	\begin{pmatrix}
	\bm{N}\cdot \boldsymbol{\sigma}-t_2f_{2}(\mathbf{k})+\delta_2\phi_{2} & -t_1f_1(\mathbf{k})-it_1\phi_1(\mathbf{k})\\
	-t_1f_1(\mathbf{k})+it_1\phi_1(\mathbf{k}) & -\bm{N}\cdot \boldsymbol{\sigma}-t_2f_{2}(\mathbf{k})-\delta_{2}\phi_{2}(\mathbf{k})
	\end{pmatrix}_{s's}
	\begin{pmatrix}
	c_{\mathbf{k}s,A} \\
	c_{\mathbf{k}s,B}
	\end{pmatrix}, \\
\end{split} \end{equation}
\begin{equation} \begin{split}
	\begin{cases}
	f_{1}(\mathbf{k})=\frac{2\sqrt{5}}{\tau}(1-\frac{\sigma}{\sqrt{5}})(1-\frac14 k^2+\frac{k^4}{64})\\
	\phi_{1}(\mathbf{k})=\frac{2}{384\sqrt{5}\tau}(1+\sigma)(k_x^5-10k_x^3k_y^2+5k_xk_y^4)\\
	{f_{2}(\mathbf{k})= \frac{-4\sigma}{\tau^2}(1-\frac14 \sigma^2k^2+\frac{\sigma^4k^4}{64})}\\
	{\phi_{2}(\mathbf{k})=\frac{\sigma^6}{480\tau^2}(k_x^5-10k_x^3k_y^2+5k_xk_y^4)}
	\end{cases}.
\end{split} \end{equation}

Similarly, it can be rewritten as
\begin{equation}
\begin{split}
    {H}_\Gamma(\mathbf{k})=-t_2f_{2}(\mathbf{k})\lambda_0+[\boldsymbol{N}\cdot \boldsymbol{\sigma}+\delta_2\phi_{2}(\mathbf{k})]\lambda_z
    -t_1f_1(\mathbf{k})\lambda_x+t_1\phi_1(\mathbf{k})\lambda_y.
\end{split}
\end{equation}

The effective Hamiltonian for the lower and higher bands in the decorated Penrose-tiling quasicrystal is given by:
\begin{equation} \begin{split}
	H_\mathrm{eff,\mp}=(m+tk^2-ak^4)\sigma_0 \mp (\bm{J}\cdot \boldsymbol{\sigma}) (5k_xk_y^4 - 10k_x^3k_y^2 + k_x^5)+O(k^6),\\
	\begin{cases}
		m=- \frac{4\sigma}{\tau^2} t_2 \mp \sqrt{N^2+\frac{20}{\tau^2}(1-\frac{\sigma}{\sqrt{5}})^2 t_1^2}\\
		t=-\frac{\sigma^3}{\tau^2}t_2 \pm \frac{5(1-\frac{\sigma}{\sqrt{5}})^2 t_1^2}{\tau^2 \sqrt{N^2+\frac{20}{\tau^2}(1-\frac{\sigma}{\sqrt{5}})^2 t_1^2}}\\
		a=-\frac{\sigma^5}{16\tau^2}t_2 \pm (\frac{15(1-\frac{\sigma}{\sqrt{5}})^2 t_1^2}{16\tau^2\sqrt{N^2+\frac{20}{\tau^2}(1-\frac{\sigma}{\sqrt{5}})^2 t_1^2}} - \frac{25(1-\frac{\sigma}{\sqrt{5}})^4 t_1^4}{2\tau^4 (N^2+\frac{20}{\tau^2}(1-\frac{\sigma}{\sqrt{5}})^2 t_1^2)^{3/2}})\\
		\bm{J} = \frac{\delta_2\bm{N}}{\sqrt{N^2+\frac{20}{\tau^2}(1-\frac{\sigma}{\sqrt{5}})^2 t_1^2}}\cdot\frac{\sigma^6}{480\tau^2}
	\end{cases}.
\end{split} \end{equation}

The altermagnetic term, which captures the spin splitting around the $\Gamma$ point is
\begin{equation} \begin{split}
	{H}_\mathrm{AM}^{(h-wave)}=(\bm{J}\cdot \boldsymbol{\sigma}) (5k_xk_y^4 - 10k_x^3k_y^2 + k_x^5).
\end{split} \end{equation}
Unlike the effective AM Hamiltonian in the ABT providing even parity, the spin-splitting in Penrose tiling manifests an unconventional odd parity, and yields an unconventional h-wave magnetism, which is compatible with the quasicrystalline symmetry.

It is worthwhile to note that $\bm{J}$ depends on the imaginary part of hopping $t_{2r/2b}=t_2\pm i\delta_2$ and the N\'eel magnetization $\bm{N}$. Therefore, this unconventional $h$-wave magnetism requires a complex hopping, which would otherwise vanish in the absence of $\delta_2$.

\subsection{Discussion about Orthogonality in Quasicrystals\label{sec:orth}}
In this section, we take the ABT as an example to discuss the orthogonality in quasicrystals, as mentioned in Eq.~\eqref{orthogonality}. Consider a 4D hyperlattice with width $L$ and bond length $a$. Since $L/a\gg1$ in  the thermodynamic limit, the Fourier transform takes a simple form:
\begin{equation} \begin{split}
	\int d\boldsymbol{\xi} \sum_{\{m_i\}} \delta(\boldsymbol{\xi}-\sum_{i}m_i \mathbf{e}_i) e^{-i\boldsymbol{\Pi}\cdot\boldsymbol{\xi}}=
	\frac{1}{a^4}\sum_{\boldsymbol{H}\in\mathcal{L}}\delta(\boldsymbol{\Pi}-\boldsymbol{H}),
\end{split} \end{equation}
where $\boldsymbol{\xi}\in \mathbb{R}^4$ is the 4D coordinate, $e_i$ are basis vectors of the hyperlattice, and $\mathcal{L}$ is the reciprocal hyperlattice spanned by $\{\mathbf{Q}_1,\mathbf{Q}_2,\mathbf{Q}_3,\mathbf{Q}_4\}=\pi\{(1,1,0,0),(0,0,-1,1),(-1,1,0,0),(0,0,-1,-1)\}$.

The projection of a vertex into the quasicrystal is determined by an acceptance window:
\begin{equation} \begin{split}
\mathcal{W}(\boldsymbol{\xi})=\mathcal{W}(\mathcal{S}_\perp \boldsymbol{\xi})= \begin{cases}
	1 &(\mathcal{S}_\perp \boldsymbol{\xi} \in W_\perp) \\
	0 &(\text{or else})
\end{cases}.
\end{split} \end{equation}
Separating this function into two subspaces and Fourier transforming yields:
\begin{equation} \begin{split}
\tilde{\mathcal{W}}(\boldsymbol{\Pi})= \delta^\parallel(\mathcal{S}\boldsymbol{\Pi}) \tilde{\mathcal{W}}_\perp(\mathcal{S}_\perp \boldsymbol{\Pi})=A_\parallel \delta^\parallel_{\mathcal{S}\boldsymbol{\Pi}} \cdot A_\perp \frac{\tilde{\mathcal{W}}_\perp(\mathcal{S}_\perp \Pi)}{A_\perp}.
\end{split} \end{equation}
Here $\boldsymbol{\Pi}\in \mathbb{R}^4$ is a 4D wave vector (not the reciprocal vectors defined earlier). $W_\perp$ is the acceptance window in perpendicular space, and $A_\parallel$ and $A_\perp$ represent the ``volume" of the quasicrystal in physical and perpendicular spaces, respectively. In the thermodynamic limit, $A_\parallel \to \infty$, while $A_\perp$ corresponds to the area of the acceptance window. In the ABT, $A_\parallel$ is the total area of the quasicrystal in physical space, and $A_\perp$ is the area of the octagonal acceptance window.

The Fourier transform of the acceptance window in perpendicular space is:
\begin{equation} \begin{split}
\tilde{\mathcal{W}}_\perp(\mathcal{S}_\perp \Pi) =\tilde{\mathcal{W}}_\perp(\boldsymbol{\Pi}_\perp)=\int_{\boldsymbol{\xi}_\perp \in W_\perp} d\boldsymbol{\xi}_\perp e^{-i\boldsymbol{\Pi}_\perp\cdot \boldsymbol{\xi}_\perp}.
\end{split} \end{equation}
Figure~\ref{fig:orth_discuss} shows the behavior of $\frac{|\tilde{\mathcal{W}}_\perp(\boldsymbol{\Pi}_\perp)|}{A_\perp}$. We estimate that this function decays inversely linearly, i.e., $|\tilde{\mathcal{W}}_\perp(\boldsymbol{\Pi}_\perp)| \sim |\boldsymbol{\Pi}_\perp a|^{-1}$.

\begin{figure}
    \centering
    \includegraphics[width=7cm]{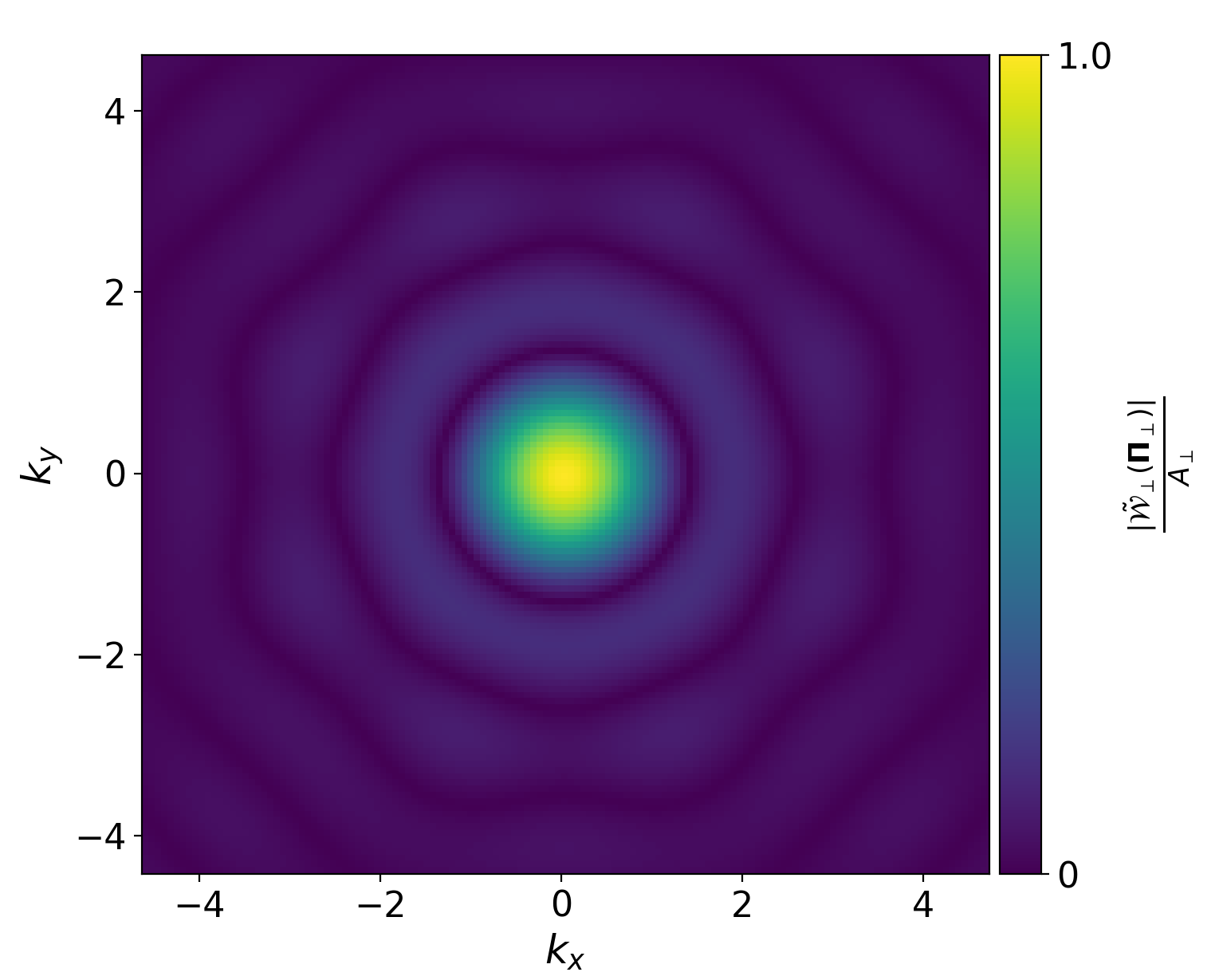}
    \caption{\label{fig:orth_discuss} Distribution of $\frac{|\tilde{\mathcal{W}}_\perp(\boldsymbol{\Pi}_\perp)|}{A_\perp}$ for the regular octagon acceptance window in ABT.
	}
\end{figure}

The term we wish to derive is:
\begin{equation} \begin{split}
\sum_{\mathbf{R}_i} e^{-i\mathbf{k}\cdot\mathbf{R}_i}=\int d\boldsymbol{\xi} e^{-i(\mathbf{k}^T,0)\cdot\boldsymbol{\xi}} \sum_{\{m_i\}} \delta(\boldsymbol{\xi}-\sum_{i}m_i \mathbf{e}_i) \mathcal{W}(\boldsymbol{\xi}).
\end{split} \end{equation}
Applying the convolution theorem, the Fourier transform of the quasicrystal screened by an arbitrary window is a sum of a countable set of delta functions with intensities modulated by $\mathcal{S}_\perp \mathbf{H}$:
\begin{equation} \begin{split}
\sum_{\mathbf{R}_i} e^{-i\mathbf{k}\cdot\mathbf{R}_i}&=\int d\boldsymbol{\pi} \frac{1}{a^4} \sum_{\boldsymbol{H}\in\mathcal{L}} \delta(\boldsymbol{\pi}-\boldsymbol{H}) \tilde{\mathcal{W}}((\mathbf{k},0)-\boldsymbol{\pi}) \\
&=\frac{A_\parallel A_\perp}{a^4}\int d\boldsymbol{\pi} \sum_{\boldsymbol{H}\in\mathcal{L}} \delta(\boldsymbol{\pi}-\boldsymbol{H}) \delta^\parallel_{\mathbf{k},\mathcal{S}\boldsymbol{\pi}} \frac{\tilde{\mathcal{W}}_\perp(\mathcal{S}_\perp ((\mathbf{k},0)-\boldsymbol{\pi}))}{A_\perp}\\
&=N\sum_{\boldsymbol{h} \in \mathcal{L}} \delta^\parallel_{\mathbf{k}-\mathcal{S}\boldsymbol{h}} \frac{\tilde{\mathcal{W}}_\perp(-\mathcal{S}_\perp\boldsymbol{h})}{A_\perp},
\end{split} \end{equation}
where $N$ is the number of vertices in the ``quasicrystal". For example, $N=N_A=N_B$ for A or B vertices in the initial acceptance window, $N=N(\mathbf{r}\in\mathcal{R}_\mathrm{NNN})$ for A/B vertices in the window of Fig.~\ref{fig:acceptance_hop}(a) and~\ref{fig:acceptance_hop}(b), and $N=N(\mathbf{r}\in\mathcal{R}_\mathrm{NN})$ for A/B vertices in the window of Fig.~\ref{fig:acceptance_hop}(c) and~\ref{fig:acceptance_hop}(d).

For $\mathcal{S}\mathbf{h}\ll1/a$, we have $\mathcal{S}_\perp\mathbf{h} \gg 1/a$. Intuitively, if we constrain $\mathcal{S}\mathbf{h}$ within a small radius $r$ in physical space, the hyperlattice vertices satisfying this condition become increasingly sparse as $r\rightarrow 0$ due to the irrational nature of quasicrystals.
Thus, for $\mathbf{h}\neq0$, $\mathcal{S}\mathbf{h}\to 0$ implies $\mathcal{S}_\perp\mathbf{h} \to \infty$.

Consequently, for Hamiltonians localized in pseudo k-space (e.g. $\mathbf{k}$ near the $\Gamma$ point), we obtain an approximate orthogonality relation:
\begin{equation} \begin{split}
\sum_{\mathbf{R}_i} e^{-i\mathbf{k}\cdot\mathbf{R}_i}\simeq N\delta^\parallel_\mathbf{k},
\end{split} \end{equation}
where we neglect higher-order delta functions, as they contribute only perturbatively or lack new physical significance.

\section{Higher-Order Topology based on Altermagnetic Quasicrystal}
Furthermore, we show that interfacing these quasicrystalline altermagnets with topological insulators or superconductors yields higher-order topological phases featuring symmetry-matched corner modes.

\begin{figure}
    \centering
        \includegraphics[width=0.85\linewidth]{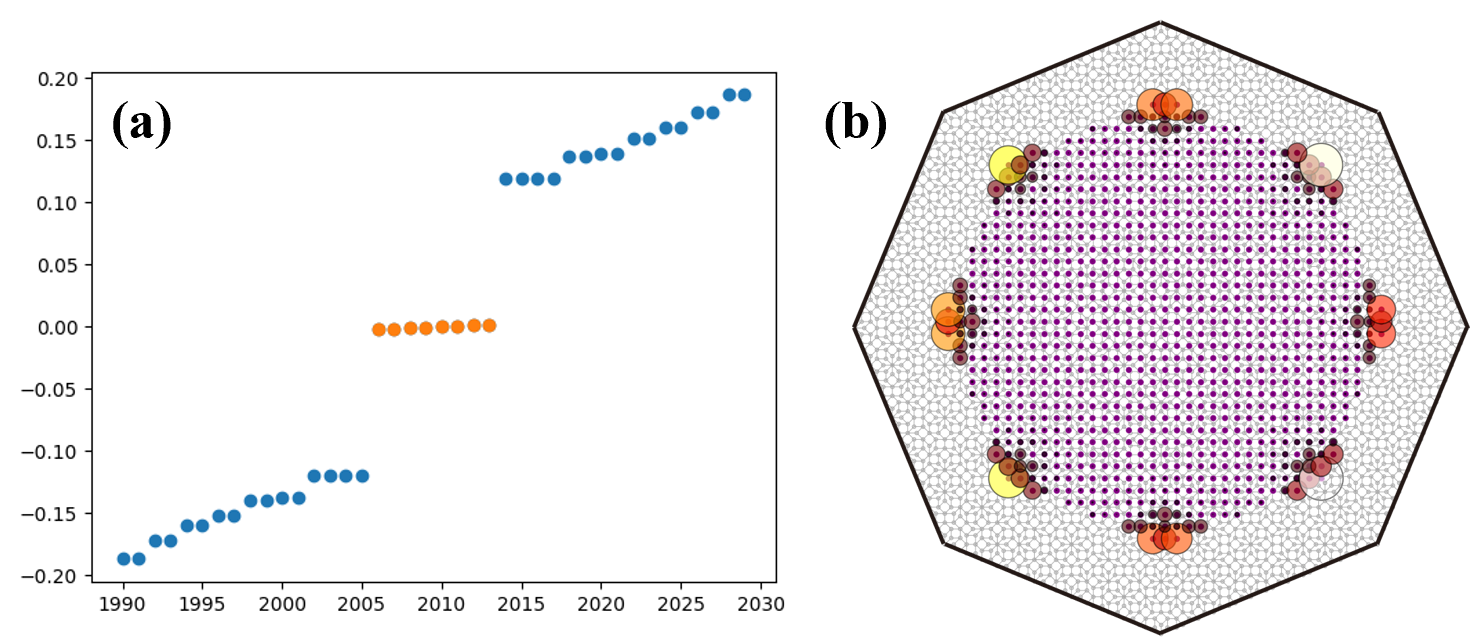}
    \caption{\label{fig:HOTI} (a) Eigenvalues of energy for HOTI in a finite octagonal sample with system size $N=1005$. The eight corner modes at $E=0 $ are marked in orange. (b) Spatial distribution of 8 corner modes for the heterostructure consisting of a 2D topological insulator (in a square lattice) and an octagonal altermagnet (in ABT quasicrystalline lattice). The finite octagonal sample contains $N=1005$ sites for the 2D TI in the square lattice. Parameters: $t=1, m=1, A=0.25, J_0=0.05$.
    }
\end{figure}

\subsection{Topological Corner Modes in Altermagnetic Quasicrystal and TI Heterostructure}
We study the route to higher-order topological insulators via the proximity effect \cite{PhysRevLett.133.106601}.
Specifically, we construct a heterostructure comprising a 2D topological insulator (TI) and our proposed quasicrystalline altermagnet. The effective Hamiltonian around the $\Gamma$ point is $\mathcal{H}_\mathrm{eff}=\mathcal{H}_\mathrm{2D-TI}+\mathcal{H}_\mathrm{AM}$:
\begin{equation} \begin{split}
\mathcal{H}_\mathrm{2D-TI}= &(m+tk^2)\kappa_x+A(k_x\sigma_x+k_y\sigma_y)\kappa_z,\\
\mathcal{H}_\mathrm{AM}=&J_0k_xk_y(k_x^2-k_y^2)\sigma_z\kappa_z,
\end{split} \end{equation}
where $\kappa_i$ and $\sigma_i$ are Pauli matrices in orbital and spin spaces, respectively.

To derive the edge theory, we rotate the coordinate system clockwise by an angle $\alpha$, enabling analysis of the dispersion along an arbitrary tangential boundary $L(\alpha)$. The transformation is expressed as
\begin{equation} \begin{split}
	\begin{pmatrix}
	k_x \\ k_y
	\end{pmatrix}
	=
	\begin{pmatrix}
	\cos \alpha & -\sin \alpha\\
	\sin \alpha & \cos \alpha
	\end{pmatrix}
	\begin{pmatrix}
	k_\parallel \\
	k_\perp
	\end{pmatrix}, \\
\end{split} \end{equation}
The transformed Hamiltonian becomes $\mathcal{H}_\mathrm{eff}(k_\parallel,k_\perp) =\mathcal{H}_0+\mathcal{H}_p$:
\begin{equation} \begin{split}
	\mathcal{H}_0 = &M(k_\parallel,k_\perp)\kappa_x+Ak_\perp(-\sin\alpha \sigma_x+\cos\alpha \sigma_y)\kappa_z,\\
	\mathcal{H}_p = &J_0k_\perp^4\sin\alpha\cos\alpha(\cos^2\alpha-\sin^2\alpha)\sigma_z\kappa_z\\
	&+Ak_\parallel(\sin\alpha \sigma_y+\cos\alpha \sigma_x)\kappa_z+O(J_0k_\parallel).
\end{split}  \label{transformation} \end{equation}
Here, we have decomposed the Hamiltonian into an unperturbed part $\mathcal{H}_0$ and a perturbation $\mathcal{H}_p$, which is reasonable for a weak strength of the altermagnetism.

For a semi-infinite plane $x_\perp \in (-\infty, 0]$ with a boundary at $x_\perp = 0$, we replace $k_\perp$ with $-i\partial_\perp$. The zero-energy solutions of $H_0$ at $E = 0$ are $$\psi_i(x_\perp) = \mathcal{N}_\perp \sin(\kappa_1x_\perp)e^{\kappa_2x_\perp}e^{ik_\parallel x_\parallel}\xi_i,$$
where $\mathcal{N}_\perp$ is given by $|\mathcal{N}_\perp|^2 = 4|\kappa_2(\kappa_1^2+\kappa_2^2)/\kappa_1^2|$ and the spinor part of the eigenvector $\xi_i$ satisfies $(-\sin \alpha \sigma_x + \cos \alpha \sigma_y)\kappa_y \xi_i = \xi_i$. We choose the basis:
$$\xi_1 = \frac{1}{2}(1,ie^{i\alpha},i,-e^{i\alpha})^T,\quad \mathrm{and}\quad
\xi_2 = \frac{1}{2}(1,-ie^{i\alpha},-i,-e^{i\alpha})^T.$$
Projecting $\mathcal{H}_0$ and $\mathcal{H}_p$ onto the basis $\psi_1,\psi_2$, yields the edge Hamiltonian:
\begin{equation} \begin{split}
	\mathcal{H}_{edge}(x_\perp,k_\parallel)=Ak_\parallel \eta_z + M_z\eta_x,
\end{split} \end{equation}
where $M_z(\alpha) \sim \sin 4\alpha$ is the perturbation term induced mass term. For a finite octagonal sample, the $g$-wave altermagnetism induces a sign-changing Dirac mass along the edges, creating eight domain walls and corresponding corner modes via the Jackiw-Rebbi mechanism \cite{PhysRevD.13.3398}. Figure~\ref{fig:HOTI} shows the calculated energy spectrum and real-space distribution of the corner modes, confirming the higher-order topology.

For Penrose tiling, analogous calculations give $M_z(\alpha) \sim \sin 5\alpha$, leading to 10 corner modes in an $h$-wave altermagnet.

\begin{figure}
    \centering
        \includegraphics[width=1\linewidth]{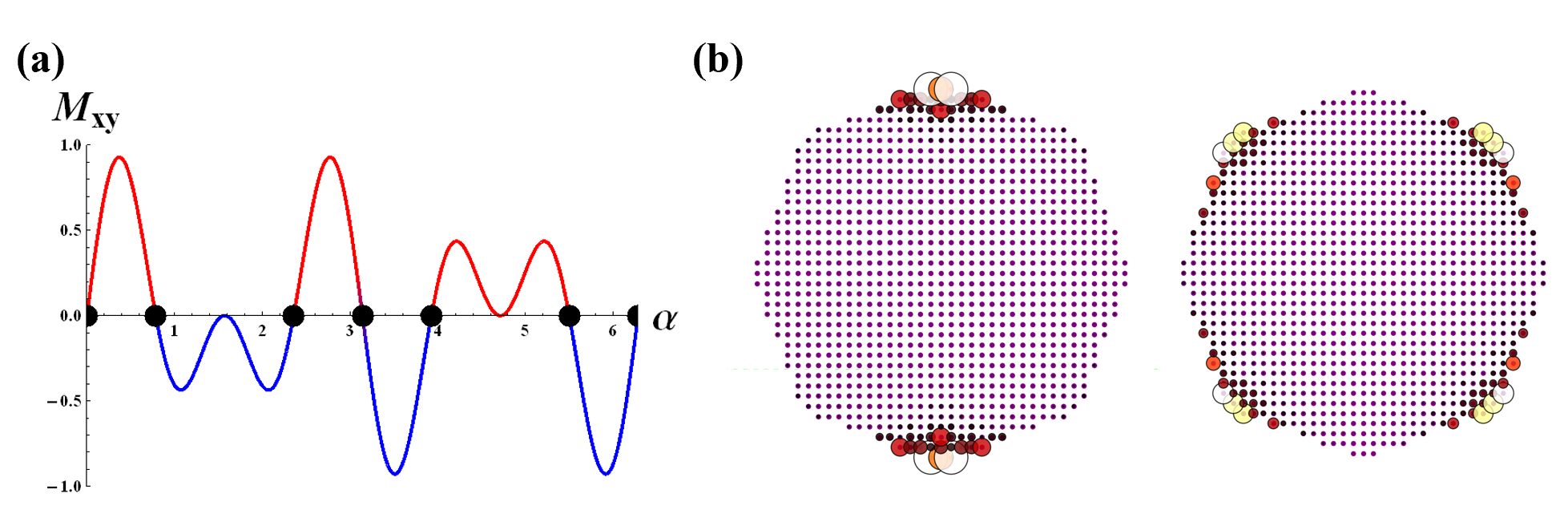}
    \caption{\label{fig:HOTSC} (a) Mass term $M_{xy}$ in Eq.~\eqref{Mxy} as a function of $\alpha$ in the presence of a in-plane N\'eel vector ($\phi=0,\theta=\pi/2$). Six black dots indicate domain-well solutions, corresponding to corner states. (b) Two of 6 MCMs in HOTSC with system size of $N=1005$ for the TSC in the square lattice. Parameters: $\Delta_0=0.5, t=1, m=1, J_0=0.05$.
	}
\end{figure}

\subsection{Majorana Corner Modes in Quasicrystalline Altermagnet-TSC heterostructure}
Tunable higher-order topological superconductors (HOTSCs) can be engineered by coupling TSCs to different magnets \cite{PhysRevB.109.224502, PhysRevLett.124.227001, PhysRevLett.121.096803,PhysRevLett.123.167001}.
Here, we propose a scheme to realize Majorana corner modes (MCMs) by proximitizing a class DIII TSC with our proposed quasicrystalline altermagnet.

The Bogoliubov-de Gennes (BdG) Hamiltonian for a 2D $p \pm ip$ superconductor is expressed as
$$H=\sum_{k}\frac{1}{2}\Psi^\dagger_\mathbf{k}\mathcal{H}(\mathbf{k})\Psi_\mathbf{k},\quad \Psi=(c_{\mathbf{k},\uparrow},c_{\mathbf{k},\downarrow},c^\dagger_{-\mathbf{k},\downarrow},-c^\dagger_{-\mathbf{k},\uparrow}),$$
with
\begin{equation} \begin{split}
	\mathcal{H}(\mathbf{k})=\epsilon(\mathbf{k})\tau_z-2\Delta_0(\sin k_x\sigma_x-\sin k_y\sigma_y)\tau_x,
\end{split} \end{equation}
where $\epsilon(\mathbf{k}) = (\mu - 2t \cos k_x - 2t \cos k_y)$, and $\tau_i, \sigma_j$ are Pauli matrices acting on the particle-hole and spin $(\uparrow, \downarrow)$ space, respectively.

The altermagnet introduces spin-splitting with momentum-dependent sign reversal:
\begin{equation} \begin{split}
	\mathcal{H}_\mathrm{AM}=J_0k_xk_y(k_x^2-k_y^2)(\boldsymbol{\sigma} \cdot \mathbf{n})\tau_z,
\end{split} \end{equation}
where $\mathbf{n} = (\sin \theta \cos \phi, \sin \theta \sin \phi, \cos \theta)$ represents the direction of the N\'{e}el vector, and $J_0$ indicates the strength of spin splitting.

In this proximitized system, the effective Hamiltonian around the $\Gamma$ point is
\begin{equation} \begin{split}
	\mathcal{H}_\mathrm{eff}=(m+tk^2)\cdot \tau_z-2\Delta_0(k_x\sigma_x-k_y\sigma_y)\tau_x
	+J_0k_xk_y(k_x^2-k_y^2)(\boldsymbol{\sigma} \cdot \mathbf{n})\tau_z,
\end{split} \end{equation}

In order to construct an effective edge theory, we rotate the coordinate system clockwise by an angle $\alpha$, as displayed in Eq.~\eqref{transformation}. For weak altermagnetism, we decompose the Hamiltonian $\mathcal{H}_\mathrm{eff}(k_\parallel,k_\perp)=\mathcal{H}_0+\mathcal{H}_p$:
\begin{equation} \begin{split}
	\mathcal{H}_0 = &M(k_\parallel,k_\perp)\tau_z+2\Delta_0k_\perp(\sin\alpha \sigma_x+\cos\alpha \sigma_y)\tau_x\\
	\mathcal{H}_p = &J_0k_\perp^4\sin\alpha\cos\alpha(\cos^2\alpha-\sin^2\alpha)(\boldsymbol{\sigma}\cdot\mathbf{n})\tau_z\\
	&+2\Delta_0k_\parallel(\sin\alpha \sigma_y-\cos\alpha \sigma_x)\tau_x+O(J_0k_\parallel).
\end{split} \end{equation}

Analogously, we consider a semi-infinite plane $x_\perp \in (-\infty, 0]$ with a boundary at $x_\perp = 0$. Two solutions of $H_0$ at zero energy are obtained as
$\psi_i(x_\perp) = \mathcal{N}_\perp \sin(\kappa_1x_\perp)e^{\kappa_2x_\perp}e^{ik_\parallel x_\parallel}\xi_i$. The spinor part $\xi_i$ satisfies
$(\sin \alpha \sigma_x + \cos \alpha \sigma_y)\tau_y \xi_i = \xi_i$. We choose the basis $\xi_i$ as $\xi_1 = \frac{1}{\sqrt{2}}(e^{i\alpha},0,0,-1)^T$
and $\xi_2 = \frac{1}{\sqrt{2}}(0,1,e^{i\alpha},0)^T$.
Upon projecting $\mathcal{H}_0$ and $\mathcal{H}_p$ onto the basis $\psi_1,\psi_2$, we obtain the edge Hamiltonian:
\begin{equation} \begin{split}
	\mathcal{H}_{edge}(x_\perp,k_\parallel)=(2\Delta_0k_\parallel+M_z) \eta_z + M_{xy}\eta_x.
\end{split} \end{equation}
The Dirac mass terms induced by the altermagnet are:
\begin{equation} \begin{split}
	M_{xy}(\alpha,\theta,\phi) &\sim \sin\theta \sin 4\alpha \cos(\alpha+\phi),\\
	M_{z}(\alpha,\theta,\phi) &\sim \cos \theta \sin 4\alpha.\label{Mxy}
\end{split} \end{equation}
which vary with boundary direction $\alpha$ and the direction of the N\'eel vector direction.
The sign reversal of $M_{xy}$ leads to the formation of bound states that are akin to the Jackiw-Rebbi mode. For in-plane N\'{e}el vector ($\theta=\pi/2$), $M_{xy}$ drives sign reversals, creating MCMs tunable via $\phi$.

For out-of-plane alignment ($\theta = 0$), only $M_z$ remains, preserving $C_{8z}\mathcal{T}$ symmetry but yielding trivial edge states. To realize symmetry-protected MCMs, alternative proximitization schemes are needed.

\section{Possible Experimental Realization of the Decoration-Induced Altermagnetism}

{In the main text, we have mathematically proposed a decoration structure based on the hyperspace projection method.
In this section, we first revisit the geometric role of the ``decoration'' in our model, and then outline a feasible experimental route to reproduce the same sublattice-symmetry-breaking pattern in real quasicrystalline systems.}

{In our model, the decoration simultaneously breaks both the eightfold rotational symmetry ($C_{8z}$) and the antiferromagnetic time-reversal symmetry ($\mathcal{T}$), while preserving their combined operation ($C_{8z}\mathcal{T}$), in the sense of ``indistinguishability''.
This preserved composite symmetry gives rise to a momentum-direction-dependent spectral splitting, a hallmark of the altermagnetic phase.}

{Geometrically, the decoration sites form a pattern consistent with the Ammann–Beenker (AB) tiling [Fig.~\ref{fig:r1}(a)], where the discrepancy between $t_{2r}$ and $t_{2b}$  originates from inequivalent local bonding environments. Specifically, $t_{2b}$  corresponds to bonds obstructed by pairs of decoration sites, while $t_{2r}$ bonds remain unobstructed. This modulation persists even when one of the paired decoration sites is removed [Fig.~\ref{fig:r1}(b)], without violating the octagonal tiling rule. Hence, the decoration pattern itself constitutes an AB-type quasicrystal lattice, suggesting that the proposed symmetry-breaking pattern is experimentally realizable through the following two routes.}

{\paragraph{AFM quasicrystal on a nonmagnetic quasicrystalline substrate.}
A heterostructure comprising an antiferromagnetically ordered octagonal quasicrystal placed on a nonmagnetic quasicrystalline substrate twisted by $\pi$/8 [Fig. ~\ref{fig:r1}(c)] naturally produces the required modulation in hopping amplitudes. The substrate acts as an effective ``decoration layer,'' providing the local quasiperiodic potential that modulates $t_{2r}$ and $t_{2b}$. Such heterostructures could be realized through van der Waals stacking of quasiperiodic layers, analogous to experimentally achieved twisted bilayer and quasicrystalline systems. Two-dimensional quasicrystal with eightfold rotational axis bas been experimentally discovered in rapidly solidified V–Ni–Si and Cr–Ni–Si alloys ~\cite{PhysRevLett.59.1010}, demonstrating the feasibility of octagonal quasiperiodic geometries. Moreover, long-range antiferromagnetic order has been observed in icosahedral quasicrystals ~\cite{s41567-025-02858-0, jacs.1c09954}, confirming the compatibility between antiferromagnetism and quasiperiodicity. Therefore, it is plausible that AFM order could emerge in octagonal quasicrystals, enabling the realization of a $C_{8z}\mathcal{T}$-preserving altermagnetic phase via interfacial modulation of hopping amplitudes.}

\begin{figure}
    \centering
        \includegraphics[width=1\linewidth]{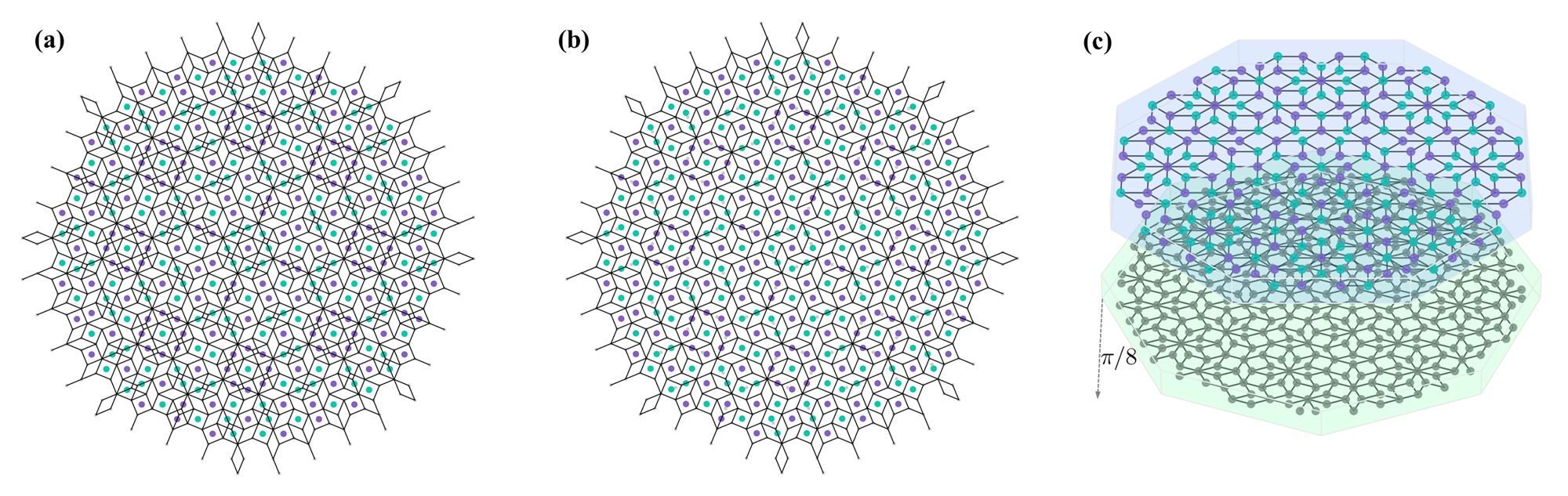}
    \caption{\label{fig:r1} Quasicrystalline pattern generated by the decoration sites. (a) The decoration pattern used in the manuscript. (b) After removing the sites that violate the octagonal tiling rule, the pattern reduces to an AB-tiling quasicrystal. (c) Illustration of the twisted quasicrystal system, comprised of a AFM layer (blue) and a non-magnetic layer with a $\pi$/8 twist (green).
	}
\end{figure}

{\paragraph{AFM octagonal quasicrystal sandwiched between two bipartite square lattices.}
Previous studies have shown that octagonal quasicrystals can form in twisted bilayer square-lattice systems with a relative twist of $\pi$/4 ~\cite{PhysRevB.103.045408, PhysRevB.106.075121, Jin:20, PhysRevB.111.195161, PhysRevB.107.235425, PhysRevB.107.144202, peng2018effect}. We propose a trilayer heterostructure in which an AFM octagonal quasicrystal is sandwiched between two nonmagnetic bipartite square lattices twisted by $\pi$/8 and $-\pi$/8, respectively [Fig.~\ref{fig:r2}(a)]. The two atomic species in the square lattices (marked in red and blue) possess different local potentials. As illustrated in Fig.~\ref{fig:r2}(b), the top view of this configuration shows that each $t_{2r}$ ($t_{2b}$) bond is approximately obstructed by a red (blue) atom, leading to a systematic modulation of inter-sublattice hopping amplitudes via the proximity effect. Such trilayer systems can be experimentally realized using rock-salt-structured materials (e.g., NaCl~\cite{peng2018effect}, PbS, SnSe, or SnTe) as substrates or capping layers with tunable lattice constants. Although the match to an ideal quasicrystalline tiling is approximate, this configuration offers a practical route to realizing quasicrystalline altermagnetism protected by $C_{8z}\mathcal{T}$ symmetry in the absence of $C_{8z}$ and $\mathcal{T}$ individually.}

{In both realization schemes, $C_{8z}$ and $P\mathcal{T}$ symmetries are broken, while $C_{8z}\mathcal{T}$ remains preserved in the sense of indistinguishability, ensuring the emergence of altermagnetic order in quasicrystals.}
\begin{figure}
    \centering
        \includegraphics[width=0.8\linewidth]{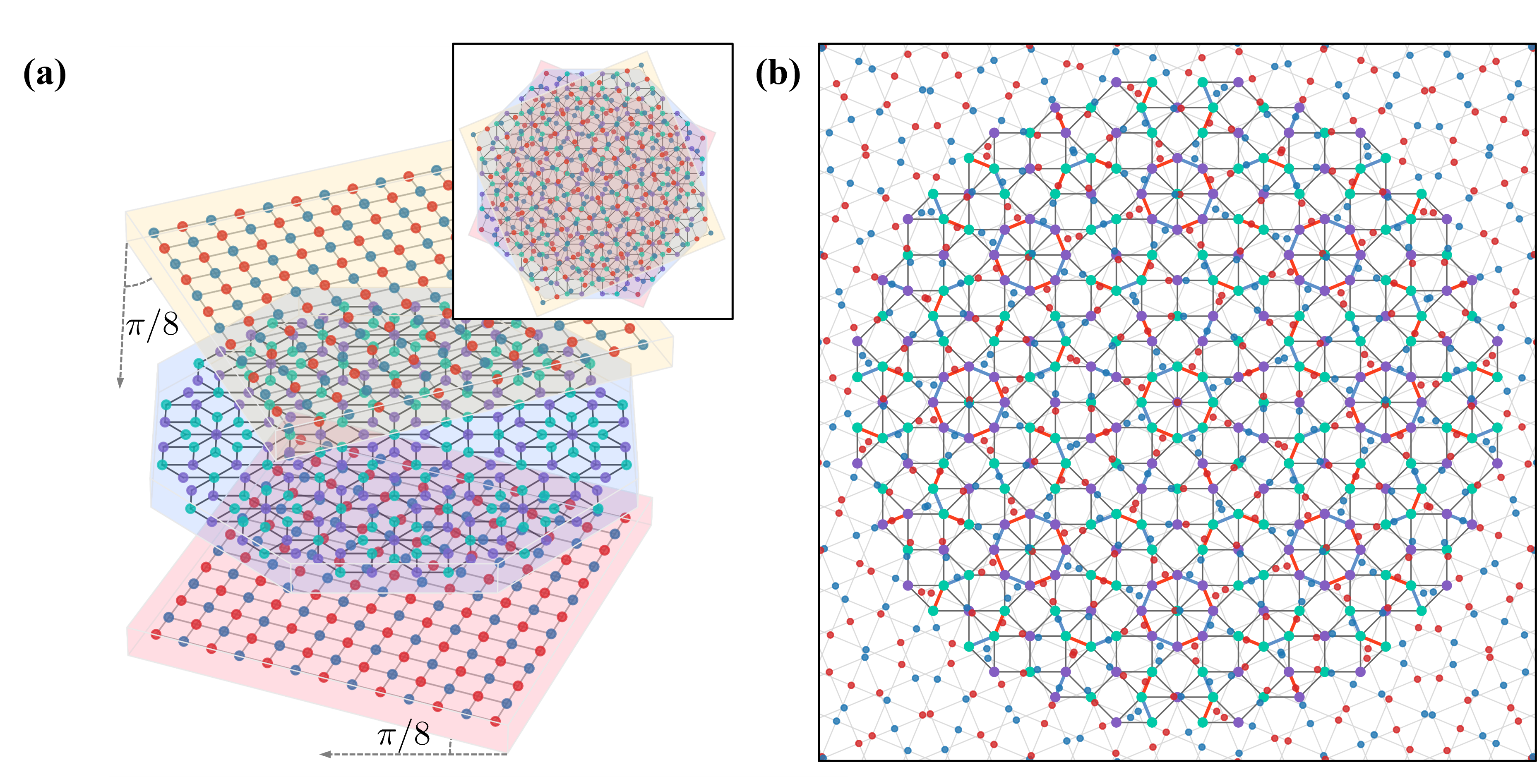}
    \caption{\label{fig:r2} (a) Illustration of the twisted trilayer system, consisting of an AFM octagonal quasicrystal (blue layer) sandwiched between two bipartite nonmagnetic layers twisted by angles ±$\pi$/8. The inset shows the top view of the system. (b) Schematic correspondence between the hopping discrepancy and the twist-induced lattice decoration: each red (blue) bond corresponds to a red (blue) atom in the twisted bipartite square lattice.
	}
\end{figure}

{Beyond solid-state materials, the same symmetry-breaking mechanism could also be explored in artificial materials such as optical-lattice systems. Recent cold-atom experiments have demonstrated long-range AFM order in the Hubbard model~\cite{xu2025neutral,Jagannathan_2013,Corcovilos:19,cryst6100124,mazurenko2017cold,doi:10.1126/science.1239873}, as well as the synthesis of artificial quasicrystalline metal–organic coordination networks~\cite{urgel2016quasicrystallinity, SHI20242464}. These advances suggest that our proposed altermagnetic quasicrystal model could also be implemented in highly controllable synthetic or quantum-simulation platforms. By engineering an eightfold quasiperiodic potential with a controlled imbalance in tunneling amplitudes ($t_{2r} \neq t_{2b}$), one could emulate the essential ingredients of our model, providing a quantum-simulation platform to investigate the $C_{8z}\mathcal{T}$-preserving altermagnetic phase. Such an implementation would not only validate the proposed mechanism, but also enable tunable exploration of altermagnetic phenomena in a controlled quasiperiodic environment.}

\end{widetext}

\providecommand{\noopsort}[1]{}\providecommand{\singleletter}[1]{#1}%

\end{document}